\documentclass{article}
\usepackage{graphicx} % Required for inserting images.
\usepackage[margin=1in,letterpaper]{geometry} % Decreases margins.
\usepackage{authblk} % Multiple authors.
\usepackage[backend=biber, style=ieee]{biblatex} %style=alphabetic
\usepackage{amsmath}
\usepackage{alphalph}
\usepackage{subcaption}
\usepackage{xcolor}
\definecolor{myPink}{HTML}{FFC0CB}
\definecolor{myBlue}{HTML}{1F77B4}
\usepackage{nicematrix}

\usepackage{lineno}
\allowdisplaybreaks

\usepackage{alphalph}

\makeatletter
\@ifundefined{c@subequation}{\newcounter{subequation}}{}
\@ifundefined{c@parentequation}{\newcounter{parentequation}}{}

\newenvironment{SafeSubequations}{%
  \refstepcounter{equation}%
  \setcounter{subequation}{0}%
}{%
  \setcounter{subequation}{0}%
}
\newcommand{\nextsbeq}{\refstepcounter{subequation}}
\makeatother

\providecommand{\keywords}[1]
{
  \small	
  \textbf{\textit{Keywords---}} #1
}
\graphicspath{{Images/}}

\begin{document}
% \linenumbers

% ------- TITLE AND AUTHOR INFORMATION -------

\title{\textbf{Minimum Carbon Trusses: Constructible Multi-Component Designs with Mixed-Integer Linear Programming}}

\author[1]{Zane Hallowell Schemmer}
\author[1]{Josephine Voigt Carstensen}
\affil[1]{Department of Civil and Environmental Engineering, Massachusetts Institute of Technology}
\date{June 27, 2025}

\maketitle

\begin{abstract} \label{sec:Abstract}
Truss optimization is a rich research field receiving renewed interest in limiting the carbon emissions of construction.
However, a persistent challenge has been to construct highly optimized and often complex designs.
This contribution formulates and solves new mixed-integer linear programs that enable consideration of the interplay between environmental impact and constructability.
Specifically, the design engineer is enabled to design with multiple materials and/or structural components, apply separate minimum and maximum cross-sectional area bounds, and constrain the complexity of the structural connections.
This is done while explicitly considering compatibility and constitutive laws.
The results demonstrate that the lowest embodied carbon designs change significantly when constructability constraints are applied, highlighting the need for an integrated optimization approach.
In one example, introducing a lower-carbon material option has almost no effect on the environmental performance, whereas another sees an improvement of nearly 29\%.
The extensibility of the formulation to design with three component types and additional constraints is demonstrated for a prestressed tensegrity example.
\end{abstract}


\section{Introduction}
\label{sec:Introduction}

The world needs to reduce greenhouse gas emissions by 43\% by 2030 to limit the global average temperature increase to the $1.5^\circ$C set forth by the Paris Agreement \cite{UNEP2023}. 
In 2022, buildings accounted for around 37\% of total global energy and process-related carbon emissions, increasing by around 9.8 GtCO$_2$ and exceeding the 2018 peak \cite{UNEP2024}.
Consequently, the carbon reduction targets for the building and construction sector are currently classified as ``more efforts needed'' \cite{IEA2023}.
The building-related CO$_2$ emissions includes emissions from the building’s use, also known as \emph{operational carbon}, and from the building’s materials, commonly referred to as the \emph{embodied carbon}.
Embodied carbon emissions stem from the production, transportation, installation, maintenance, and disposal of building materials.
While the operational carbon of buildings in the US has decreased in recent years due to increased awareness and energy performance standards, significant changes are yet to be seen for embodied carbon emissions \cite{de2017measuring}. 
In 2022, the global production of construction materials accounted for 7--9\% of the total carbon emissions summing up to 3.7 GtCO$_2$ \cite{UNEP2024}.
Since structural elements typically make up the largest portion of the embodied carbon in a building \cite{kaethner2012embodied}, several studies on embodied carbon mitigation strategies highlight the potential associated with reducing the consumption of structural materials by increasing the use of structural optimization \cite{pomponi2016embodied,shi2013performance,hakkinen2015reducing,fang2023reducing,alaux2024whole}.

When using structural optimization, the largest design benefits are typically seen when using automated approaches, where the design problem is cast as a formal optimization problem and solved using a mathematical program.
This can be done using one of the three categories of structural optimization; namely, sizing-, shape-, and topology optimization \cite{bendsoe2013topology}. 
Topology optimization typically offers the most significant performance improvements since, unlike size- and shape optimization, it does not demand the design engineer to predetermine a layout for the final design \cite{EO01}.
It only demands the designer to define the space the structure is allowed to occupy, along with relevant loads and boundary conditions.
Within the iterative process of the mathematical program, the design's performance is evaluated numerically.
Topology optimization approaches exist for both discrete (truss or frame elements) and continuum finite element discretizations \cite{sigmund2013topology}, where truss or frame element topology optimization is sometimes also referred to as layout optimization \cite{bendsoe1994optimization,rozvany1995layout}. 

A recent white paper published by the Structural Engineering Institute's Sustainability Committee under the American Society of Civil Engineers explicitly suggests increased use of topology optimization as a way to avoid over-design and reduce construction material consumption \cite{webster2020achieving}.
This recommendation comes after a recent increase in research that applies topology optimization to achieve more environmentally friendly structures \cite{li2017novel,wang2024topology,zhang2024material}. 
For buildings and civil infrastructure, examples include experimental works that reduce the concrete consumption of reinforced concrete beams, demonstrating an environmental savings potential of 25-30\% \cite{liu2020experimental,wethyavivorn2022topology,pressmair2023contribution}.
These works predominately rely on the reduction of the structural weight of concrete to reduce the embodied carbon of the structure.
Although this is straightforward when reducing the weight of a single material, the interplay between environmental benefit, constructability constraints, and structural performance is significantly more involved for multi-material structures.
Consequently, structural optimization that addresses the material selection problem is gaining interest \cite{chandrasekhar2022integrating,lee2024differentiable}.
Although the topology optimization literature on multi-material design is rich, relatively few frameworks have been developed to facilitate efficient design for low-embodied carbon structures.
Ching and Carstensen \cite{CC22} consider material selection in topology optimization when explicitly minimizing the environmental impact of timber-steel trusses. 
Kundu and Zhang \cite{kundu2025sustainability} suggest a continuum topology optimization approach to minimize the sum of the economic and environmental impact of steel and biomass structures.
Both of these existing works are subject to stress constraints, design with two candidate materials and solve the design problem using gradient-based optimization.
Although carbon savings are demonstrated numerically, the resulting structures in both cases are complex and difficult to realize.
This is despite the fact that it is well recognized that constructability concerns pose a major barrier in the adoption of structural optimization by practicing structural engineers \cite{smith2024reducing}.
Consequently, this work presents a topology optimization framework that explicitly lowers the embodied carbon of multi-component and multi-material structures while imposing several realistic constructability constraints.
For this paper, components are defined as structural members that have more specific behavior than what is implied by their material properties alone.
For instance, steel cables should only be loaded in tension, despite the fact that steel as a material has a nonzero compressive strength.

\begin{figure}[ht]
    \centering
    \begin{subfigure}[b]{0.28125\textwidth}
        \centering
        \includegraphics[width=\linewidth, trim={0cm 0cm 0cm 0cm}, clip]{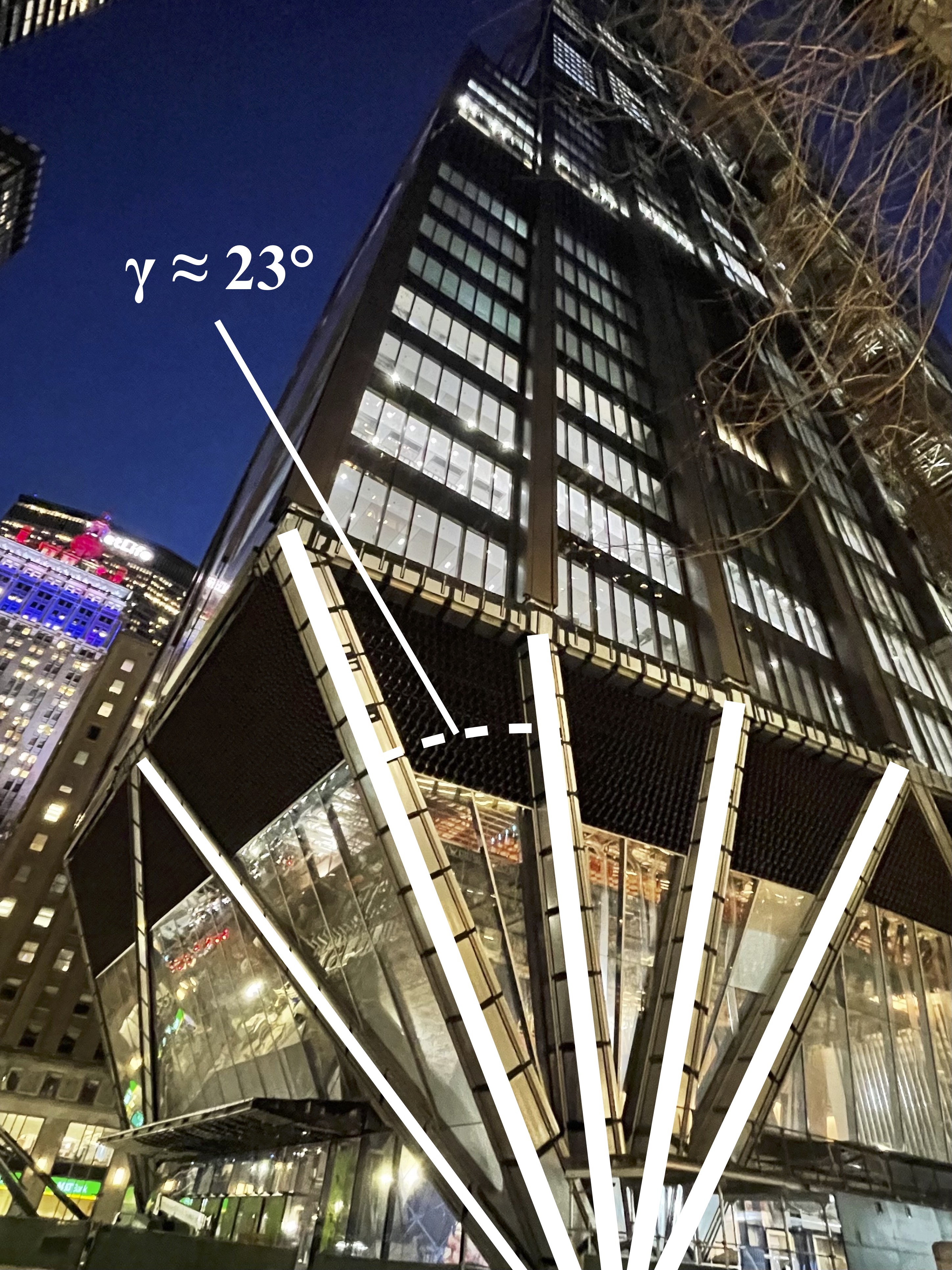}
        \caption{}
        \label{subfig:Constraint_Example_Steel}
    \end{subfigure} 
    \quad
    \begin{subfigure}[b]{0.5\textwidth}
        \centering
        \includegraphics[width=\linewidth, trim={0cm 0cm 0cm 0cm}, clip]{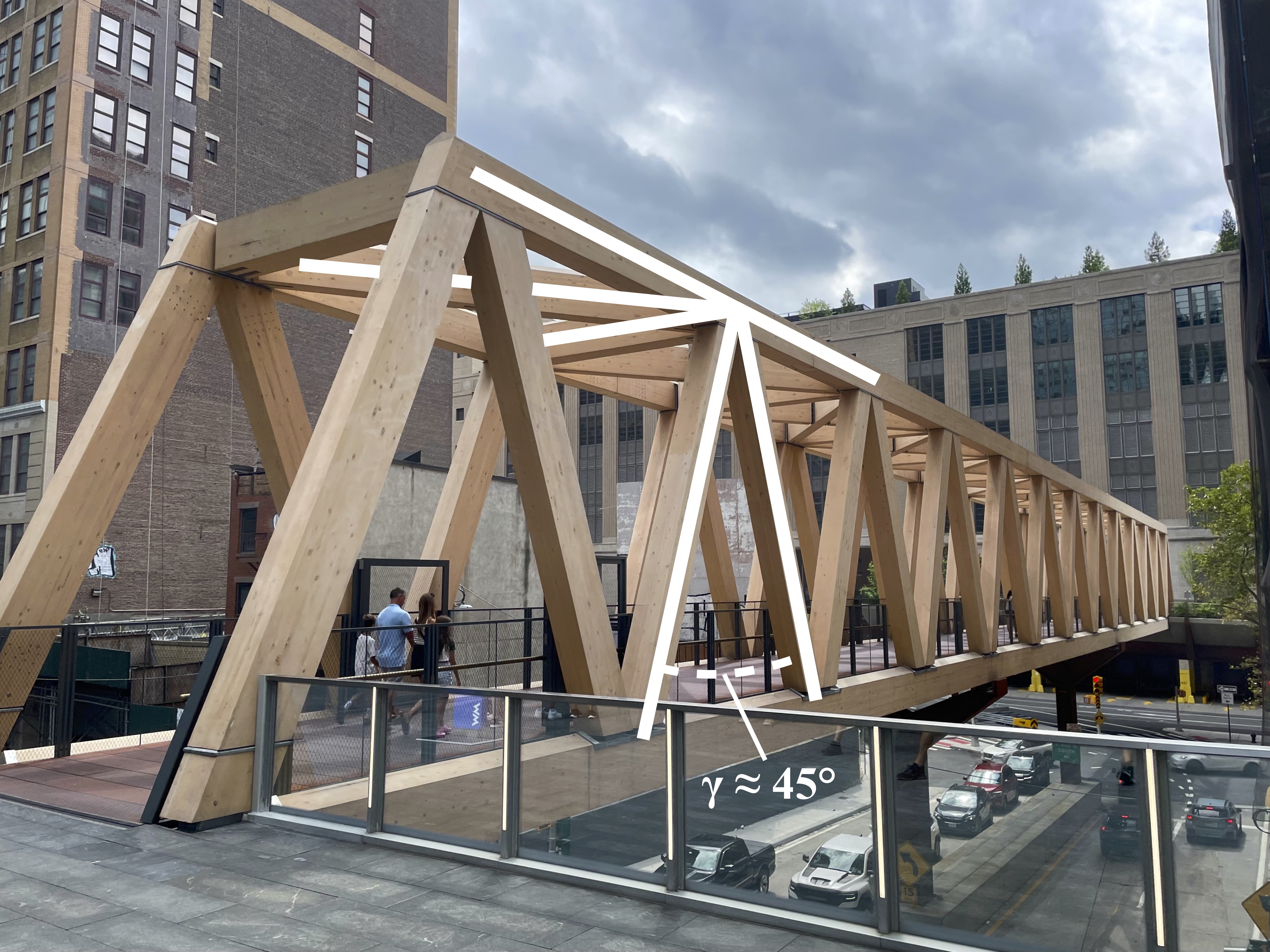}
        \caption{}
        \label{subfig:Constraint_Example_Timber}
    \end{subfigure}
    \caption{Examples of existing complex truss nodes in building and civil infrastructure.
    (a) shows a steel node connecting five large sections at the bottom of a skyscraper on 270 Park Avenue, NY.
    (b) gives the Moynihan Connector bridge, also in Manhattan, NY, which is a mass timber pedestrian bridge with up to seven members connecting at each node.}
    \label{fig:Constraint_Real_Examples}
\end{figure}

This contribution focuses on low-carbon truss design using the ground structure approach by Dorn et al. \cite{dorn1964automatic}.
The approach is well-suited for topology optimization of large-scale structures \cite{DT20} and has been shown to outperform continuum-based topology optimization in cases where the design is highly constrained by constructability requirements \cite{carstensen2023improving}, as is typically the case for civil construction.
The formal definition for constructability states that ``constructability is the optimum use of construction knowledge and experience in planning, design, procurement, and field operations to achieve overall project objectives'' \cite{CII}.
For the design phase, constructability requires the design consideration of issues such as simplicity, labor skill, and availability  \cite{jergeas2001benefits}.
Herein, \emph{constructability of an optimized design} is considered to be affected by (i) the extent to which post-processing of a design is required prior to construction, and (ii) the complexity of the structural layout.
In this work, the latter is largely assessed through the complexity of the structural nodes, where nodes that connects many members are viewed as having higher complexity.
In addition, the angle between connecting members affects the structural complexity as a node connecting a few members can be viewed as more complex than a node with more members, joined with larger angular spacing.
The number of connecting elements as well as the angular spacing can be dictated both by the planned node fabrication process (e.g., casting versus welding or bolting of steel joints) and the material selection.
Fig. \ref{fig:Constraint_Real_Examples} shows two connections from civil infrastructure in steel and timber, respectively.
In Fig. \ref{subfig:Constraint_Example_Steel}, a large steel truss directs the weight of a skyscraper to the foundation on 270 Park Ave in Manhattan, NY.
The node appears complex even though it only connects five members, as the smallest angle between members is about $\gamma = 23^\circ$.
The Moynihan Connector bridge (also Manhattan, NY) is shown in Fig. \ref{subfig:Constraint_Example_Timber}.
It is a mass timber pedestrian bridge where the structural nodes join up to seven elements. However, the nodes appear less complex as the minimum angular spacing is $\gamma = 45^\circ$.

Previous works have addressed different aspects of constructability of single-material truss topology-optimized structures. 
Most efforts have been dedicated to explicitly eliminating post-processing needs when planning construction by using only standard cross-sections that are commercially available, solving the so-called discrete truss topology optimization problem.
The reader is referred to Stolpe \cite{stolpe2016truss} for a full review on this topic.
Other works on improving constructability concepts include Asadpoure et al. \cite{asadpoure2015incorporating} that add a cost-per-member consideration to the objective function, giving the designer the ability to implicitly limit the complexity of the truss when defined as the number of members in the optimized design. 
Torii et al. \cite{torii2016design} expands this idea by defining complexity as the number of bars and nodes in the final design.
Both of these approaches rely on nonlinear objective formulations and require a final post-processing step to enforce a minimum size of the structural members. 
Asadpoure et al. \cite{AHT20} suggest that this final post-processing step can be eliminated by introducing an additional set of design variables that are encouraged to take on near 0--1 values.
These design variables signify if a certain member is active (when close to 1) in the final solution and allow the designer to explicitly define a lower bound for the cross-sectional area.
Lu and Xie \cite{lu2023reducing} also tackle the idea of reducing structural complexity, but do it through a new, expanded post-processing technique that uses clustering to avoid having many different cross sections and lengths.

Recently, researchers have started to assess the structural complexity through the nodes or joints of the truss. 
Lee et al. \cite{lee2022spherical} suggest quantifying the complexity of spatial trusses by the variation in nodal force demands.
Van Mellaert et al. \cite{van2016global} use a Mixed Integer Linear Programming (MILP) formulation to constrain the node complexity for planar steel truss sizing problems with fixed joint connectivity.
They apply Eurocode joint constraints and show that accounting for the node complexity eliminates post-processing and reduces the overall material consumption by 15\%.
The formulation of truss topology optimization as MILPs has historically been avoided since the problems are typically are more difficult to solve, requiring more computational resources than gradient-based methods \cite{RL21}.
However, advanced solvers have seen significant advancements in recent years \cite{Kro+19}, prompting increased research interest. 
Several works have developed formulations for minimum weight of single-material design subject to stress-constraints \cite{grossmann1992mixed,bollapragada2001optimal,2006_faustino,RS08,shahabsafa2018novel}.
Extensions have been suggested to address buckling constraints \cite{2014_mela,cai2024topology} and to cast the classic compliance design problem as a MILP \cite{stolpe2007reformulation}.
Recent works use MILP formulations for ground structure-based topology optimization to capture additive manufacturing constraints \cite{carstensen2023improving,RL21}. 
Most relevant to this contribution is the work by Fairclough and Gilbert \cite{FG20} that leverages a MILP formulation to expand the idea of constraining the node complexity to single-material truss topology optimization. 
Their formulation allows the designer to limit both the number of joints in the structure and/or the minimum angle between connected members. 

The extension of MILP truss topology optimization to design scenarios with multiple discrete component (or material) options has focused on frameworks for tensegrity structures.
Tensegrity structures are characterized by having a force flow through the structure with ``Discontinuous Compression, Continuous Tension'' as defined by Snelson \cite{snelson2012art}. 
Resulting structures typically appear as cable nets with ``floating'' compressive struts.
To achieve tensegrity using truss topology optimization, Ehara and Kanno \cite{ehara2010topology} introduce a new set of discrete (strictly 0--1) design variables to identify if members are tensile or compressive components (cables or struts).
Kanno \cite{K13} later recast the design problem for discrete cross-sectional areas under external loads as a MILP subject to compliance, stress and tensegrity constraints. 
Kanno then extends to consider self-weight \cite{kanno2012topology}, and constraints on member lengths to promote symmetry of the obtained structural solutions \cite{kanno2013exploring}.
Liu and Paulino \cite{liu2019tensegrity} suggests a MILP formulation that maximizes the summation of self-equilibrating forces which implicitly encourages larger angles between connected members at joints.
Work by Xu and Wang considers varying discrete cross-section sizes, in addition to considering buckling, large deformations, and global stability constraints for mixed-integer linear programs \cite{xu2018topology,wang2025topology,wang2020topology}.
These authors also extend their work to consider nonlinear terms and constraints in a mixed-integer nonlinear (MINLP) formulation for active tensegrity designs \cite{wang2024topology_tens}.
Nanayakkara et al. \cite{nanayakkara2020simple} suggest a simplified truss topology optimization approach for tensegrity design that address the self-stress load case in a post-processing step.
To the best of the authors knowledge, Nanayakkara et al. \cite{nanayakkara2020simple} are the first to propose a MILP tensegrity formulation that allows the final design to have varying continuous variable cross-sectional areas.
However, this formulation does not consider prestressing during the optimization stage.

Levering ideas from tensegrity topology optimization, this contribution significantly expands upon the state-of-the-art in low-carbon truss topology optimization.
The current work enables design with multiple discrete component types or materials \emph{while} explicitly enforcing constructability constraints on the node complexity. 
This allows the structural designer to identify low-carbon constructible design solutions with variable cross-sections.
To eliminate post-processing, the minimum size of constructible cross-sectional areas can be prescribed for each component type.
The designer can additionally explicitly constrain the minimum angle between connected members \emph{and} the number of members connecting at each node.
To illustrate the versatility of the presented approach, an extension to low-carbon design of multi-component tensegrity structures is also provided.

\section{Methods} \label{sec:Methods}
As mentioned, this work uses the ground structure approach of Dorn et al. \cite{dorn1964automatic} to design truss topologies that minimize the structure's embodied carbon.
The following simple relation by De Wolfe et al. \cite{de2017measuring} is used to quantify the environmental performance:

\begin{equation} \label{eq:GWP}
Embodied \ Carbon = \sum_j Q_j C_j
\end{equation}

Through Eq. \eqref{eq:GWP}, the embodied carbon is computed by summing up over the used structural materials, where $Q_j$ is the quantity of material $j$, and $C_j$ is the location-specific embodied carbon coefficient.
The material quantity is herein evaluated in terms of volume, and the embodied carbon coefficient is typically provided in units of CO$_2$e (CO$_2$ equivalent) per unit volume.
In this work, the embodied carbon calculation only includes the cradle-to-gate carbon contributions of the structural elements within the superstructure of a design.
Consequently, the carbon emissions from construction machinery, the fabrication of structural connections, and foundations are ignored.
The carbon impact of maintenance and repairs and end-of-life disposal are also not included.

\begin{figure}[ht]
    \centering
    \begin{subfigure}[b]{0.32\textwidth}
        \centering
        \includegraphics[width=\linewidth, trim={3cm 2cm 3cm 2cm}, clip]{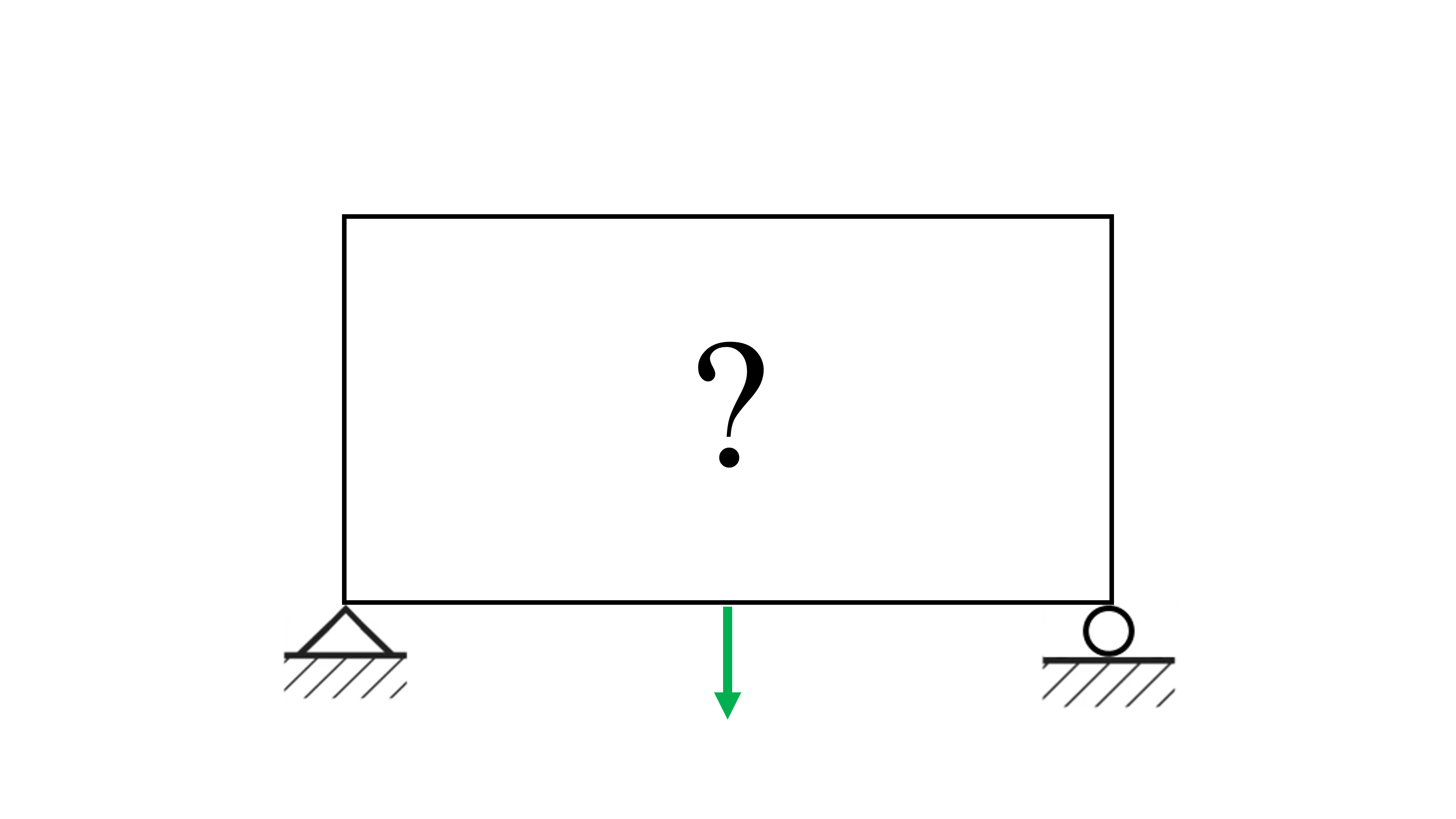}
        \caption{}
        \label{subfig:Examples_Design_Space}
    \end{subfigure}
    \begin{subfigure}[b]{0.32\textwidth}
        \centering
        \includegraphics[width=\linewidth, trim={3cm 2cm 3cm 2cm}, clip]{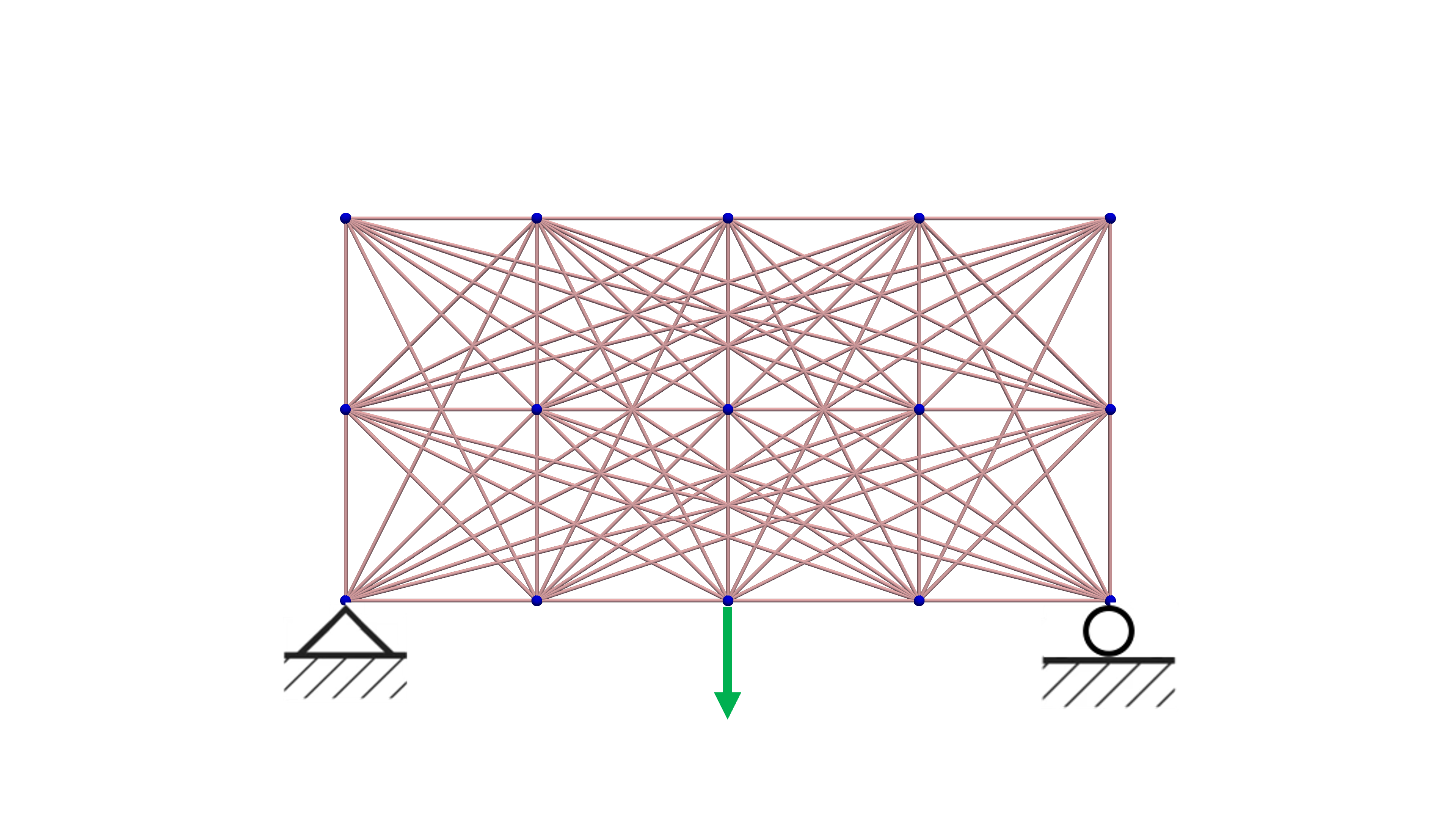}
        \caption{}
        \label{subfig:Examples_Ground_Structure}
    \end{subfigure}
    \begin{subfigure}[b]{0.32\textwidth}
        \centering
        \includegraphics[width=\linewidth, trim={3cm 2cm 3cm 2cm}, clip]{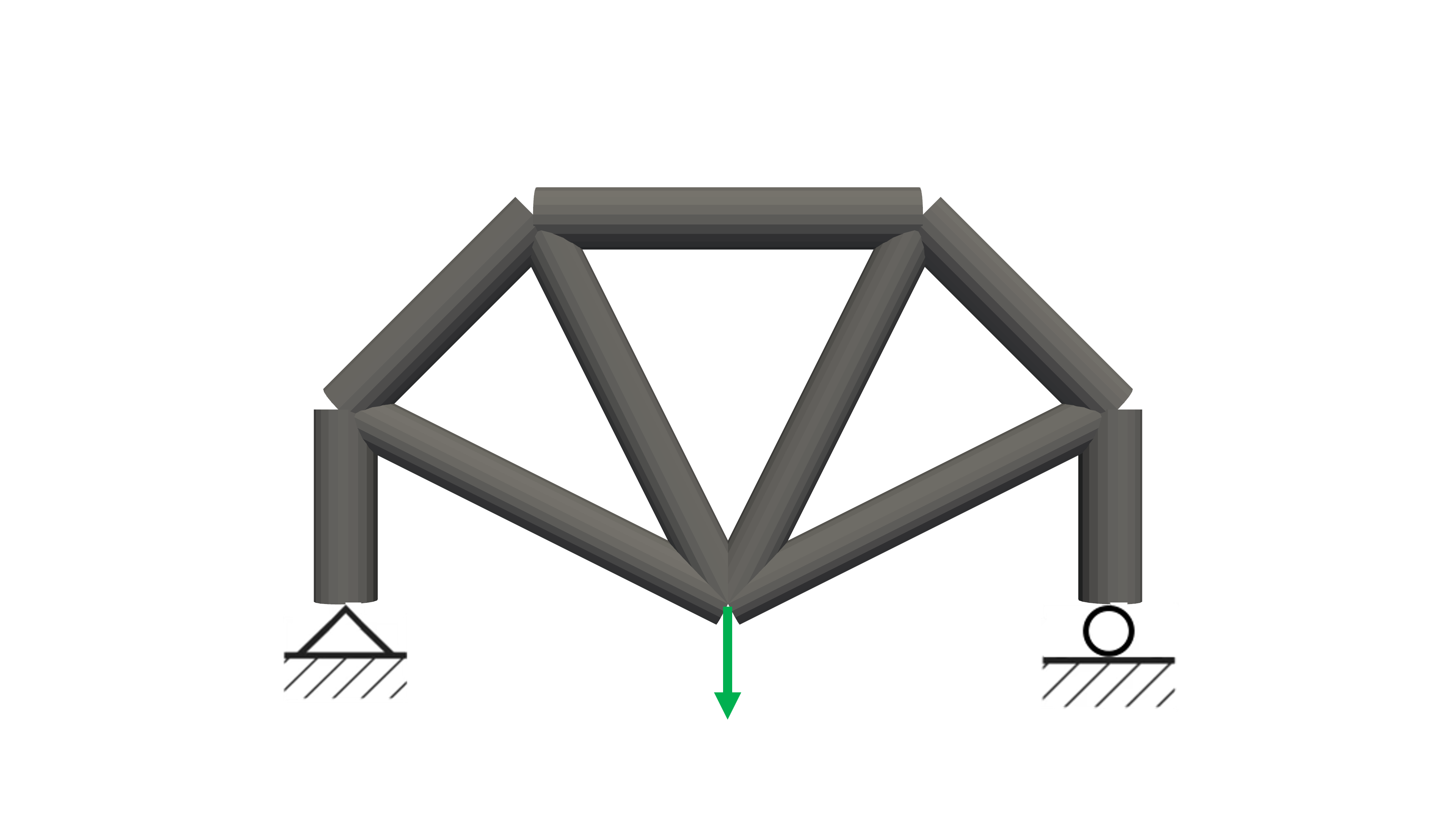}
        \caption{}
        \label{subfig:Mesh_Example_Optimization}
    \end{subfigure}
    \caption{Example of the ground structure approach to topology optimization.
    (a) shows a user-defined design space with an applied load and boundary conditions.
    (b) gives the design space when populated with a set of candidate truss members, also known as the ground structure.
    (c) shows the optimal truss design.}
    \label{fig:Mesh_Examples}
\end{figure}

The ground structure approach requires the user to define a design domain with applied loads and boundary conditions, as shown in Fig. \ref{subfig:Examples_Design_Space}.
The user must also define a distribution of nodes and candidate truss members. 
This work utilizes a total connectivity mesh, which connects every node in the design space with every other node in the design space, as shown in Fig. \ref{subfig:Examples_Ground_Structure}.
This gives the design algorithm the most freedom to choose an optimal design, but can lead to having overlapping members in the ground structure.
To avoid overlapping members in the final solution, this work removes overlapping members in the ground structure with the added benefit of reducing the number of design variables in the optimization problem.
Once a mesh is constructed, a generalized sizing problem is solved where the area of the candidate members is allowed to approach zero, shown in Fig. \ref{subfig:Mesh_Example_Optimization}.

\subsection{Conventional Linear Problem Formulation for Single-Material Trusses} \label{subsec:LinearFormulation}
The contributions in this paper are based on the rich traditions of optimal design of linear truss problems.
The minimum weight design of a single-material truss with continuously variable cross-sectional areas under static equilibrium and stress constraints is not new and has been extensively discussed since the 1960s \cite{dorn1964automatic,hemp1973optimum,rozvany1995layout,ohsaki2016optimization}.
The conventional linear problem formulation does not allow for the consideration of multiple component or material options for each candidate member, and the consideration of constructability constraints is not possible.
Nevertheless, a short introduction is provided here for completeness. 
The reader is referred to one of the several educational papers on the topic for more detail \cite{HGS19,zegard2014grand,zegard2015grand3}. 
The linear formulation for single-material low-carbon design of trusses is taken as:

\begin{subequations} \label{eq:LinearForm}
\begin{align}
    & \underset{A_i,\: q_{ih}}{\text{minimize}}
    && \sum_{i=1}^{N} l_{i}A_{i}C
    \label{eq:LinearFormObj} \\
    & \text{subject to}
    && \sum_{i=1}^{N} B_{ki}q_{ih} = F_{kh}
    && \forall \ k=1,...\,,N^F,\: \forall \ h=1,...\,,L^C
    \label{eq:LinearFormEquilibrium} \\
    &&& q_{ih} \leq A_{i}\sigma^{t}
    && \forall \ i=1,...\,,N,\: \forall \ h=1,...\,,L^C
    \label{eq:LinearFormTension} \\
    &&& q_{ih} \geq - A_{i}\sigma^{c}
    && \forall \ i=1,...\,,N,\: \forall \ h=1,...\,,L^C
    \label{eq:LinearFormCompression} \\
    &&& A_{i} \leq A^{\max}
    && \forall \ i=1,...\,,N
    \label{eq:LinearFormMaxArea} \\
    &&& A_{i} \geq 0
    && \forall \ i=1,...\,,N
    \label{eq:LinearFormMinArea}
\end{align}
\end{subequations}

Eq. \eqref{eq:LinearForm} has two types of variables: $A_{i}$, the cross-sectional areas of each candidate member $i$, and $q_{ih}$, the internal forces within each member $i$ for each load case $h$.
According to the standard sign convention, a member $i$ subjected to compression will have a negative $q_{ih}$, while a member in tension will have a positive $q_{ih}$.
Both types of design variables are indexed by $i$ ranging from 1 to $N$, where $N$ is the total number of candidate truss members in the ground structure.
The force variable is also indexed by $h$ ranging from 1 to $L^C$, where $L^C$ represents the total number of load cases that the ground structure is subjected to.
The objective function in Eq. \eqref{eq:LinearFormObj} minimizes the total embodied carbon of the structure by multiplying the length of a member, $l_i$, with the member's cross-sectional area and the embodied carbon coefficient of the material it is made of.
Since this formulation can only consider one material, the embodied carbon is simply a scaled multiple of the structural weight.

Structural equilibrium is enforced through constraint Eq. \eqref{eq:LinearFormEquilibrium}, which ensures that the sum of internal forces within the truss members, $q_{ih}$, are equal to the externally applied forces, $F_{kh}$.
Here, $k$ indexes the structural degrees of freedom that range from 1 to the total degrees of freedom, $N^{F}$. 
For planer trusses, each member has two degrees of freedom at each node, whereas 3D truss members have three degrees of freedom at each node.
In Eq. \eqref{eq:LinearFormEquilibrium}, the internal and external forces are balanced by multiplying the internal forces with the entries in the equilibrium matrix, $B_{ki}$.
The equilibrium matrix is a property of the geometry of the initial ground structure.
It projects the internal force of each member into the global degrees of freedom.
For a full description of how $B_{ki}$ is constructed, the reader is referred to He et al. \cite{HGS19}.

The linear truss design problem in Eq. \eqref{eq:LinearForm} is additionally constrained by Eqs. \eqref{eq:LinearFormTension} and \eqref{eq:LinearFormCompression} that set compressive and tensile stress limits for each candidate truss member $i$.
For trusses, the stress within a member is computed by $\sigma_{ih} = q_{ih} / A_i$.
To ensure that the member is not stressed past the yield point of its material, $\sigma^{y}$, this work adds a factor of safety $\varphi$ to reduce the maximum stress.
The compressive and tensile stresses limits are thus taken as $\sigma^{c} = \varphi \sigma^{y,c}$ and $\sigma^{t} = \varphi \sigma^{y,t}$.
For multi-material and indeterminate designs, this safety factor will also provide a margin of error when correcting forces that are relaxed by McCormick constraints, which will be discussed later.

Finally, the cross-sectional area design variables in Eq. \eqref{eq:LinearForm} are bound by Eqs. \eqref{eq:LinearFormMaxArea} and \eqref{eq:LinearFormMinArea}.
The cross-sectional area of all candidate members is thus allowed to approach zero.
When this happens for a given member, it essentially becomes inactive in the final design solution. 

\subsection{Mixed-Integer Linear Problem Formulations} \label{subsec:MILPFormulation}

\subsubsection{Multiple Component or Material Options} \label{subsubsec:MILPMultiMat}

\begin{figure}[b!]
    \centering
        \begin{subfigure}[b]{\textwidth}
        \centering
        \includegraphics[scale = 0.9]{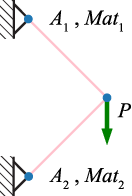}
        \caption{Design problem.}
        \label{subfig:Cantilever_Sizing_Example_Timber}
    \end{subfigure} \\[12 pt]
    \begin{subfigure}[b]{0.6\textwidth}
        \centering
        \includegraphics[scale = 0.9]{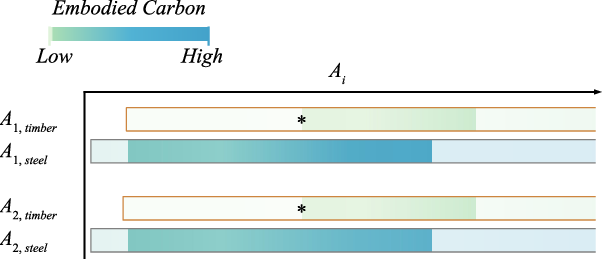}
        \caption{$P$ = 1 kN}
        \label{subfig:Cantilever_Sizing_Example_Var_Graph_Timber}
    \end{subfigure}
        \begin{subfigure}[b]{0.3\textwidth}
        \centering
        \includegraphics[scale = 0.9]{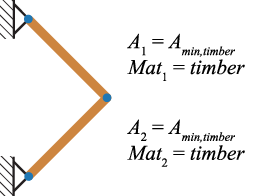}
        \caption{$P$ = 1 kN}
        \label{subfig:Cantilever_Sizing_Example_Timber+Steel}
    \end{subfigure} \\[18 pt]
    \begin{subfigure}[b]{0.6\textwidth}
        \centering
        \includegraphics[scale = 0.9]{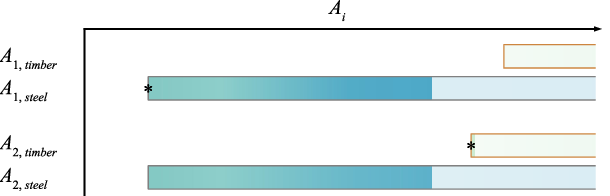}
        \caption{$P$ = 100 kN}
        \label{subfig:Cantilever_Sizing_Example_Var_Graph_Timber+Steel}
    \end{subfigure}
        \begin{subfigure}[b]{0.3\textwidth}
        \centering
        \includegraphics[scale = 0.9]{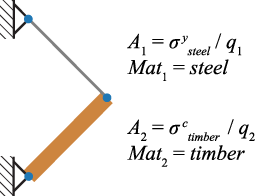}
        \caption{$P$ = 100 kN}
        \label{subfig:Cantilever_Sizing_Example}
    \end{subfigure}
    \caption{Illustrative two-member sizing and material selection example for the problem shown in (a).
    The design options and lowest carbon solutions are given in (b-c) for $P = 1$ kN, and in (d-e) for $P=100$ kN.
    The plots in (b) and (d) give members 1 and 2 when made of timber and steel along the $y$-axis and the cross-sectional area increases along the $x$-axis.
    The bars indicate the cross-sectional areas possible when designing for crushing where the color map gives the corresponding embodied carbon.
    The color within the bars are faded where the design options violate the designers constructability limits.
    The stars indicate the lowest carbon design solutions as shown in (c) and (d).}
    \label{fig:Cantilever_Member_Sizing_Problem}
\end{figure}

This work extends the embodied carbon minimization problem in Eq. \eqref{eq:LinearForm} to include component or material selection of all candidate members.  
As indicated by Eq. \eqref{eq:GWP} and previously mentioned, the consideration of multiple material options makes the interplay between constructability, environmental, and structural performance more involved.

Fig. \ref{fig:Cantilever_Member_Sizing_Problem} illustrates the added complexity for a planar two-bar truss sizing example where the designer seeks to find the lowest carbon solution.
For simplicity, buckling is ignored.
Each member has two design variables, the cross-sectional area $A_i$, and if they are made of $j = timber$ or $j = steel$ (Fig. \ref{subfig:Cantilever_Sizing_Example_Timber}).
To ease construction, the minimum and maximum cross-sectional areas are limited by $A_{timer}^{\min}$, $A_{timer}^{\max}$,  $A_{steel}^{\min}$, and $A_{steel}^{\max}$.
For the current example, all members are made of circular rods where the minimum and maximum radii are taken as 25--45 mm and 5--40 mm for timber and steel, respectively.
The truss members are designed based on the compressive and tensile yield stress limits of the material $\sigma_{j}^{y}$.
The values used for this examples are provided later in Tab. \ref{tab:mat_probs}.
Consequently, the cross-sectional area depends on the applied load, $P$, and is different for the two material options.
Additionally, for timber, different yield limits apply in tension and compression. 
Fig. \ref{subfig:Cantilever_Sizing_Example_Var_Graph_Timber} maps the embodied carbon when $P = 1$ kN.
Along the $y$-axis, the figure gives each of the two truss members when made of timber and of steel.
The cross-sectional area increases along the $x$-axis from $A_i = 0$ mm$^2$ to $A_i = 60^2 \pi$ mm$^2$.
The bars in the plot indicate the cross-sectional areas that are possible when designing for compressive or tensile yielding ($A_i=q_i/\sigma_{j}^{y}$) and the color map within them shows the corresponding embodied carbon as calculated by Eq. \eqref{eq:GWP}.
The color within the bars are faded for the design options that violate the constructability limits.
In line with intuition, Fig. \ref{subfig:Cantilever_Sizing_Example_Var_Graph_Timber} shows that, for the lowest carbon design solution under $P=1$ kN (indicated with stars), both members made of timber with $A_1=A_2 = A_{timber}^{\min}$ (Fig. \ref{subfig:Cantilever_Sizing_Example_Timber+Steel}).
Fig. \ref{subfig:Cantilever_Sizing_Example_Var_Graph_Timber+Steel} shows how the design options change when the applied load is increased to $P = 100$ kN.
For both members, the higher applied load increases the area needed to withstand yielding. 
As expected, the increase in area demand is greater for timber than for steel. 
The timber option for member one, which is in tension, now requires a cross-sectional area that is larger than $A_{timer}^{\max}$.
The entire $A_{1,timer}$ bar is therefore faded as it is infeasible. 
Consequently, steel must be used for member 1 with the smallest carbon impact at $A_1 = \sigma_{steel}^y/q_1$, while timber still provides the lowest feasible carbon solution for member 2 (Fig. \ref{subfig:Cantilever_Sizing_Example}).  

Inspired by ideas from single-material tensegrity design, this work addresses the component or material selection problem through the formulation of a novel MILP problem that explicitly minimizes the embodied carbon. 
This work thus enriches the linear problem such that each candidate member can be made of one of $M$ multiple component options. 
These options refer to structural elements that differ in their stress limits, either because they are made of different materials or because the type of element dictates a difference in behavior.
An example of the latter includes steel cables and struts, where the struts can withstand compressive loads but the cables cannot. 
As mentioned, with the exception of Nanayakkara et al. \cite{nanayakkara2020simple}, all existing multi-component literature formulated as a MILP considers design with discrete cross-sectional areas.
This restriction is not applied in the current work.
Additionally, the existing literature on tensegrity topology optimization is limited to design with $M=2$ single-material component options, whereas the current approach leaves the number of options (and thus candidate materials) to the discretion of the designer.
To achieve the minimum embodied carbon objective when designing with multiple component options, the following relation is used:

\begin{subequations} \label{eq:BasicMILP}
\begin{align}
    & \underset{A_{ij},\: q_{ih},\: z_{ij}}{\text{minimize}}
    && \sum_{i=1}^{N} \sum_{j=1}^{M} l_{i}A_{ij}C_{j}
    \label{eq:BasicMILPObj} \\
    & \text{subject to}
    && \sum_{i=1}^{N} B_{ki}q_{ih} = F_{kh}
    && \forall \ k=1,...\,,N^F,\: \forall \ h=1,...\,,L^C
    \label{eq:BasicMILPFormEquilibrium} \\
    &&& q_{ih} \leq \sum_{j=1}^{M} A_{ij}\sigma_j^{t}
    && \forall \ i=1,...\,,N,\: \forall \ h=1,...\,,L^C
    \label{eq:BasicMILPTension} \\
    &&& q_{ih} \geq - \sum_{j=1}^{M} A_{ij}\sigma_j^{c}
    && \forall \ i=1,...\,,N,\: \forall \ h=1,...\,,L^C
    \label{eq:BasicMILPCompression} \\
    &&& \sum_{j=1}^{M} z_{ij} \leq 1
    && \forall \ i=1,...\,,N
    \label{eq:BasicMILPUniqueMaterial} \\
    &&& A_{ij} \leq A_{ij}^{\max}z_{ij}
    && \forall \ i=1,...\,,N,\: \forall \ j=1,...\,,M
    \label{eq:BasicMILPMaxArea} \\
    &&& A_{ij} \geq 0
    && \forall \ i=1,...\,,N,\: \forall \ j=1,...\,,M
    \label{eq:BasicMILPMinArea} \\
    &&& z_{ij} \in \{0,1\}
    && \forall \ i=1,...\,,N,\: \forall \ j=1,...\,,M
    \label{eq:BasicMILPBinary}
\end{align}
\end{subequations}

The mixed-integer formulation used herein relies on the introduction of additional discrete binary (strict 0--1) variables, $z_{ij}$, that determine which members should exist in the final design.
When $z_{ij} = 1$, candidate member $i$ of element (or material) type $j$ exists in the final design.
Opposite, if $z_{ij} = 0$, the candidate member will not appear in the final solution.   
The corresponding cross-sectional areas $A_{ij}$ are left as continuous design variables.

The constraints in Eq. \eqref{eq:BasicMILP} largely resemble the conventional linear formulation from Eq. \eqref{eq:LinearForm}.
However, the maximum area constraint in Eq. \eqref{eq:BasicMILPMaxArea} has seen a noticeable modification.
It has now become a linking constraint that forces the binary decision variable, $z_{ij}$, to identify if a member has a non-zero cross-sectional area, $A_{ij}$.
Another key difference lies in the introduction of two constraints.
These constraints are Eq. \eqref{eq:BasicMILPUniqueMaterial} and the binary condition on $z_{ij}$ as specified in Eq. \eqref{eq:BasicMILPBinary}.
Eq. \eqref{eq:BasicMILPUniqueMaterial} is included to ensure that when $M \geq 2$, only one material or component option is assigned to a given member in the ground structure.
Note that when $M=1$, the herein developed formulation in Eq. \eqref{eq:BasicMILP} is the same as in Fairclough and Gilbert \cite{FG20}.

\paragraph{Compatibility Conditions and Constitutive Law ---}
For single material design problems, minimizing the material volume (and embodied carbon) is known to lead to statically determinate solutions. 
Consequently, the application of compatibility conditions and consideration of the material response are not needed.
However, when expanding to multi-material design, it is no longer guaranteed that the optimized designs are statically determinate. 
The assumption of elastic-perfectly plastic material response that is used in single material design can still be used when considering multiple material or component options.
To ensure structural forces are calculated correctly for statically indeterminate structures, the explicit consideration of compatibility conditions and constitutive laws are necessary.
For trusses, these constraints are defined as:

\begin{subequations} \label{eq:OGCompatibility}
\begin{align}
    & \sum_{k=1}^{N^F} A_{ik}^{f}u_{kh} = v_{ih}
    && \forall \ i=1,...\,,N,\: \forall \ h=1,...\,,L^C
    \label{eq:OGCompatibilityKinematic} \\
    & \sum_{j=1}^{M} \left( \frac{E_{j}A_{ij}}{l_{i}} \right) v_{ih} = q_{ih}
    && \forall \ i=1,...\,,N,\: \forall \ h=1,...\,,L^C
    \label{eq:OGCompatibilityElongations}
\end{align}
\end{subequations}

Here, two new design variables are introduced: $u_{kh}$, the displacements at each degree of freedom $k$ for each load case $h$, and $v_{ih}$, the elongations of each member $i$ for each load case $h$.
In Eq. \eqref{eq:OGCompatibilityKinematic}, the kinematic matrix (which is the transpose of the equilibrium matrix) is represented by entries $A_{ik}^{f}$ for each member $i$ and degree of freedom $k$.
In the constitutive relation in Eq. \eqref{eq:OGCompatibilityElongations}, the Young's modulus for material $j$ is denoted $E_j$.
The relation is seen to multiply two design variables, $A_{ij}$ and $v_{ih}$, and the formulation in Eq. \eqref{eq:OGCompatibilityElongations} thus changes the optimization problem from a linear formulation to a non-convex bilinear formulation.
In cases where members are assigned an area from a library of predefined cross-sectional areas, Xu et al. \cite{xu2018topology} identify how Eq. \eqref{eq:OGCompatibilityElongations} can be reformulated to maintain a linear program.
However, the suggested reformulation does not hold for the variable cross-sectional areas considered herein.

In this work, an extension to design with variable cross-sectional areas is developed by implementing a McCormick envelope to approximate the constitutive law in Eq. \eqref{eq:OGCompatibilityElongations} while maintaining a linear problem formulation.
Topology optimization is cited by Leyffer and Manns as a real-world application where McCormick envelopes could be used to provide tight relaxations for mixed-integer PDE constrained optimization problems \cite{leyffer2025mccormick}.
To the best of the authors' knowledge, McCormick envelopes have not previously been implemented for truss problems similar to the one studied here.
In this work, the McCormick envelope is defined as:

\begin{subequations} \label{eq:McCormickEnvelope}
\begin{align}
    & q_{ih} \leq \frac{E_j}{l_i} \left( A_{ij}^{\max}v_{ih} + v_{ij}^{\min}A_{ij} - A_{ij}^{\max}v_{ij}^{\min} \right) + \mathcal{M}\left( 1 - z_{ij} \right)
    && \forall \ i=1,...\,,N,\: \forall \ j=1,...\,,M,\: \forall \ h=1,...\,,L^{C}
    \label{eq:McCormick_Over1} \\
    & q_{ih} \leq \frac{E_j}{l_i} \left( A_{ij}^{\min}v_{ih} + v_{ij}^{\max}A_{ij} - A_{ij}^{\min}v_{ij}^{\max} \right) + \mathcal{M}\left( 1 - z_{ij} \right)
    && \forall \ i=1,...\,,N,\: \forall \ j=1,...\,,M,\: \forall \ h=1,...\,,L^{C}
    \label{eq:McCormick_Over2} \\
    & q_{ih} \geq \frac{E_j}{l_i} \left( A_{ij}^{\min}v_{ih} + v_{ij}^{\min}A_{ij} - A_{ij}^{\min}v_{ij}^{\min} \right) - \mathcal{M}\left( 1 - z_{ij} \right)
    && \forall \ i=1,...\,,N,\: \forall \ j=1,...\,,M,\: \forall \ h=1,...\,,L^{C}
    \label{eq:McCormick_Under1} \\
    & q_{ih} \geq \frac{E_j}{l_i} \left( A_{ij}^{\max}v_{ih} + v_{ij}^{\max}A_{ij} - A_{ij}^{\max}v_{ij}^{\max} \right) - \mathcal{M}\left( 1 - z_{ij} \right)
    && \forall \ i=1,...\,,N,\: \forall \ j=1,...\,,M,\: \forall \ h=1,...\,,L^{C}
    \label{eq:McCormick_Under2}
\end{align}
\end{subequations}

In Eq. \eqref{eq:McCormickEnvelope}, the constants $v_{ij}^{\min}$ and $v_{ij}^{\max}$ represent the minimum and maximum allowable elongation for each element $i$ if it is made of material $j$.
For most of this work, these bounds are defined as the minimum and maximum allowable elongation for an element to remain elastic under the compressive and tensile stress limits $\sigma^c$ and $\sigma^t$.
The constant $\mathcal{M}$ needs to be sufficiently large to ensure that members not selected for the final design, which have a cross-sectional area of zero, are allowed to deform as much as they want.
Since there is not an explicitly tight bound for this, an exact value for $\mathcal{M}$ is difficult to determine and is not easily computationally implemented as a big M constraint.
Because of this, the constraints in Eq. \eqref{eq:McCormickEnvelope} are computationally implemented as indicator constraints.
Writing them as indicator constraints removes the final $\mathcal{M}\left(1-z_{ij}\right)$ and only activates the remainder of the constraint if $z_{ij}=1$.

The McCormick envelope in Eq. \eqref{eq:McCormickEnvelope} relaxes the relation between a member's cross-sectional area $\left(A_{ij}\right)$, its elongation $\left(v_{ih}\right)$, and its internal force $\left(q_{ih}\right)$.
It produces an exact solution for all variables in the following cases:
\begin{enumerate}
    \item Every member in the final design has a cross-sectional area equivalent to either its minimum or maximum area bound as the McCormick envelope provides an exact solution at its extreme bounds.
    \item The minimum and maximum area bounds are equivalent $\left(A_{ij}^{\min} = A_{ij}^{\max}\right)$.
    This special case reduces a member's cross-sectional area from a variable to a constant and simplifies the non-convex bilinear constitutive law to a linear function.
    This is equivalent to the formulation used by Xu et al. \cite{xu2018topology}.
    \item The optimized structure is statically determinate and made from a single material.
\end{enumerate}

Outside these cases, the McCormick envelope still produces exact solutions for some variables, while others have small inaccuracies.
Because the objective is directly a function of cross-sectional area, this variable is assumed to be accurate.
For multi-material statically determinate trusses, the element forces will always be exact as only a single load path being possible. 
For single material structures, all members have the same stiffness, making the member elongations exact. 
However, when a truss is made of multiple materials, the elongations may be inaccurate because they no longer have to be proportional under the same load.
This freedom can allow for multiple kinematically admissible solutions to exist while satisfying compatibility conditions and the McCormick relaxation of the constitutive law, but violating a strict enforcement of the constitutive law.

Inaccuracies with the element forces and elongations decrease when the bounds on the cross-sectional areas and the elongations are tighter.
Since elongations have small, tight bounds to stay linear-elastic, the authors have found that this approximation is generally adequately precise.
Any errors in the force and elongation variables can be corrected as a post-processing step by resolving the structure with the optimized areas.

\begin{figure}[h!]
    \centering
    \begin{subfigure}[b]{0.32\textwidth}
        \centering
        \includegraphics[width=\linewidth, trim={0cm 0cm 0cm 0cm}, clip]{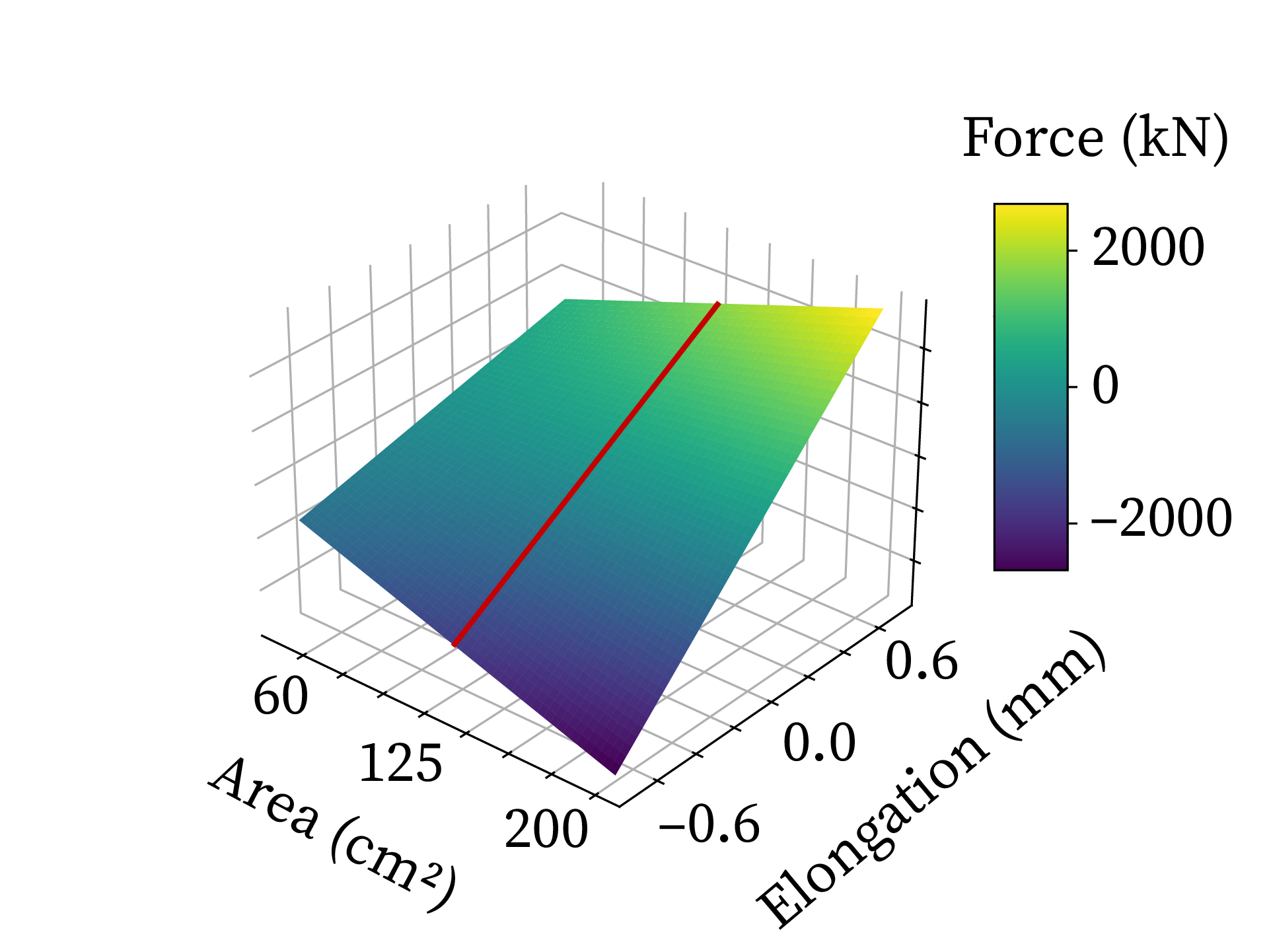}
        \caption{}
        \label{subfig:Constitutive_Law_Example_Plot}
    \end{subfigure}
    \begin{subfigure}[b]{0.32\textwidth}
        \centering
        \includegraphics[width=\linewidth, trim={0cm 0cm 0cm 0cm}, clip]{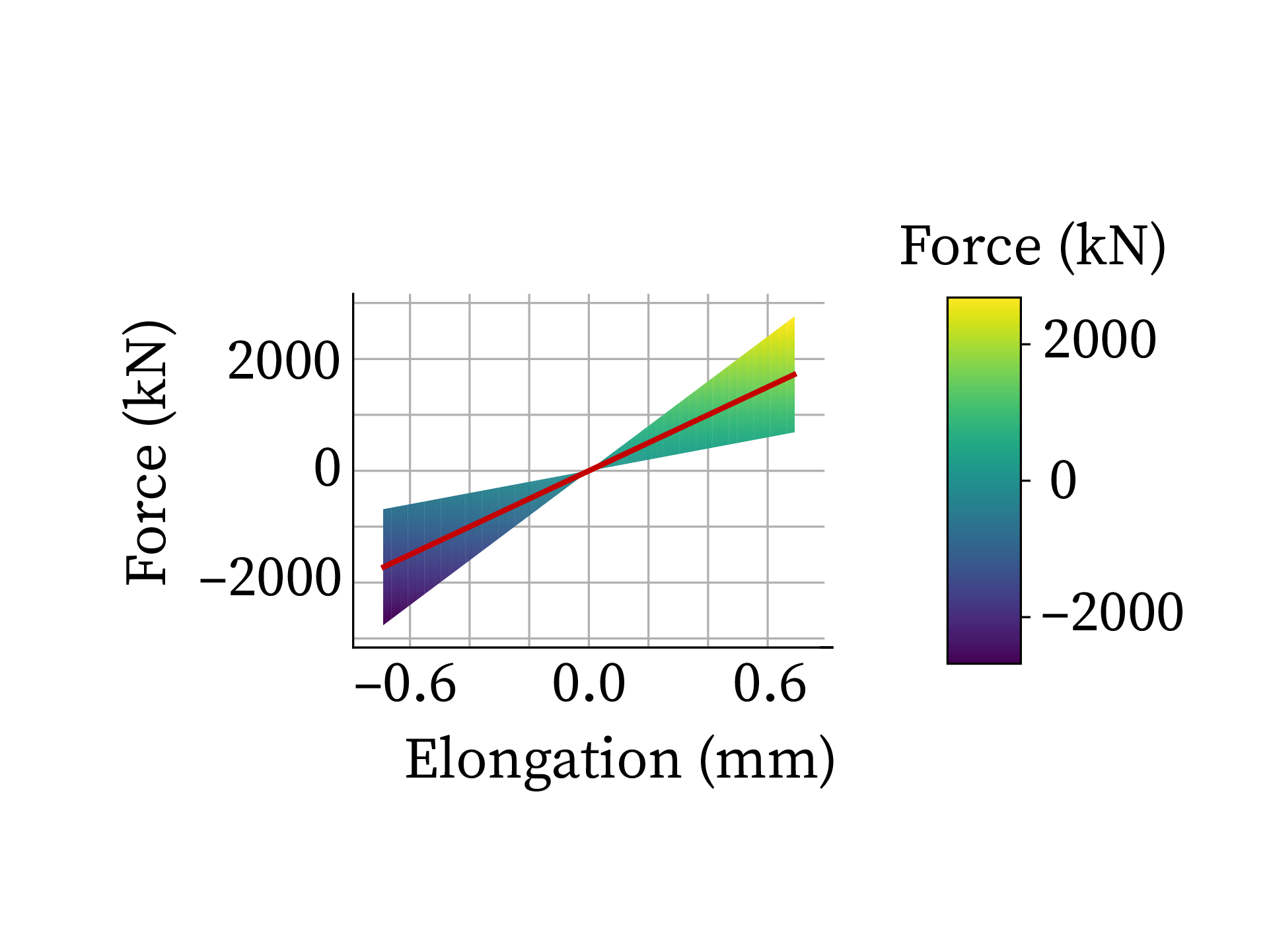}
        \caption{}
        \label{subfig:Constitutive_Law_Example_Plot_Side}
    \end{subfigure}
    \begin{subfigure}[b]{0.32\textwidth}
        \centering
        \includegraphics[width=\linewidth, trim={0cm 0cm 0cm 0cm}, clip]{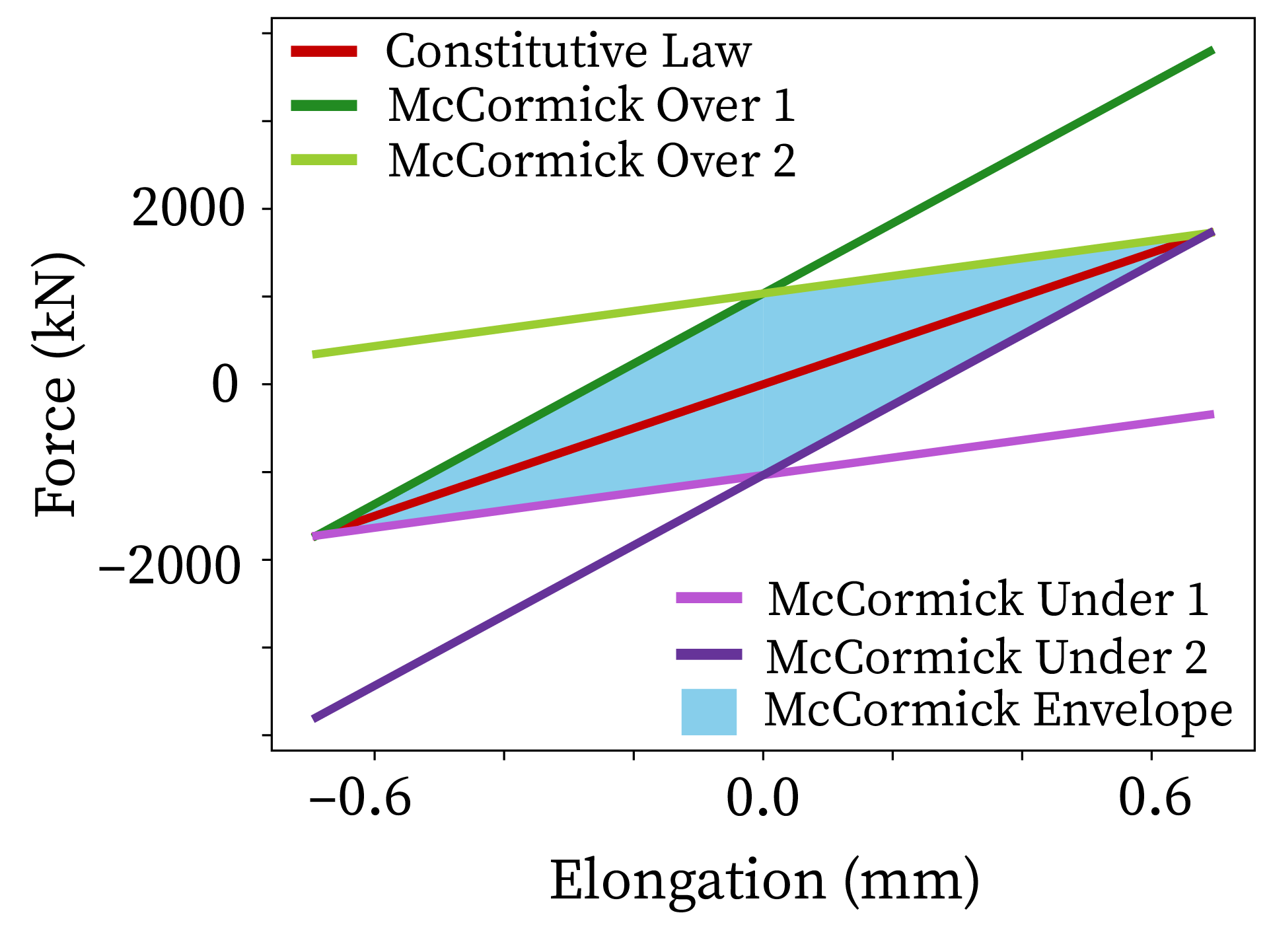}
        \caption{}
        \label{subfig:McCormick_Envelope_Example_Plot}
    \end{subfigure}
    \caption{
    Example visualization of the constitutive law and corresponding McCormick constraints.
    (a) shows the exact surface for all feasible variable values, within their corresponding bounds, under the constitutive truss law.
    (b) shows the same surface when viewed in the force-elongation plane with positive area coming out of the page.
    The color map in both (a) and (b) represents the member force for a corresponding area and elongation.
    The light yellow color represents large tensile forces while the dark purple represents large compressive forces.
    A slice of the surface is taken at $A=125$ cm$^2$, shown by the red line in (a) and (b).
    The exact same slice is plotted in (c) as a red line.
    Additionally, the slices of all four McCormick planes, taken at $A=125$ cm$^2$, are plotted as well.
    The teal highlighted area is the McCormick envelope.
    }
    \label{fig:McCormick_Example}
\end{figure}

A visual numeric example of the McCormick envelope used in this paper is provided in Fig. \ref{fig:McCormick_Example}.
This example looks at the variables associated with a single member.
Here, it is assumed the material of the member has an elastic modulus of $E = 200,000$ MPa, a length of $l = 1$ meter, and a minimum and maximum allowable elongation of $v^{\min} = -0.69$ mm and $v^{\max} = 0.69$ mm.

Fig. \ref{subfig:Constitutive_Law_Example_Plot} illustrates the bilinear nature of the constitutive truss law in three-dimensions.
When considering both the area and elongation as variables, Fig. \ref{subfig:Constitutive_Law_Example_Plot_Side} shows that the set is non-convex.
This is because there are numerous points that cannot be connected by a line segment that remains entirely inside the set.
However, when either the area or elongation variable are held constant, as shown by the red line slice at $A=125$ cm$^2$, the relation between the remaining variable and the force is linear.

Fig. \ref{subfig:McCormick_Envelope_Example_Plot} shows a slice of the four McCormick constraint planes at $A=125$ cm$^2$.
The two over-estimators, shown in green, correspond to Eqs. \eqref{eq:McCormick_Over1} and \eqref{eq:McCormick_Over2}, respectively.
The two under-estimators, shown in purple, correspond to Eqs. \eqref{eq:McCormick_Under1} and \eqref{eq:McCormick_Under2}, respectively.
Anywhere point inside the teal highlighted area is part of the McCormick envelope.
The exact relation for the constitutive law when $A=125$ cm$^2$ is plotted in red.
It is clear that the McCormick envelope produces solutions that do not have to abide by a strict enforcement of the constitutive law.
For a result to be feasible under a strict enforcement of the constitutive law, it must fall directly on the red line.
Because the McCormick envelope contains numerous points that are not on the red line, the optimizer has the freedom to choose any of these points as a ``feasible'' force value, even though they are not physically correct
However, it can be seen that these errors are smaller towards the variable bounds.
When looking at the value of the forces, plotted along the vertical axis in Fig. \ref{subfig:McCormick_Envelope_Example_Plot}, it can be seen that the McCormick over-estimator and under-estimator lines are most similar to the constitutive law near the minimum and maximum value of the elongations.
At the variable bounds (the exact minimum and maximum values), the McCormick envelope falls directly on the constitutive law, producing exact results.
Thus, having tight variable bounds is an efficient way to ensure accuracy within the McCormick envelope.

Eq. \eqref{eq:BasicMILP_Compatibility} summarizes the full mixed-integer linear formulation used to encapsulate multiple material options.
It considers compatibility and a McCormick relaxation of the constitutive law.

\begin{subequations} \label{eq:BasicMILP_Compatibility}
\begin{align}
    & \underset{A_{ij},\: q_{ih},\: u_{kh},\: v_{ih},\: z_{ij}}{\text{minimize}}
    && \sum_{i=1}^{N} \sum_{j=1}^{M} l_{i}A_{ij}C_{j}
    \label{eq:BasicMILPObj_Compatibility} \\
    & \text{subject to}
    && \sum_{i=1}^{N} B_{ki}q_{ih} = F_{kh}
    && \forall \ k,\: \forall \ h
    \label{eq:BasicMILPFormEquilibrium_Compatibility} \\
    &&& \sum_{k=1}^{N^F} A_{ik}^{f}u_{kh} = v_{ih}
    && \forall \ i,\: \forall \ h
    \label{eq:BasicMILPKinematic_Compatibility} \\
    &&& q_{ih} \leq \frac{E_j}{l_i} \left( A_{ij}^{\max}v_{ih} + v_{ij}^{\min}A_{ij} - A_{ij}^{\max}v_{ij}^{\min} \right) + \mathcal{M}\left( 1 - z_{ij} \right)
    && \forall \ i,\: \forall \ j,\: \forall \ h
    \label{eq:BasicMILPMcCormick_Over1_Compatibility} \\
    &&& q_{ih} \leq \frac{E_j}{l_i} \left( A_{ij}^{\min}v_{ih} + v_{ij}^{\max}A_{ij} - A_{ij}^{\min}v_{ij}^{\max} \right) + \mathcal{M}\left( 1 - z_{ij} \right)
    && \forall \ i,\: \forall \ j,\: \forall \ h
    \label{eq:BasicMILPMcCormick_Over2_Compatibility} \\
    &&& q_{ih} \geq \frac{E_j}{l_i} \left( A_{ij}^{\min}v_{ih} + v_{ij}^{\min}A_{ij} - A_{ij}^{\min}v_{ij}^{\min} \right) - \mathcal{M}\left( 1 - z_{ij} \right)
    && \forall \ i,\: \forall \ j,\: \forall \ h
    \label{eq:BasicMILPMcCormick_Under1_Compatibility} \\
    &&& q_{ih} \geq \frac{E_j}{l_i} \left( A_{ij}^{\max}v_{ih} + v_{ij}^{\max}A_{ij} - A_{ij}^{\max}v_{ij}^{\max} \right) - \mathcal{M}\left( 1 - z_{ij} \right)
    && \forall \ i,\: \forall \ j,\: \forall \ h
    \label{eq:BasicMILPMcCormick_Under2_Compatibility} \\
    &&& q_{ih} \leq \sum_{j=1}^{M} A_{ij}\sigma_j^{t}
    && \forall \ i,\: \forall \ h
    \label{eq:BasicMILPTension_Compatibility} \\
    &&& q_{ih} \geq - \sum_{j=1}^{M} A_{ij}\sigma_j^{c}
    && \forall \ i,\: \forall \ h
    \label{eq:BasicMILPCompression_Compatibility} \\
    &&& \sum_{j=1}^{M} z_{ij} \leq 1
    && \forall \ i
    \label{eq:BasicMILPUniqueMaterial_Compatibility} \\
    &&& A_{ij} \leq A_{ij}^{\max}z_{ij}
    && \forall \ i,\: \forall \ j
    \label{eq:BasicMILPMaxArea_Compatibility} \\
    &&& A_{ij} \geq 0
    && \forall \ i,\: \forall \ j
    \label{eq:BasicMILPMinArea_Compatibility} \\
    &&& z_{ij} \in \{0,1\}
    && \forall \ i,\: \forall \ j
    \label{eq:BasicMILPBinary_Compatibility}
\end{align}
\end{subequations}

\subsubsection{Constructability Constraints} \label{subsubsec:MILPManufacturingConstraints}
The power of the mixed-integer formulation lies in its ability to capture highly specific design constraints.
The formulation herein thus allows the designer to control several aspects related to constructability.
Namely, post-processing of hairlike thin members is eliminated through the explicit application of minimum cross-sectional area requirements.
Additionally, several aspects related to the complexity of the nodes or joints in the final design are controlled.

\paragraph{Minimum Cross-Sectional Area Requirements ---}
As discussed by Asadpoure et al. \cite{AHT20}, a crucial practical requirement for constructability is a minimum size limit for the members that appear in the final design.
This minimum size is typically different for elements of different materials and for elements of different types.
Examples include rods that can be fabricated smaller when made of steel as opposed to timber, and cables that typically can be made smaller than rods.
The requirement means that if a candidate member $i$ of type $j$ is active ($z_{ij}=1$), then its cross-sectional area should be greater than some specified limit.
Asadpoure et al. \cite{AHT20} have suggested an approach that eliminate hair-like elements in nonlinear continuous truss topology optimization problems.
In contrast, this contribution formulates a novel set of equations that allows the implementation of minimum area requirements in MILP truss topology optimization problems.

The approach used herein is illustrated for a truss structure with a single steel member in Fig. \ref{fig:Conceptual_Area_Range}.
Fig. \ref{subfig:DesignAreaDomain} shows the desired jump discontinuity in the cross-sectional area.
If the candidate member does not exist in the final design ($z_{1,\:steel}=0$), it should be assigned a cross-sectional area of zero ($A_{1,\:steel}=0$).
In contrast, if the candidate member is an active element in the final design ($z_{1,\:steel}=1$), the cross-sectional area must be chosen in the range $A_{1,\:steel}^{\min}$ to $A_{1,\:steel}^{\max}$.
To accomplish this, two variables are used to describe the member's area.
Fig. \ref{subfig:Aij_and_aij_Domain} shows the variation of these two variables.
Notice that they are both continuous. 
Here, $a_{1,\:steel}$ will be referred to as the intermediate area and $A_{1,\:steel}$ is the final cross-sectional area that appears in the design.
The final cross-sectional area varies continuously between 0 and $A_{1,\:steel}^{\max}$, whereas the intermediate area varies between $A_{1,\:steel}^{\min} \leq a_{1,\:steel} \leq A_{1,\:steel}^{\max}$.
The set of linear relations in Eq. \eqref{eq:AreaCon} are formulated to capture the minimum area constraint for the single steel member example in Fig. \ref{fig:Conceptual_Area_Range}: 

\begin{figure}[h!]
    \centering
    \begin{subfigure}[b]{0.45\textwidth}
        \centering
        \includegraphics[scale=1]{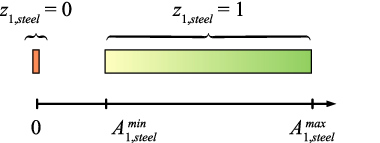}
        \caption{}
        \label{subfig:DesignAreaDomain}
    \end{subfigure}
    \begin{subfigure}[b]{0.45\textwidth}
        \centering
        \includegraphics[scale=1]{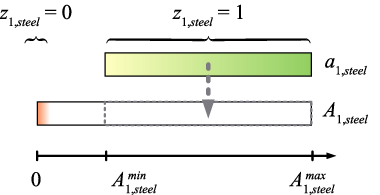}
        \caption{}
        \label{subfig:Aij_and_aij_Domain}
    \end{subfigure}
    \caption{Illustration on a single steel bar example of how the definition of two design variables allows the incorporation of a minimum area requirement for elements that are active in the final design.
    The desired jump discontinuity in the cross-sectional area is shown in (a).
    (b) shows how the intermediate area $a_{1,\:steel}$ and the final cross-sectional area $A_{1,\:steel}$ are combined to achieve the desired requirement.
    Both $a_{1,\:steel}$ and $A_{1,\:steel}$ vary continuously.}
    \label{fig:Conceptual_Area_Range}
\end{figure}

\begin{subequations} \label{eq:AreaCon}
\begin{align}
    & A_{1,\:steel} \leq A_{1,\:steel}^{\max}z_{1,\:steel}
    \label{eq:AreaConMaxA} \\
    & A_{1,\:steel} \geq a_{1,\:steel} - A_{1,\:steel}^{\max}\left(1 - z_{1,\:steel}\right)
    \label{eq:AreaConBigM} \\
    & A_{1,\:steel} \geq 0
    \label{eq:AreaConMinA} \\
    & a_{1,\:steel} \leq A_{1,\:steel}^{\max}
    \label{eq:AreaConMax_a} \\
    & a_{1,\:steel} \geq A_{1,\:steel}^{\min}
    \label{eq:AreaConMin_a}
\end{align}
\end{subequations}

The continuous intermediate area $a_{1,\:steel}$ is bounded by Eqs. \eqref{eq:AreaConMax_a} and \eqref{eq:AreaConMin_a}. 
The continuous final cross-sectional area $A_{1,\:steel}$ can vary between 0 and $A_{1,\:steel}^{\max}z_{1,\:steel}$ from Eqs. \eqref{eq:AreaConMaxA} and \eqref{eq:AreaConMinA}.
This ensures that, if steel member 1 is inactive in the final design ($z_{1,\:steel}=0$), the final cross-sectional area is forced to be zero $\left( 0 \leq A_{1,\:steel} \leq 0 \right)$.
In contrast, if steel member 1 is active in the final design $\left( z_{1,\:steel}=1 \right)$, the upper bound for the final cross-sectional area is non-zero and has the user-specified value of $A_{steel}^{\max}$.
The lower bound on $A_{1,\:steel}$ is altered by Eq. \eqref{eq:AreaConBigM} such that $A_{1,\:steel} \geq a_{1,\:steel} \geq A_{1,\:steel}^{\min}$, pushing the final cross-sectional area above the minimum requirement as desired.

\paragraph{Limiting the Complexity of Nodal Connections ---}
An equally important limiting factor for the constructability of a topology-optimized truss solution is the complexity of the joints. 
In this work, the complexity of a joint is determined both by the number of elements it connects and by the angles between these elements.

\begin{figure}[h!]
    \centering
    \begin{subfigure}[b]{0.45\textwidth}
        \centering
        \includegraphics[scale=0.9]{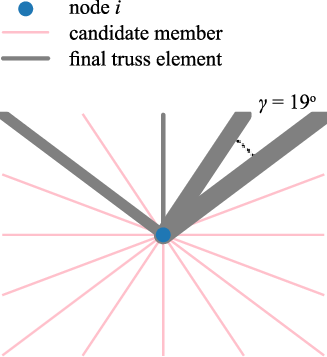}
        \caption{}
        \label{subfig:MaxNodeStick}
    \end{subfigure}
    \begin{subfigure}[b]{0.45\textwidth}
        \centering
        \includegraphics[scale=0.9]{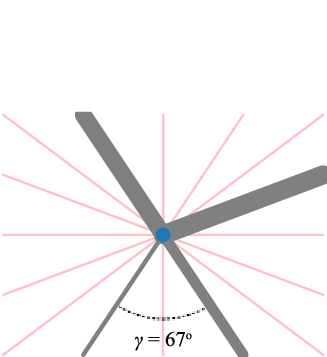}
        \caption{}
        \label{subfig:MinAngleStick}
    \end{subfigure}
    \caption{Schematic of joint complexities based on members' angles of separation.
    Both (a) and (b) illustrate a connection between 4 members.
    The joint depicted in (a) is considered more complex to construct due to the shallow angle of separation between connecting members, while the joint depicted in (b) is considered simpler to construct due to a wide angle of separation between connecting members.}
    \label{fig:Stick_Constructability_Constraints}
\end{figure}

Fig. \ref{fig:Stick_Constructability_Constraints} illustrates how the angles between connected members affect the nodal complexity for two examples of a node in a 2D ground structure.
For both examples, the node connects 16 candidate truss members.
Only four of these appear as elements in the final design.
In Fig. \ref{subfig:MaxNodeStick}, the four selected design elements connect at relatively shallow angles where the smallest of these is only $\gamma = 19^\circ$.
Structural connections are generally easier to design when elements have sufficiently large angles between them.
This is the case for the example in Fig. \ref{subfig:MinAngleStick}, where the smallest angle between elements in the connection is $\gamma = 67^\circ$.
In this work, the sufficiently large angle is a user-defined parameter that can be adjusted to fit the specific constructability requirements of a given project.

This work presents a constraint formulation that allows the design engineer to specify both the number of elements that can connect at a node and the angle between connecting elements. 
While Fairclough and Gilbert \cite{FG20} present constraints to limit the angle between connecting elements, there has not been work to directly mitigate the total number of members connecting at a joint to a set value to the best of the authors' knowledge.
To accomplish this, this work defines a preallocated nodal constraint matrix with entries $n_{pi}$.
The nodal constraint matrix contains binary information that maps how the candidate truss members are connected to each of the $N^{NP}$ nodes in the design space. 
If the node $p$ is a joint for member $i$, then $n_{pi}=1$.
If the node $p$ is not connected to member $i$, then $n_{pi}=0$.

\begin{figure}[htp]
    \centering
    \includegraphics[scale = 1]{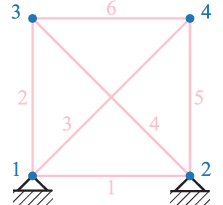}
    \caption{Simple ground structure used to show the construction of each entry in the nodal and relation constraint matrices, $n_{pi}$ from Eq. \eqref{eq:n_{pi}_ExampleConstruction} and $\Gamma_{xi}$ from Eq. \eqref{eq:Gamma_{xi}_IntersectExampleConstruction}.}
    \label{fig:Basic_Truss_Example}
\end{figure}

An example construction for each entry of the nodal constraint matrix, $n_{pi}$, for the simple ground structure in Fig. \ref{fig:Basic_Truss_Example} is shown in Eq. \eqref{eq:n_{pi}_ExampleConstruction}.
This ground structure is composed of four nodes and six candidate members.
Consequently, the nodal constraint matrix will take on a $4\times 6$ size.
At node 1, candidate members 1, 2, and 3 connect to the joint.
This means $n_{1,\:1} = n_{1,\:2} = n_{1,\:3} = 1$.
Since there are no other candidate members connecting at this joint, the rest of the entries in row 1 are populated with zeros ($n_{1,\:4} = n_{1,\:5} = n_{1,\:6} = 0$).
The entries in rows 2--4 of $n_{pi}$ are populated by looping over the nodes 2--4 of the ground structure and assigning 1 in the columns that corresponds to connected candidate members.

\begin{equation} \label{eq:n_{pi}_ExampleConstruction}
    n_{pi} =
    \begin{pNiceMatrix}[first-row,first-col]
        & \textcolor{myPink}{i=1} & \textcolor{myPink}{i=2} & \textcolor{myPink}{i=3} & \textcolor{myPink}{i=4} & \textcolor{myPink}{i=5} & \textcolor{myPink}{i=6} \\
    \textcolor{myBlue}{p=1} &  1  &  1  &  1  &  0  &  0  &  0  \\
    \textcolor{myBlue}{p=2} &  1  &  0  &  0  &  1  &  1  &  0  \\
    \textcolor{myBlue}{p=3} &  0  &  1  &  0  &  1  &  0  &  1  \\
    \textcolor{myBlue}{p=4} &  0  &  0  &  1  &  0  &  1  &  1
    \end{pNiceMatrix}
\end{equation}

The construction of the nodal constraint matrix allows the novel formulation of the constraint in Eq. \eqref{eq:MaxNodeCon} that limits the number of elements connected at node $p$:

\begin{align} \label{eq:MaxNodeCon}
     & \sum_{i=1}^{N} \sum_{j=1}^{M} n_{pi}z_{ij} \leq \eta_{p}^{\max}
     && \forall \  p=1,...\,,N^{NP}
\end{align}
where, $\eta_{p}^{\max}$ is the maximum number of elements that the designer allows to connect at a node $p$.

The number of members connecting at a joint can also be component or material-dependent.
For example, it may be feasible to have a large number of small steel cable members that consolidate to one point.
The same cannot be said, however, for large timber sections.
To control the total number of members connecting at joint $p$ by component or material, the constraint in Eq. \eqref{eq:MaxNodeCon} is slightly altered:

\begin{align} \label{eq:MaxNodeConPerMaterial}
    & \sum_{i=1}^{N}  n_{pi}z_{ij} \leq \eta_{pj}^{\max}
    && \forall \ p=1,...\,,N^{NP},\: \forall \ j=1,...\,,M
\end{align}

Notice how Eqs. \eqref{eq:MaxNodeCon} and \eqref{eq:MaxNodeConPerMaterial} are nearly identical, except that Eq. \eqref{eq:MaxNodeConPerMaterial} is applied for each component type $j$ instead of summing up over the $M$ component types.

To ensure that the nodal constraints are applied to all nodes in the structure, it is important to prevent intersecting members in the ground structure from coexisting in the final design.
An example of intersecting members can be seen in Fig. \ref{fig:Basic_Truss_Example}, where candidate members 3 and 4 cross in the middle of the ground structure where no nodes are defined.
Previous work has shown that using a mixed-integer formulation for this problem can easily prevent a set of two intersecting members from coexisting by ensuring that, at most, only one exists in the final design.
This is typically done by generating one constraint for each pair of members in the design space that intersect \cite{FG20}, \cite{K13}, \cite{cai2024topology}.

Instead of adding constraint pairs that identify what members cannot exist at the same time, this work presents a new formulation to prevent intersecting members.
Intersections are avoided by generating entries in a preallocated symmetric relation matrix, $\Gamma_{xi}$, that identify the entire set of members that intersect.
This reduces the total number of constraints in the problem, but can be a looser problem formulation than that defined in previous literature \cite{FG20}.
A looser problem formulation can mean slower computation times.
However, the construction of this relation matrix is more easily extendable to more complex existence relationships between members.
This will be discussed in more detail with an application to tensegrity structures in section \ref{subsubsec:MILPTensegrityConstraints}.
The formulation used in this paper is defined in Eq. \eqref{eq:MinAngleCon}. Here, both $x$ and $i$ range from 1 to $N$, where $N$ is the total number of candidate members in the ground structure.

\begin{align} \label{eq:MinAngleCon}
    & \sum_{i=1}^{N} \sum_{j=1}^{M} \Gamma_{xi}z_{ij} \leq \Gamma_{xx}
    && \forall \ x=1,...\,,N
\end{align}

An example construction of the entries in the relation matrix, $\Gamma_{xi}$, for the simple ground structure in Fig. \ref{fig:Basic_Truss_Example} is shown in Eq. \eqref{eq:Gamma_{xi}_IntersectExampleConstruction}.
The relation matrix is an $N\times N$ matrix populated with zeros where candidate members can coexist.
If a candidate member intersects with another member in the ground structure where no nodes are defined, a value of 1 is assigned.
The diagonal entries ($\Gamma_{xi}$ where $i=x$) are set to equal the sum of all other entries in that row ($\Gamma_{xx} = \sum_{i = 1}^N \Gamma_{xi}$).
For the example in Fig. \ref{fig:Basic_Truss_Example}, candidate members $x = 1$ and $x = 2$ do not intersect with any other members.
Consequently, the entries $\Gamma_{xi}$ in the first two rows remain zero. 
If element 1 is active in the final solution $\left( z_{1,\:j}=1 \right)$, Eq. \eqref{eq:MinAngleCon} states that $\sum_{i=1}^N \Gamma_{xi}z_{ij}=0 \leq 0$ no matter what 0--1 values the other $z_{ij}$ variables take on.
Since the constraint is satisfied, there is thus no restriction on coexistence with other candidate members for element 1.
In contrast, candidate members $x = 3$ and $x = 4$ intersect and cannot coexist. 
Consequently, in row $x=3$ of the relation matrix, a 1 is placed in column $i=4$ $\left( \Gamma_{3,\:4} = 1 \right)$.
Likewise, a 1 is placed in row $x = 4$, column $i = 3$ $\left( \Gamma_{4,\:3} = 1 \right)$, making the relation matrix symmetric.
The sum of entries in rows 3 and 4 each add up to 1, making the diagonal entries $\Gamma_{3,\:3}=\Gamma_{4,\:4}=1$.
This prevents both candidate members 3 and 4 from being active at the same time $\left( z_{3,j}=z_{4,j}=1 \right)$ since Eq. \eqref{eq:MinAngleCon} will give an invalid $\sum_{i=1}^N \Gamma_{xi}z_{ij} = 1 + 1 \leq 1$ for both $x = 3$ and $x = 4$.

\begin{equation} \label{eq:Gamma_{xi}_IntersectExampleConstruction}
    \Gamma_{xi} =
    \begin{pNiceMatrix}[first-row,first-col]
        & \textcolor{myPink}{i=1} & \textcolor{myPink}{i=2} & \textcolor{myPink}{i=3} & \textcolor{myPink}{i=4} & \textcolor{myPink}{i=5} & \textcolor{myPink}{i=6} \\
    \textcolor{myPink}{x=1} &  0  &  0  &  0  &  0  &  0  &  0  \\
    \textcolor{myPink}{x=2} &  0  &  0  &  0  &  0  &  0  &  0  \\
    \textcolor{myPink}{x=3} &  0  &  0  &  1  &  1  &  0  &  0  \\
    \textcolor{myPink}{x=4} &  0  &  0  &  1  &  1  &  0  &  0  \\
    \textcolor{myPink}{x=5} &  0  &  0  &  0  &  0  &  0  &  0  \\
    \textcolor{myPink}{x=6} &  0  &  0  &  0  &  0  &  0  &  0
    \end{pNiceMatrix}
\end{equation}

The benefit of this formulation is that the entries in the relation matrix, $\Gamma_{xi}$, can be generated for any constructability constraint involving a relationship between multiple candidate members.
When multiple relational constraints need to be considered, the entries $\Gamma_{xi}$ simply need to be updated, keeping the total number of constraints the same no matter how many relational constraints are considered.

One example of this is implementing a constraint that relies on the angle relationship between members.
The entries in the relation matrix $\Gamma_{xi}$ can be updated to ensure a sufficiently large angles between connecting members in addition to preventing intersecting members.
As mentioned earlier, ensuring large angles between connecting members has been done in previous work \cite{FG20}.
However, this was accomplished by comparing pairs of members as opposed to comparing one member to all other members.

Eq. \eqref{eq:Gamma_{xi}_AngleExampleConstruction} gives an example construction of the relation matrix for the simple ground structure in Fig. \ref{fig:Basic_Truss_Example} to prevent intersecting members and angles smaller than $\gamma^{\min} = 50^\circ$.
To start, the relation matrix from Eq. \eqref{eq:Gamma_{xi}_IntersectExampleConstruction} is taken as a baseline to prevent intersecting members.
Next, any index that remains zero is checked to see if the angle allows for co-existence.
For member $x = 1$, both members $i = 3$ and $i = 4$ share an angle of $\gamma=45^\circ$.
Since this is less than the prescribed $\gamma^{\min} = 50^\circ$, candidate member 1 cannot be active if members 3 or 4 are also active.
Consequently, the zeros in row $x = 1$ in columns $i = 3$ and $i = 4$ are replaced by ones $\left( \Gamma_{1,\:3} = \Gamma_{1,\:4} = 1 \right)$.
Because members $i = 2$ and $i = 5$ connect to member $x=1$ with an angle of $90^\circ$, the entries in columns $i=2$ and $i=5$ remain 0 $\left( \Gamma_{1,\:2} = \Gamma_{1,\:5} = 0 \right)$.
Finally, member $i=6$ does not connect to member $x = 1$ and does not need to be checked for a potential angle conflict.
The rest of the rows are filled in the same manner.
The diagonal values are initialized at zero and updated to reflect the new sum of the row entries.

\begin{equation} \label{eq:Gamma_{xi}_AngleExampleConstruction}
    \Gamma_{xi} =
    \begin{pNiceMatrix}[first-row,first-col]
        & \textcolor{myPink}{i=1} & \textcolor{myPink}{i=2} & \textcolor{myPink}{i=3} & \textcolor{myPink}{i=4} & \textcolor{myPink}{i=5} & \textcolor{myPink}{i=6} \\
    \textcolor{myPink}{x=1} &  2  &  0  &  1  &  1  &  0  &  0  \\
    \textcolor{myPink}{x=2} &  0  &  2  &  1  &  1  &  0  &  0  \\
    \textcolor{myPink}{x=3} &  1  &  1  &  5  &  1  &  1  &  1  \\
    \textcolor{myPink}{x=4} &  1  &  1  &  1  &  5  &  1  &  1  \\
    \textcolor{myPink}{x=5} &  0  &  0  &  1  &  1  &  2  &  0  \\
    \textcolor{myPink}{x=6} &  0  &  0  &  1  &  1  &  0  &  2
    \end{pNiceMatrix}
\end{equation}

\paragraph{MILP Formulation with Constructability Constraints ---}
Summarizing the above, this contribution has developed a MILP truss topology optimization problem for minimizing the embodied carbon using multiple element options while applying constructability constraints. The full problem formulation is taken as:

\begin{subequations} \label{eq:MultiMaterialMILP}
\begin{align}
    & \underset{A_{ij},\: a_{ij},\: q_{ih},\:u_{kh},\:v_{ih},\: z_{ij}}{\text{minimize}}
    && \sum_{i=1}^{N} \sum_{j=1}^{M} l_{i}A_{ij}C_{j}
    \label{eq:MultiMaterialMILPObj} \\
    & \text{subject to}
    && \sum_{i=1}^{N} B_{ki}q_{ih} = F_{kh}
    && \forall \ k,\: \forall \ h
    \label{eq:MultiMaterialMILPEquilibrium} \\
    &&& 
    \sum_{k=1}^{N^F} A_{ik}^{f}u_{kh} = v_{ih}
    && 
    \forall \ i,\: \forall \ h
    \label{eq:MultiMaterialMILPCompatibility} \\
    &&& 
    q_{ih} \leq \frac{E_j}{l_i} \left( A_{ij}^{\max}v_{ih} + v_{ij}^{\min}A_{ij} - A_{ij}^{\max}v_{ij}^{\min} \right) + \mathcal{M}\left( 1 - z_{ij} \right)
    && 
    \forall \ i,\: \forall \ j,\: \forall \ h
    \label{eq:MultiMaterialMILPMcCormickOver1} \\
    &&& 
    q_{ih} \leq \frac{E_j}{l_i} \left( A_{ij}^{\min}v_{ih} + v_{ij}^{\max}A_{ij} - A_{ij}^{\min}v_{ij}^{\max} \right) + \mathcal{M}\left( 1 - z_{ij} \right)
    && 
    \forall \ i,\: \forall \ j,\: \forall \ h
    \label{eq:MultiMaterialMILPMcCormickOver2} \\
    &&& 
    q_{ih} \geq \frac{E_j}{l_i} \left( A_{ij}^{\min}v_{ih} + v_{ij}^{\min}A_{ij} - A_{ij}^{\min}v_{ij}^{\min} \right) - \mathcal{M}\left( 1 - z_{ij} \right)
    && 
    \forall \ i,\: \forall \ j,\: \forall \ h
    \label{eq:MultiMaterialMILPMcCormickUnder1} \\
    &&& 
    q_{ih} \geq \frac{E_j}{l_i} \left( A_{ij}^{\max}v_{ih} + v_{ij}^{\max}A_{ij} - A_{ij}^{\max}v_{ij}^{\max} \right) - \mathcal{M}\left( 1 - z_{ij} \right)
    && 
    \forall \ i,\: \forall \ j,\: \forall \ h
    \label{eq:MultiMaterialMILPMcCormickUnder2} \\
    &&& q_{ih} \leq \sum_{j=1}^{M} A_{ij}\sigma_j^{t}
    && \forall \ i,\: \forall \ h
    \label{eq:MultiMaterialMILPTension} \\
    &&& q_{ih} \geq - \sum_{j=1}^{M} A_{ij}\sigma_j^{c}
    && \forall \ i,\: \forall \ h
    \label{eq:MultiMaterialMILPCompression} \\
    &&& \sum_{j=1}^{M} z_{ij} \leq 1
    && \forall \ i
    \label{eq:MultiMaterialMILPUniqueMaterial} \\
    &&& \sum_{i=1}^{N} \sum_{j=1}^{M} n_{pi}z_{ij} \leq \eta_{p}^{\max}
    && \forall \ p
    \label{eq:MultiMaterialMILPNode} \\
    &&& \sum_{i=1}^{N} n_{pi}z_{ij} \leq \eta_{pj}^{\max}
    && \forall \ p,\: \forall \ j
    \label{eq:MultiMaterialMILPNodePerMaterial} \\
    &&& \sum_{i=1}^{N} \sum_{j=1}^{M} \Gamma_{xi}z_{ij} \leq \Gamma_{xx}
    && \forall \ x
    \label{eq:MultiMaterialMILPAngle} \\
    &&& A_{ij} \leq A_{ij}^{\max}z_{ij}
    && \forall \ i,\: \forall \ j
    \label{eq:MultiMaterialMILPMaxArea} \\
    &&& A_{ij} \geq a_{ij} - A_{ij}^{\max}\left(1 - z_{ij}\right)
    && \forall \ i,\: \forall \ j
    \label{eq:MultiMaterialMILPBigM} \\
    &&& A_{ij} \geq 0
    && \forall \ i,\: \forall \ j
    \label{eq:MultiMaterialMILPMinArea} \\
    &&& a_{ij} \leq A_{ij}^{\max}
    && \forall \ i,\: \forall \ j
    \label{eq:MultiMaterialMILPa_Max} \\
    &&& a_{ij} \geq A_{ij}^{\min}
    && \forall \ i,\: \forall \ j
    \label{eq:MultiMaterialMILPa_Min} \\
    &&& z_{ij} \in \{0,1\}
    && \forall \ i,\: \forall \ j
    \label{eq:MultiMaterialMILPBinary}
\end{align}
\end{subequations}

When comparing Eq. \eqref{eq:MultiMaterialMILP} to the multi-component MILP design formation without constructability constraints in Eq. \eqref{eq:BasicMILP}, a noticeable difference is the added continuous design variables $a_{ij}$.
These intermediate cross-sectional area variables are introduced to ensure all active members respect a minimum across-sectional area constraint, as defined in Eq. \eqref{eq:AreaCon}.
The minimum cross-sectional area constraints are applied in Eqs. \eqref{eq:MultiMaterialMILPMaxArea}, \eqref{eq:MultiMaterialMILPBigM}, \eqref{eq:MultiMaterialMILPa_Max}, and \eqref{eq:MultiMaterialMILPa_Min} for all $N$ candidate members and all $M$ component options.

The connection complexity is restricted by limiting the total number of connected elements to $\eta_{p}^{\max}$ in Eq. \eqref{eq:MaxNodeCon} and the total number of elements of each component type to $\eta_{pj}^{\max}$ in Eq. \eqref{eq:MaxNodeConPerMaterial}.
These constraints are applied in Eqs. \eqref{eq:MultiMaterialMILPNode} and \eqref{eq:MultiMaterialMILPNodePerMaterial} for all $N^{NP}$ nodes in the ground structure.
Additionally, coexistence of intersecting elements or elements that join at angles that are separated by less than $\gamma^{\min}$ is prohibited by Eq. \eqref{eq:MinAngleCon}.
This is applied as a constraint in the MILP in Eq. \eqref{eq:MultiMaterialMILPAngle}.

Note that the nodal constraint matrix, whose entries are defined as $n_{pi}$ in Eqs. \eqref{eq:MultiMaterialMILPNode} and \eqref{eq:MultiMaterialMILPNodePerMaterial}, and the relation matrix, whose entries are defined as $\Gamma_{xi}$ in Eq. \eqref{eq:MultiMaterialMILPAngle}, are only dependent upon the initial configuration of the ground structure. 
As they do not change during the optimization, they are preallocated before the start of the optimization process.

\subsubsection{Prestressing Constraints} \label{subsubsec:MILPPrestressingConstraints}

Accounting for prestressing forces is important for a variety of structural applications, including tensegrity structures, which will be discussed in more detail later.
To incorporate prestressing, the total internal force in a member must be updated to account for both the elastic force contribution due to applied external loads and the initial prestressing.
This is described in by:

\begin{align} \label{eq:PrestressForceSum}
     & q_{ih} = q^e_{ih} + q^0_i
     && \forall \ i=1,...\,,N,\: \forall \ h=1,...\,,L^C
\end{align}

Eq. \eqref{eq:PrestressForceSum} introduces two new design variables: $q^e_{ih}$, the internal elastic force in each member $i$ due to applied external loads for each load case $h$, and $q^0_i$, the initial prestressing force in each member $i$.
The prestressing force is not indexed over each load case $h$ because a structure can realistically only be prestressed with one force in each member.
It can be noted that the elastic force in each member is what was previously defined by the McCormick envelope in Eq. \eqref{eq:McCormickEnvelope}.
Thus, this equation has to be modified slightly to specify that these bounds only apply to the elastic forces in each member.
This modification results in the following relations:

\begin{subequations} \label{eq:McCormickEnvelopePrestress}
\begin{align}
    & q^e_{ih} \leq \frac{E_j}{l_i} \left( A_{ij}^{\max}v_{ih} + v_{ij}^{\min}A_{ij} - A_{ij}^{\max}v_{ij}^{\min} \right) + \mathcal{M}\left( 1 - z_{ij} \right)
    && \forall \ i=1,...\,,N,\: \forall \ j=1,...\,,M,\: \forall \ h=1,...\,,L^{C}
    \label{eq:McCormick_Over1Prestress} \\
    & q^e_{ih} \leq \frac{E_j}{l_i} \left( A_{ij}^{\min}v_{ih} + v_{ij}^{\max}A_{ij} - A_{ij}^{\min}v_{ij}^{\max} \right) + \mathcal{M}\left( 1 - z_{ij} \right)
    && \forall \ i=1,...\,,N,\: \forall \ j=1,...\,,M,\: \forall \ h=1,...\,,L^{C}
    \label{eq:McCormick_Over2Prestress} \\
    & q^e_{ih} \geq \frac{E_j}{l_i} \left( A_{ij}^{\min}v_{ih} + v_{ij}^{\min}A_{ij} - A_{ij}^{\min}v_{ij}^{\min} \right) - \mathcal{M}\left( 1 - z_{ij} \right)
    && \forall \ i=1,...\,,N,\: \forall \ j=1,...\,,M,\: \forall \ h=1,...\,,L^{C}
    \label{eq:McCormick_Under1Prestress} \\
    & q^e_{ih} \geq \frac{E_j}{l_i} \left( A_{ij}^{\max}v_{ih} + v_{ij}^{\max}A_{ij} - A_{ij}^{\max}v_{ij}^{\max} \right) - \mathcal{M}\left( 1 - z_{ij} \right)
    && \forall \ i=1,...\,,N,\: \forall \ j=1,...\,,M,\: \forall \ h=1,...\,,L^{C}
    \label{eq:McCormick_Under2Prestress}
\end{align}
\end{subequations}

Note that, the only difference between Eq. \eqref{eq:McCormickEnvelope} and Eq. \eqref{eq:McCormickEnvelopePrestress} is that the left side of each inequality is updated to explicitly represent the elastic element force, $q^e_{ih}$, as opposed to the total element force, $q_{ih}$.

To design prestressed structures using a mixed-integer formulation, it is necessary to introduce an additional binary variable that determines if members are in compression or tension. 
Using a sign function on the total internal force variables $q_{ih}$ can also be done, but makes the problem nonlinear and hence not solvable as a MILP.
This work defines a new binary design variable, $z_{i}^{(2)}$, that identifies a member is in compression when $z_{i}^{(2)}$ = 1. 
It is assigned to each candidate truss member $i$ using the following constraint formulation developed herein: 

\begin{subequations} \label{eq:TensegrityForce}
\begin{align}
    & q_{ih} \leq \left(\sigma_{\max}^{t}A_{\max}^{\max} + \frac{\sigma_{\max}^{t}A_{\max}^{\max}\varepsilon}{1-\varepsilon}\right)\left(1-\varepsilon-z_{i}^{(2)}\right)
    && \forall \ i=1,...\,,N,\: \forall \ h=1,...\,,L^C
    \label{eq:TensegrityForce0orTension} \\
    & q_{ih} \geq -\sigma_{\max}^{c}A_{\max}^{\max}z_{i}^{(2)}
    && \forall \ i=1,...\,,N,\: \forall \ h=1,...\,,L^C
    \label{eq:TensegrityForceCompression}
\end{align}
\end{subequations}
where the maximum tensile and compressive stress limits and the maximum allowable cross-sectional area across all component options are defined as:

\begin{subequations} \label{eq:TensegrityForceLimits}
\begin{align}
    & \sigma_{\max}^{t} = \max\left\{ \sigma_j^t \right\}\\
    & \sigma_{\max}^{c} = \max\left\{ \sigma_j^c \right\} \\
    & A_{\max}^{\max} = \max\left\{ A_{ij}^{\max} \right\}.
\end{align}
\end{subequations}

Note that there is no $j$ index in Eq. \eqref{eq:TensegrityForce} because a member experiences compression independent of its component or material type.
This set of inequalities ensures that $z_i^{(2)}$ only on takes on a value of one for members in compression.
When $z_i^{(2)}=1$, Eq. \eqref{eq:TensegrityForce} reduces to $-\sigma_{\max}^cA_{\max}^{\max} \leq q_{ih} \leq -\sigma^t_{\max}A^{\max}_{\max}\varepsilon/(1-\varepsilon)$, where all stress cross-sectional areas are positive constants.
This inequality only holds when $q_{ih}$ is negative and thus a compressive force.
Opposite, when $z_i^{(2)}=0$, the inequalities reduce to $0\leq q_{ih} \leq \sigma_{\max}^{t}A_{\max}^{\max}$.
Here, $\varepsilon$ is a small constant that is included to ensure that the binary variables are set to zero for zero force members ($q_{ih}=0$).
It is set to $1\times 10^{-6}$ in this work.
This is important because most candidate truss members are not used in the final design, and thus have zero force going through them.

To ensure that the prestressed structure is in a state of self equilibrium without any external loads being applied, Eq. \eqref{eq:PrestressEquilibrium} is included in the formulation.
This equation uses the same equilibrium matrix, $B_{ki}$, as before.
The right side being set to zero indicates that there are no external forces being applied.

\begin{align} \label{eq:PrestressEquilibrium}
     & \sum_{i=1}^N B_{ki}q^0_i = 0
     && \forall \ k=1,...\,,N^{F}
\end{align}

Eq. \eqref{eq:MinPrestressing} is also added to the formulation to verify that each member is prestressed to a minimum amount defined by the designer.
Compressive members are prestressed in accordance with Eq. \eqref{eq:MinPrestressingCompression}, while tensile members are prestressed in accordance with Eq. \eqref{eq:MinPrestressingTension}.
The constants $q^{0_c}$ and $q^{0_t}$ are user specified values that denote the minimum compressive and tensile prestressing forces, respectively.

\begin{subequations} \label{eq:MinPrestressing}
\begin{align}
    & q^0_i \leq -q^{0_c}z^{(2)}_i + A^{\max}_{\max}\sigma^t_{\max}\left(1-z^{(2)}_i\right) 
    && \forall \ i=1,...\,,N
    \label{eq:MinPrestressingCompression} \\
    & q^0_i \geq \left(q^{0_t} + \sigma^c_{\max}A^{\max}_{\max}\right)\left(\sum_{j=1}^M z_{ij}-z^{(2)}_i\right) - \sigma^c_{\max}A^{\max}_{\max}
    && \forall \ i=1,...\,,N
    \label{eq:MinPrestressingTension}
\end{align}
\end{subequations}

The final condition that needs to be met during prestressing is preventing members from being over stressed.
This is achieved by including Eq. \eqref{eq:MaxPrestressing} in the problem formulation.
These constraints follow the same logic first defined for the stress constraints in Eq. \eqref{eq:LinearFormTension} and Eq. \eqref{eq:LinearFormCompression}.

\begin{subequations} \label{eq:MaxPrestressing}
\begin{align}
    & q^0_i \leq \sum_{j=1}^M A_{ij}\sigma^t_j
    && \forall \ i=1,...\,,N
    \label{eq:MaxPrestressingCompression} \\
    & q^0_i \geq -\sum_{j=1}^M A_{ij}\sigma^c_j
    && \forall \ i=1,...\,,N
    \label{eq:MaxPrestressingTension}
\end{align}
\end{subequations}

An updated full problem formulation, including all prestressing and tensegrity constraints, can be found at the end of the following section in Eq. \eqref{eq:Tensegrity_Full_Formulation}.

\subsubsection{Tensegrity Constraints} \label{subsubsec:MILPTensegrityConstraints}
To show the versatility of the current formulation, the problem is extended to the design of low-carbon tensegrity structures.
This work expands upon the contributions of Nanayakkara et al. \cite{nanayakkara2020simple} and Kanno \cite{kanno2013exploring} to consider variable cross-sections with a three component library.
Additionally, the implementation used in this paper allows for compression members to be stacked, as long as they fall in the same line, to create longer continuous members.
Doing so reduces the initial size of the ground structure by eliminating overlapping members.

\begin{figure}[h!]
    \centering
    \begin{subfigure}[b]{0.45\textwidth}
        \centering
        \includegraphics[scale = 0.9]{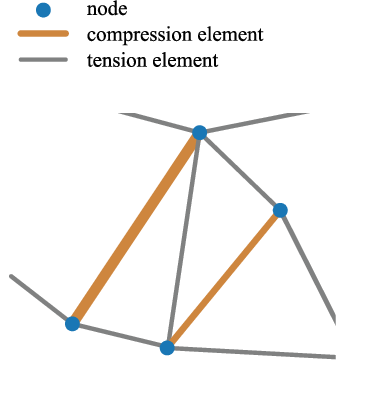}
        \caption{}
        \label{subfig:Tensegrity_Original}
    \end{subfigure}
    \begin{subfigure}[b]{0.45\textwidth}
        \centering
        \includegraphics[scale = 0.9]{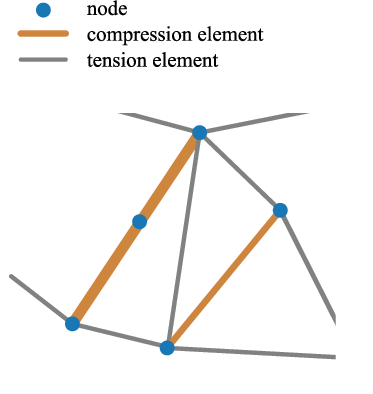}
        \caption{}
        \label{subfig:Tensegrity_New}
    \end{subfigure}
    \caption{Illustration of the (a) conventional, and (b) slightly modified tensegrity definition used in this work.
    In the conventional tensegrity definition, only a single member in compression is allowed at each node.
    This work uses a slightly altered definition that allows the existence of special case nodes with only two co-linear compression elements.}
    \label{fig:Tensegrity}
\end{figure}

The strict definition of tensegrity states that a maximum of one compression member can exist per node, as shown in Fig. \ref{subfig:Tensegrity_Original}.
To allow solving on a reduced ground structure with no overlapping candidate members, a slightly altered definition is suggested herein. 
This work allows two compression members to coexist at a node only if they lie in the same line and can thus be merged into a single longer element, as shown in Fig. \ref{subfig:Tensegrity_New}.
At these special-case nodes that join two aligned compression elements, it is important to ensure that no other elements exist.

\begin{figure}[b!]
    \centering
    \includegraphics[scale = 1]{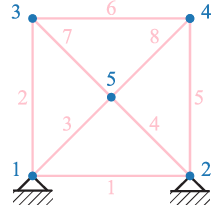}
    \caption{Simple ground structure used to illustrate the construction of each entry in the tensegrity relation constraint matrices, $\Gamma_{xi}^{(2)}$ from Eq. \eqref{eq:Gamma_{xi}^{(2)}_TensegrityExampleConstruction}, $J_{gxi}$ from Eq. \eqref{eq:J_{gxi}_TensegrityExampleConstruction}, and $\Upsilon_{gxi}$ from Eq. \eqref{eq:Y_{gxi}_TensegrityExampleConstruction}.}
    \label{fig:Basic_Tensegrity_Truss_Example}
\end{figure}

To achieve this slightly altered definition of tensegrity, a new violation relationship matrix with entries $\Gamma_{xi}^{(2)}$ is created.
An example construction of each entry $\Gamma_{xi}^{(2)}$ for the simple ground structure in Fig. \ref{fig:Basic_Tensegrity_Truss_Example} is shown in Eq. \eqref{eq:Gamma_{xi}^{(2)}_TensegrityExampleConstruction}.
Setting up the entries $\Gamma_{xi}^{(2)}$ resembles how the entries $\Gamma_{xi}$ are constructed for the minimum angle constraint in Eq. \eqref{eq:MinAngleCon}.
For an element $x$, all off-diagonal row entries are set to 1 for elements $i$ that share a node with $x$.
For the current example, element $x=1$ shares a node with elements $i=2$, $i=3$, $i=4$, and $i=5$. Consequently, the off-diagonals in row 1 are defined as $\Gamma_{1,2}^{(2)}=\Gamma_{1,3}^{(2)}=\Gamma_{1,4}^{(2)}=\Gamma_{1,5}^{(2)}=1$.
If elements $x$ and $i$ are co-linear, the entry remains zero but is flagged as important $\Gamma_{xi}^{(2)} = 0^*$.
This is the case for elements $x=3$ and $i=8$ and why $\Gamma_{3,8}^{(2)} = 0^*$.
This adjustment allows two compression members to exist at the same node if they are co-linear.
Diagonal entries $\Gamma_{xx}^{(2)}$ are filled the same way as entries $\Gamma_{xx}$ in Eq. \eqref{eq:MinAngleCon}, by summing up over all components in the row.

\begin{equation} \label{eq:Gamma_{xi}^{(2)}_TensegrityExampleConstruction}
    \Gamma_{xi}^{(2)} =
    \begin{pNiceMatrix}[first-row,first-col]
        & \textcolor{myPink}{i=1} & \textcolor{myPink}{i=2} & \textcolor{myPink}{i=3} & \textcolor{myPink}{i=4} & \textcolor{myPink}{i=5} & \textcolor{myPink}{i=6} & \textcolor{myPink}{i=7} & \textcolor{myPink}{i=8} \\
    \textcolor{myPink}{x=1} &  4  &  1  &  1  &  1  &  1  &  0  &  0  &  0  \\
    \textcolor{myPink}{x=2} &  1  &  4  &  1  &  0  &  0  &  1  &  1  &  0  \\
    \textcolor{myPink}{x=3} &  1  &  1  &  4  &  1  &  0  &  0  &  1  &0^{*}\\
    \textcolor{myPink}{x=4} &  1  &  0  &  1  &  4  &  1  &  0  &0^{*}&  1  \\
    \textcolor{myPink}{x=5} &  1  &  0  &  0  &  1  &  4  &  1  &  0  &  1  \\
    \textcolor{myPink}{x=6} &  0  &  1  &  0  &  0  &  1  &  4  &  1  &  1  \\
    \textcolor{myPink}{x=7} &  0  &  1  &  1  &0^{*}&  0  &  1  &  4  &  1  \\
    \textcolor{myPink}{x=8} &  0  &  0  &0^{*}&  1  &  1  &  1  &  1  &  4 \\
    \end{pNiceMatrix}
\end{equation}

With entries $\Gamma_{xi}^{(2)}$ defined, this work ensures that each nodes has only one compression member, unless it is a special-case node with two co-linear compression elements using the following constraint formulation:

\begin{align} \label{eq:Tensegrity_Violation_1}
    & \sum_{i=1}^{N} \Gamma_{xi}^{(2)} z_{i}^{(2)} \leq \Gamma_{xx}^{(2)}
    && \forall \ x=1,...\,,N
\end{align}
where $\Gamma_{xx}^{(2)}$ is the diagonal entries from $\Gamma_{xi}^{(2)}$.

With the tensegrity constraint in Eq. \eqref{eq:Tensegrity_Violation_1}, tension members can still exist at the special-case nodes that join aligned compression elements.
To prevent this, a new relationship matrix equation is added to the formulation.
Here, it is convenient to define the entries in two relation matrices as $J_{gxi}$ and $\Upsilon_{gxi}$.
The connection to other members at each node of every element is stored in the entries $J_{gxi}$, whereas entries $\Upsilon_{gxi}$ describes how many other members are connected at each node of every element and if co-linear elements exist at the node.
The $g$ subscript in $J_{gxi}$ and $\Upsilon_{gxi}$ refers to the start end end node of each candidate member.
The starting node of member $x$ is denoted $g=1$, whereas the end node is denoted $g=2$.
In this work, all candidate members are defined such that the start node is always smaller than the end node.

The construction of the entries $J_{gxi}$ from their initiation as zero matrix is illustrated in Eq. \eqref{eq:J_{gxi}_TensegrityExampleConstruction} for the ground structure in Fig. \ref{fig:Basic_Tensegrity_Truss_Example}.
Member $x=1$ in the structure is connected to members $i=2$ and $i=3$ at node $g=1$.
Therefore $J_{1,\,1,\,2} = J_{1,\,1,\,3} = 1$ is defined in Eq. \eqref{eq:J_{gxi}_TensegrityExampleConstructiona}.
At the opposite end of the member $x=1$ at node $g=2$, the member is connected with members $i=4$ and $i=5$.
In Eq. \eqref{eq:J_{gxi}_TensegrityExampleConstructionb}, $J_{2,\,1,\,4} = J_{2,\,1,\,5} = 1$ is consequently defined.
This procedure is used for all elements.
The only time this fill pattern changes is when two members share a node and fall in the same line.
This happens for member $x=3$ at node $g=2$, that connects to members $i=4$, $i=7$, and $i=8$.
Here, member $i=8$ is in the same line as member $x=3$.
For the connection between $x=3$ and $i=8$, the matrix entry remains zero and is flagged as important ($J_{2,\,3,\,8}=0^*$) as in $\Gamma_{xi}^{(2)}$. 

\begin{subequations} \label{eq:J_{gxi}_TensegrityExampleConstruction}
    \begin{align}
    J_{1,\,xi} =
    \begin{pNiceMatrix}[first-row,first-col]
        & \textcolor{myPink}{i=1} & \textcolor{myPink}{i=2} & \textcolor{myPink}{i=3} & \textcolor{myPink}{i=4} & \textcolor{myPink}{i=5} & \textcolor{myPink}{i=6} & \textcolor{myPink}{i=7} & \textcolor{myPink}{i=8} \\
    \textcolor{myPink}{x=1} &  0  &  1  &  1  &  0  &  0  &  0  &  0  &  0  \\
    \textcolor{myPink}{x=2} &  1  &  0  &  1  &  0  &  0  &  0  &  0  &  0  \\
    \textcolor{myPink}{x=3} &  1  &  1  &  0  &  0  &  0  &  0  &  0  &  0  \\
    \textcolor{myPink}{x=4} &  1  &  0  &  0  &  0  &  1  &  0  &  0  &  0  \\
    \textcolor{myPink}{x=5} &  1  &  0  &  0  &  1  &  0  &  0  &  0  &  0  \\
    \textcolor{myPink}{x=6} &  0  &  1  &  0  &  0  &  0  &  0  &  1  &  0  \\
    \textcolor{myPink}{x=7} &  0  &  1  &  0  &  0  &  0  &  1  &  0  &  0  \\
    \textcolor{myPink}{x=8} &  0  &  0  &  0  &  0  &  1  &  1  &  0  &  0
    \end{pNiceMatrix} \label{eq:J_{gxi}_TensegrityExampleConstructiona}\\
    J_{2,\,xi} =
    \begin{pNiceMatrix}[first-row,first-col]
        & \textcolor{myPink}{i=1} & \textcolor{myPink}{i=2} & \textcolor{myPink}{i=3} & \textcolor{myPink}{i=4} & \textcolor{myPink}{i=5} & \textcolor{myPink}{i=6} & \textcolor{myPink}{i=7} & \textcolor{myPink}{i=8} \\
    \textcolor{myPink}{x=1} &  0  &  0  &  0  &  1  &  1  &  0  &  0  &  0  \\
    \textcolor{myPink}{x=2} &  0  &  0  &  0  &  0  &  0  &  1  &  1  &  0  \\
    \textcolor{myPink}{x=3} &  0  &  0  &  0  &  1  &  0  &  0  &  1  &0^{*}\\
    \textcolor{myPink}{x=4} &  0  &  0  &  1  &  0  &  0  &  0  &0^{*}&  1  \\
    \textcolor{myPink}{x=5} &  0  &  0  &  0  &  0  &  0  &  1  &  0  &  1  \\
    \textcolor{myPink}{x=6} &  0  &  0  &  0  &  0  &  1  &  0  &  0  &  1  \\
    \textcolor{myPink}{x=7} &  0  &  0  &  1  &0^{*}&  0  &  0  &  0  &  1  \\
    \textcolor{myPink}{x=8} &  0  &  0  &0^{*}&  1  &  0  &  0  &  1  &  0
    \end{pNiceMatrix} \label{eq:J_{gxi}_TensegrityExampleConstructionb}
    \end{align}
\end{subequations}

Similarly, the construction of the entries $\Upsilon_{gxi}$ for the same ground structure is illustrated in Eq. \eqref{eq:Y_{gxi}_TensegrityExampleConstruction}.
As before, this matrix is initialized with zeros.
The only entries that are updated are the diagonals and any locations that were flagged during the construction of the entries $J_{gxi}$.
At each node $g$ of every member $x$, the diagonal entry ($i=x$) and any flagged indices where $J_{gxi} = 0^*$ are populated with the sum of the corresponding row in the relation matrix $J_{gxi}$ ($\Upsilon_{gxx} = \Upsilon_{gxi^*} = \sum_{i=1}^N J_{gxi}$).
At node $g=2$ of member $x=3$, the sum of the row in the relation matrix is $\sum_{i=1}^N J_{2,\,3,\,i} = 2$.
Thus, the diagonal entry of row 3 is $\Upsilon_{2,\,3,\,3} = 2$ and the flagged entry is $\Upsilon_{2,\,3,\,8} = 2$.

\begin{subequations} \label{eq:Y_{gxi}_TensegrityExampleConstruction}
    \begin{align}
    \Upsilon_{1,\,xi} =
    \begin{pNiceMatrix}[first-row,first-col]
        & \textcolor{myPink}{i=1} & \textcolor{myPink}{i=2} & \textcolor{myPink}{i=3} & \textcolor{myPink}{i=4} & \textcolor{myPink}{i=5} & \textcolor{myPink}{i=6} & \textcolor{myPink}{i=7} & \textcolor{myPink}{i=8} \\
    \textcolor{myPink}{x=1} &  2  &  0  &  0  &  0  &  0  &  0  &  0  &  0  \\
    \textcolor{myPink}{x=2} &  0  &  2  &  0  &  0  &  0  &  0  &  0  &  0  \\
    \textcolor{myPink}{x=3} &  0  &  0  &  2  &  0  &  0  &  0  &  0  &  0  \\
    \textcolor{myPink}{x=4} &  0  &  0  &  0  &  2  &  0  &  0  &  0  &  0  \\
    \textcolor{myPink}{x=5} &  0  &  0  &  0  &  0  &  2  &  0  &  0  &  0  \\
    \textcolor{myPink}{x=6} &  0  &  0  &  0  &  0  &  0  &  2  &  0  &  0  \\
    \textcolor{myPink}{x=7} &  0  &  0  &  0  &  0  &  0  &  0  &  2  &  0  \\
    \textcolor{myPink}{x=8} &  0  &  0  &  0  &  0  &  0  &  0  &  0  &  2
    \end{pNiceMatrix} \\
    \Upsilon_{2,\,xi} =
    \begin{pNiceMatrix}[first-row,first-col]
        & \textcolor{myPink}{i=1} & \textcolor{myPink}{i=2} & \textcolor{myPink}{i=3} & \textcolor{myPink}{i=4} & \textcolor{myPink}{i=5} & \textcolor{myPink}{i=6} & \textcolor{myPink}{i=7} & \textcolor{myPink}{i=8} \\
    \textcolor{myPink}{x=1} &  2  &  0  &  0  &  0  &  0  &  0  &  0  &  0  \\
    \textcolor{myPink}{x=2} &  0  &  2  &  0  &  0  &  0  &  0  &  0  &  0  \\
    \textcolor{myPink}{x=3} &  0  &  0  &  2  &  0  &  0  &  0  &  0  &  2  \\
    \textcolor{myPink}{x=4} &  0  &  0  &  0  &  2  &  0  &  0  &  2  &  0  \\
    \textcolor{myPink}{x=5} &  0  &  0  &  0  &  0  &  2  &  0  &  0  &  0  \\
    \textcolor{myPink}{x=6} &  0  &  0  &  0  &  0  &  0  &  2  &  0  &  0  \\
    \textcolor{myPink}{x=7} &  0  &  0  &  0  &  2  &  0  &  0  &  2  &  0  \\
    \textcolor{myPink}{x=8} &  0  &  0  &  2  &  0  &  0  &  0  &  0  &  2
    \end{pNiceMatrix}
    \end{align}
\end{subequations}

Once the entries $J_{gxi}$ and $\Upsilon_{gxi}$ have been constructed, the updated tensegrity constraint that allows co-linear compression elements to co-exist at nodes with no other active elements is defined as:

\begin{align} \label{eq:Tensegrity_Violation_2}
    & \sum_{i=1}^{N} \Upsilon_{gxi} z_{i}^{(2)} + \sum_{i=1}^N \sum_{j=1}^M J_{gxi}z_{ij} \leq 2\Upsilon_{gxx}
    && \forall \ x=1,...\,,N,\: \forall \ g=1,\,2
\end{align}
where $\Upsilon_{gxx}$ is the diagonal entries of $\Upsilon_{gxi}$.
The first term of Eq. \eqref{eq:Tensegrity_Violation_2} checks if two co-linear members are, in fact, in compression by multiplying the entries $\Upsilon_{gxi}$ with the entries of the compression variable, $z_i^{(2)}$. Recall here that if the member is either zero force or in tension, then  $z_i^{(2)}=0$.
The second term in Eq. \eqref{eq:Tensegrity_Violation_2} identifies other active elements with $z_{ij}=1$ through multiplication with the entries $J_{gxi}$.
The right-hand side of Eq. \eqref{eq:Tensegrity_Violation_2} ensures that the constraint is always satisfied at nodes that have no co-linear elements. 
When co-linear elements exist at a node, the right-hand side guarantees that, if the co-linear elements are active and in compression, no other elements can exist. 
This is achieved by multiplying the diagonal term with a factor 2 because it is checking if two members are co-linear.

\paragraph{MILP Formulation with Prestressing and Tensegrity Constraints ---}
When all of the prestressing and tensegrity constraints are added, the problem formulation is expanded to the following form:

\begin{SafeSubequations} \label{eq:Tensegrity_Full_Formulation}
\begin{align}
    \nextsbeq & \underset{A_{ij},\: a_{ij},\: q_{ih},\:q^0_i,\:q^e_{ih},\:u_{kh},\:v_{ih},\: z_{ij},\:z_i^{(2)}}{\text{minimize}}
    && \sum_{i=1}^{N} \sum_{j=1}^{M} l_{i}A_{ij}C_{j}
    \label{eq:Tensegrity_Full_Objective} \\
    \nextsbeq & \text{subject to}
    && \sum_{i=1}^{N} B_{ki}q_{ih} = F_{kh}
    && \forall \ k,\: \forall \ h
    \label{eq:Tensegrity_Full_Equilibrium} \\
    \nextsbeq &&& 
    \sum_{i=1}^N B_{ki}q^0_i = 0
    && 
    \forall \ i
    \label{eq:Tensegrity_Full_Equilibrium_Prestress} \\
    \nextsbeq &&& 
    \sum_{k=1}^{N^F} A_{ik}^{f}u_{kh} = v_{ih}
    && 
    \forall \ i,\: \forall \ h
    \label{eq:Tensegrity_Full_Compatibility} \\
    \nextsbeq &&& 
    q_{ih} = q^e_{ih} + q^0_i
    && 
    \forall \ i,\: \forall \ h
    \label{eq:Tensegrity_Full_Prestress_Force_Sum} \\
    \nextsbeq &&& 
    q^e_{ih} \leq \frac{E_j}{l_i} \left( A_{ij}^{\max}v_{ih} + v_{ij}^{\min}A_{ij} - A_{ij}^{\max}v_{ij}^{\min} \right) + \mathcal{M}\left( 1 - z_{ij} \right)
    && 
    \forall \ i,\: \forall \ j,\: \forall \ h
    \label{eq:Tensegrity_Full_McCormick_Over_1} \\
    \nextsbeq &&& 
    q^e_{ih} \leq \frac{E_j}{l_i} \left( A_{ij}^{\min}v_{ih} + v_{ij}^{\max}A_{ij} - A_{ij}^{\min}v_{ij}^{\max} \right) + \mathcal{M}\left( 1 - z_{ij} \right)
    && 
    \forall \ i,\: \forall \ j,\: \forall \ h
    \label{eq:Tensegrity_Full_McCormick_Over_2} \\
    \nextsbeq &&& 
    q^e_{ih} \geq \frac{E_j}{l_i} \left( A_{ij}^{\min}v_{ih} + v_{ij}^{\min}A_{ij} - A_{ij}^{\min}v_{ij}^{\min} \right) - \mathcal{M}\left( 1 - z_{ij} \right)
    && 
    \forall \ i,\: \forall \ j,\: \forall \ h
    \label{eq:Tensegrity_Full_McCormick_Under_1} \\
    \nextsbeq &&& 
    q^e_{ih} \geq \frac{E_j}{l_i} \left( A_{ij}^{\max}v_{ih} + v_{ij}^{\max}A_{ij} - A_{ij}^{\max}v_{ij}^{\max} \right) - \mathcal{M}\left( 1 - z_{ij} \right)
    && 
    \forall \ i,\: \forall \ j,\: \forall \ h
    \label{eq:Tensegrity_Full_McCormick_Under_2} \\
    \nextsbeq &&& 
    q^0_i \leq -q^{0_c}z^{(2)}_i + A^{\max}_{\max}\sigma^t_{\max}\left(1-z^{(2)}_i\right)
    && 
    \forall \ i
    \label{eq:Tensegrity_Full_Min_Prestressing_Compression} \\
    \nextsbeq &&& 
    q^0_i \geq \left(q^{0_t} + \sigma^c_{\max}A^{\max}_{\max}\right)\left(\sum_{j=1}^M z_{ij}-z^{(2)}_i\right) - \sigma^c_{\max}A^{\max}_{\max}
    && 
    \forall \ i
    \label{eq:Tensegrity_Full_Min_Prestressing_Tension} \\
    \nextsbeq &&& 
    q^0_i \leq \sum_{j=1}^M A_{ij}\sigma^t_j
    && 
    \forall \ i
    \label{eq:Tensegrity_Full_Max_Prestressing_Compression} \\
    \nextsbeq &&& 
    q^0_i \geq -\sum_{j=1}^M A_{ij}\sigma^c_j
    && 
    \forall \ i
    \label{eq:Tensegrity_Full_Max_Prestressing_Tension} \\
    \nextsbeq &&& q_{ih} \leq \sum_{j=1}^{M} A_{ij}\sigma_j^{t}
    && \forall \ i,\: \forall \ h
    \label{eq:Tensegrity_Full_Tension_Limit} \\
    \nextsbeq &&& q_{ih} \geq - \sum_{j=1}^{M} A_{ij}\sigma_j^{c}
    && \forall \ i,\: \forall \ h
    \label{eq:Tensegrity_Full_Compression_Limit} \\
    \nextsbeq &&& \sum_{j=1}^{M} z_{ij} \leq 1
    && \forall \ i
    \label{eq:Tensegrity_Full_Unique_Material} \\
    \nextsbeq &&& \sum_{i=1}^{N} \sum_{j=1}^{M} n_{pi}z_{ij} \leq \eta_{p}^{\max}
    && \forall \ p
    \label{eq:Tensegrity_Full_Node} \\
    \nextsbeq &&& \sum_{i=1}^{N} \sum_{j=1}^{M} \Gamma_{xi}z_{ij} \leq \Gamma_{xx}
    && \forall \ x
    \label{eq:Tensegrity_Full_Angle} \\
    \nextsbeq &&& \sum_{i=1}^{N} \Gamma_{xi}^{(2)}z_{i}^{(2)} \leq \Gamma_{xx}^{(2)}
    && \forall \ x
    \label{eq:Tensegrity_Full_Tens_1_Mem} \\
    \nextsbeq &&& \sum_{i=1}^{N} \Upsilon_{gxi} z_{i}^{(2)} + \sum_{i=1}^N \sum_{j=1}^M J_{gxi}z_{ij} \leq 2\Upsilon_{gxx}
    && \forall \ x,\: \forall \ g
    \label{eq:Tensegrity_Full_Tens_2_Mem} \\
    \nextsbeq &&& q_{ih} \leq \left(\sigma_{\max}^{t}A_{\max}^{\max} + \frac{\sigma_{\max}^{t}A_{\max}^{\max}\varepsilon}{1-\varepsilon}\right)\left(1-\varepsilon-z_{i}^{(2)}\right)
    && \forall \ i,\: \forall \ h
    \label{eq:Tensegrity_Full_0_or_Tension} \\
    \nextsbeq &&& q_{ih} \geq -\sigma_{\max}^{c}A_{\max}^{\max}z_{i}^{(2)}
    && \forall \ i,\: \forall \ h
    \label{eq:Tensegrity_Full_Compression_Var} \\
    \nextsbeq &&& A_{ij} \leq A_{ij}^{\max}z_{ij}
    && \forall \ i,\: \forall \ j
    \label{eq:Tensegrity_Full_Max_Area} \\
    \nextsbeq &&& A_{ij} \geq a_{ij} - A_{ij}^{\max}\left(1 - z_{ij}\right)
    && \forall \ i,\: \forall \ j
    \label{eq:Tensegrity_Full_BigM} \\
    \nextsbeq &&& A_{ij} \geq 0
    && \forall \ i,\: \forall \ j
    \label{eq:Tensegrity_Full_Min_Area} \\
    \nextsbeq &&& a_{ij} \leq A_{ij}^{\max}
    && \forall \ i,\: \forall \ j
    \label{eq:Tensegrity_Full_a_Max} \\
    \nextsbeq &&& a_{ij} \geq A_{ij}^{\min}
    && \forall \ i,\: \forall \ j
    \label{eq:Tensegrity_Full_a_Min} \\
    \nextsbeq &&& z_{ij} \in \{0,1\}
    && \forall \ i,\: \forall \ j
    \label{eq:Tensegrity_Full_Binary_z_{ij}} \\
    \nextsbeq &&& z_i^{(2)} \in \{0,1\}
    && \forall \ i
    \label{eq:Tensegrity_Full_Binary_z_i^{(2)}}
\end{align}
\end{SafeSubequations}

When comparing Eq. \eqref{eq:Tensegrity_Full_Formulation} to the multi-component MILP formulation with constructability constraints in Eq. \eqref{eq:MultiMaterialMILP}, a noticeable difference is the addition of the binary variable $z_i^{(2)}$ that detects if members are in compression.
The assignment of this variable is achieved through the constraints defined in Eqs. \eqref{eq:Tensegrity_Full_0_or_Tension} and \eqref{eq:Tensegrity_Full_Compression_Var}.
These constraints are applied for all $N$ candidate members across all $L^C$ load cases and are independent of their material.
The other new variables, $q^0_i$ and $q^e_{ih}$, are continuous and represent the prestressing force and elastic force due to applied external loads, respectively.
The total force in a member is now defined as the sum of these two variables, as noted in Eq. \eqref{eq:Tensegrity_Full_Prestress_Force_Sum}.
Since the McCormick relaxation applies to the elastic force, Eqs. \eqref{eq:Tensegrity_Full_McCormick_Over_1}, \eqref{eq:Tensegrity_Full_McCormick_Over_2}, \eqref{eq:Tensegrity_Full_McCormick_Under_1}, and \eqref{eq:Tensegrity_Full_McCormick_Under_2} now bound the elastic force variable, $q^e_{ih}$, not the total force variable, $q_{ih}$.
Equilibrium of the prestressing forces is ensured through Eq. \eqref{eq:Tensegrity_Full_Equilibrium_Prestress}.
The bounds on the prestressing forces are defined by Eqs. \eqref{eq:Tensegrity_Full_Min_Prestressing_Compression}, \eqref{eq:Tensegrity_Full_Min_Prestressing_Tension}, \eqref{eq:Tensegrity_Full_Max_Prestressing_Compression}, and \eqref{eq:Tensegrity_Full_Max_Prestressing_Tension}.

\begin{figure}[h!]
    \centering
    \includegraphics[width=0.95\linewidth, trim={1.5cm 1.5cm 1.5cm 1.5cm}, clip]{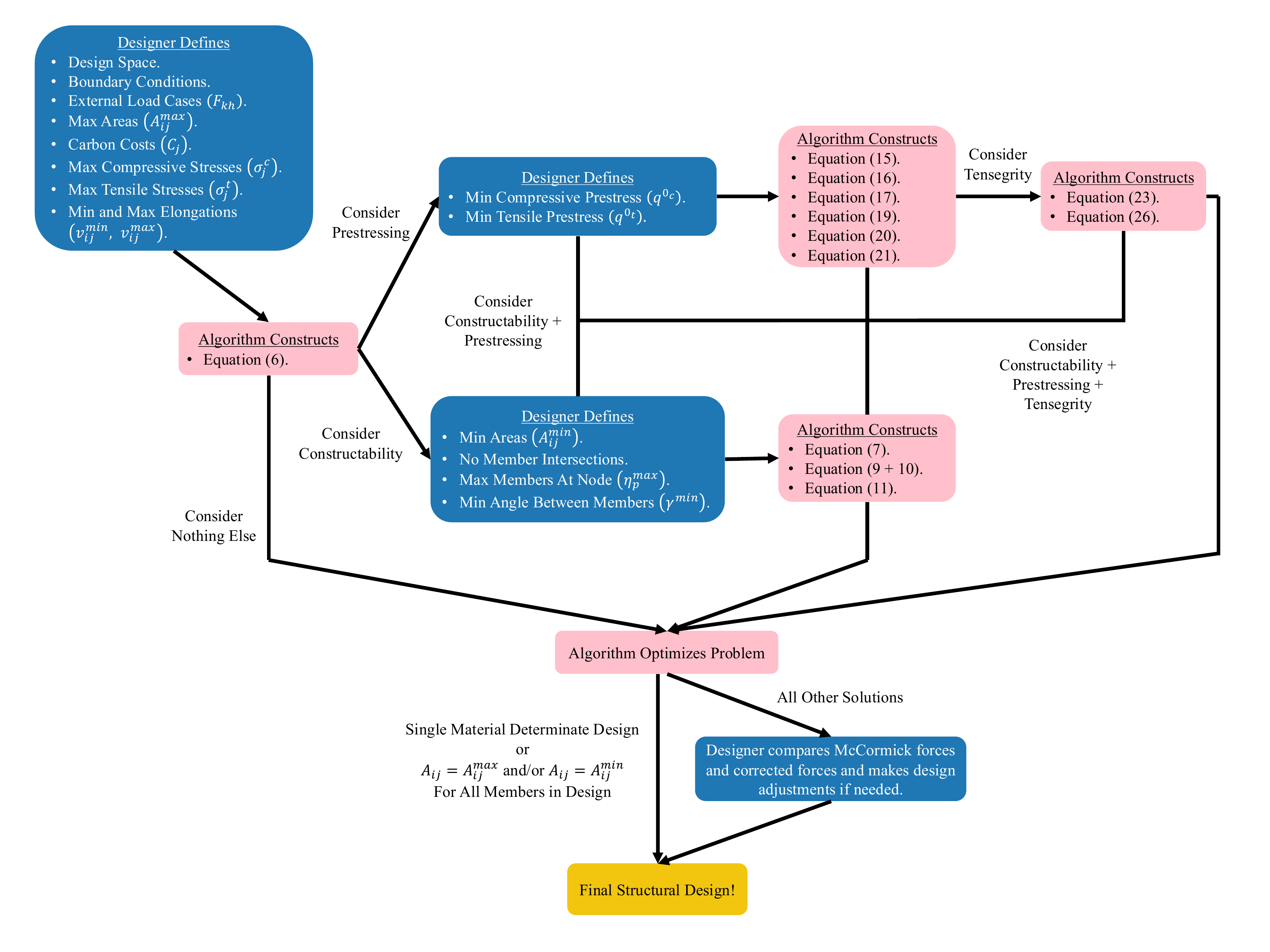}
    \caption{Optimization workflow and implementation of all design constraints.}
    \label{fig:Flow_Chart}
\end{figure}

Additionally, it can be noted that Eq. \eqref{eq:MultiMaterialMILPNodePerMaterial} is not transferred over.
This is because a nodal limit per component is not particularly useful when the tensegrity constraint severely limits the number of compressive components entering a node.
Since the tensile members in tensegrity structures are typically cables, which are allowed to have more complex overall connection designs, Eq. \eqref{eq:Tensegrity_Full_Node}, which limits the total number of members per node, is in this work deemed sufficient in reducing the number of active members at a node.
The tensegrity principles are upheld through the construction of the entries in the relationship matrices $\Gamma_{xi}^{(2)}$, $J_{gxi}$, and $\Upsilon_{gxi}$ in Eqs. \eqref{eq:Gamma_{xi}^{(2)}_TensegrityExampleConstruction}, \eqref{eq:J_{gxi}_TensegrityExampleConstruction}, and \eqref{eq:Y_{gxi}_TensegrityExampleConstruction}, respectively.
These ensure that only one compression member exists at a node, unless two compression members fall in the same line.
In this special case, both compression members can coexist to allow the stacking of compression members and the use of a reduced ground structure.
However, if two compression members share a node, no other members are allowed to exist at that shared node.
These rules are enforced by the constraints listed in Eqs. \eqref{eq:Tensegrity_Full_Tens_1_Mem} and \eqref{eq:Tensegrity_Full_Tens_2_Mem}. 
The optimization workflow and construction of all constraints is visually illustrated in Fig. \ref{fig:Flow_Chart}

\section{Results and Discussion}
\label{sec:Results}

The presented MILP formulations are used to solve low-carbon design problems for several 2D and 3D truss examples. 
The commercial solver Gurobi 12.0.1 \cite{gurobi} is used to solve all problems on a 2023 MacBook Pro with an Apple M3 Max chip containing a 16-core CPU, 40-core GPU, and 48 GB of RAM.

The formulation herein allows the design engineer to predefine the material properties they wish to use for all candidate components or materials within the design. 
The formulation is demonstrated with up to three component options: timber or steel bars (or struts), or steel cables.
The timber bars are assigned properties that mimic Douglas fir.
Steel bars and cables are assigned properties that mimic Grade 50 steel.
As mentioned, the maximum allowable stress for each material is reduced by a factor of safety of $\varphi = 2.5$ to ensure that all materials remain in the linear elastic range of their mechanical properties.
This also provides plenty of flexibility to correct any forces that were computed with error under the McCormick envelope approximating the constitutive material law.
The embodied carbon coefficient and tensile and compressive stress limits used are listed in Tab. \ref{tab:mat_probs}.
For tensegrity designs, timber and steel bars are assigned zero tensile capacity.
This is done to better illustrate the force flow through the final designs.

\begin{table}[http]
    \centering
\begin{tabular}{l c c c c} \hline
    & $\pmb{C}$ & $\pmb{\sigma^t}$ & $\pmb{\sigma^c}$ & $\pmb{E}$ \\ 
    & kgCO$_2$e/m$^3$ & MPa & MPa & MPa \\ \hline 
    Timber bar & 269.5 & 3.861 & 4.552 & 13,100 \\
    Steel bar & 11,382.5 & 138 & 138 & 200,000 \\
    Steel cable & 11,382.5 & 138 & 0 & 200,000 \\ \hline
\end{tabular}
    \caption{Embodied carbon coefficients, stress limits, and elastic moduli used for the numerical results.}
    \label{tab:mat_probs}
\end{table}

\subsection{Single-Material Design with Constructability Constraints} \label{subsec:Results_1_Material_Design}

\begin{figure}[ht]
    \centering
    \includegraphics[width=0.4\linewidth, trim={65cm 25cm 65cm 25cm}, clip]{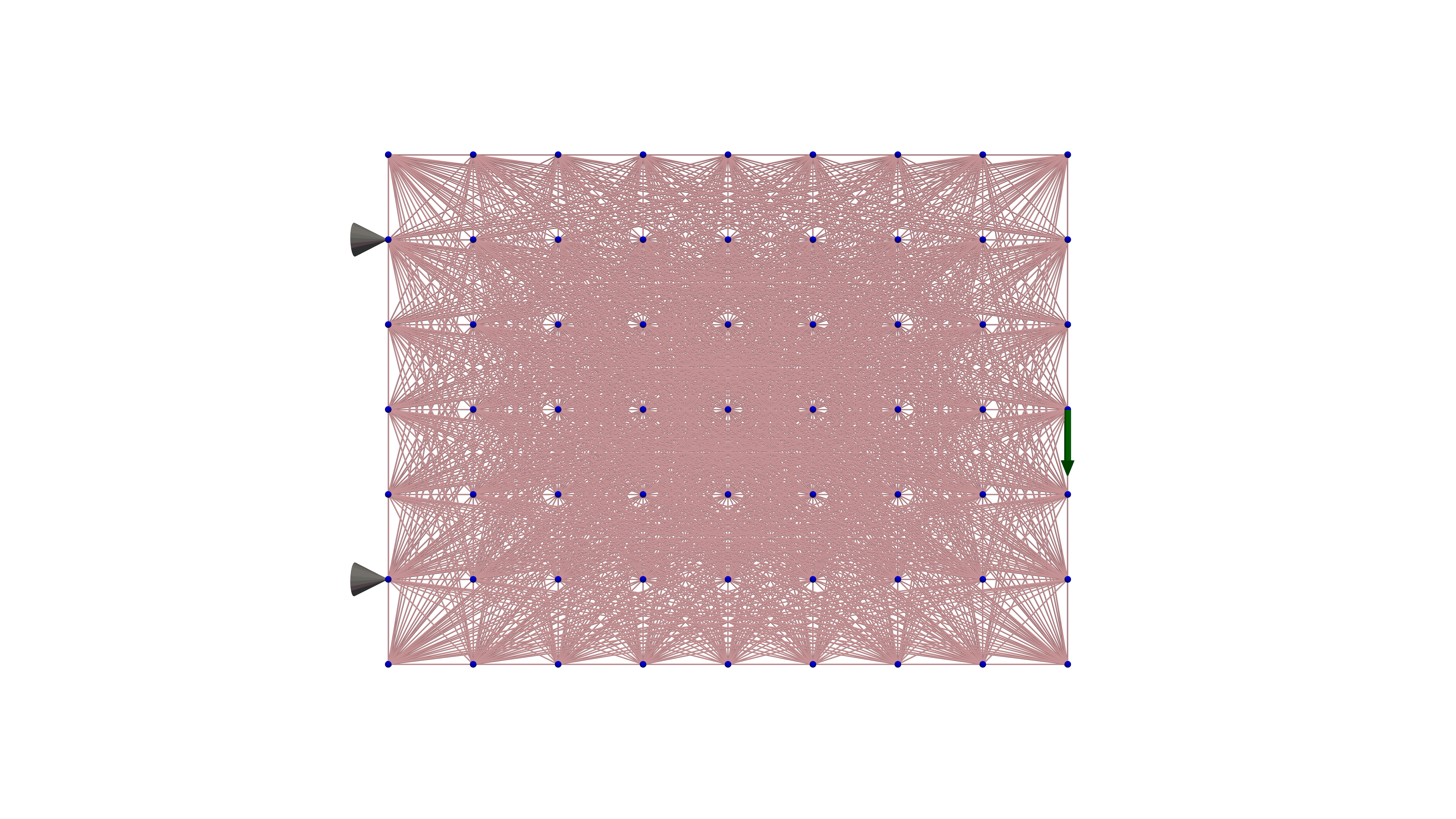}
    \caption{2D ground structure with 1,234 candidate members for cantilever Michell truss.}
    \label{fig:Result_Michell_Cantilever_Steel_1_Design_Space+Ground_Structure}
\end{figure}

Before extending to multi-material or component design, the effect of imposing the constructability constraints that eliminate hair-like cross-sections and reduce the nodal complexity is investigated.
For this, a 2D rectangular design space measuring 8 meters wide and 6 meters tall is considered.
It is supported by two pins along its left edge and a point load of 1,000 kN is applied at the center node on the right edge.
The boundary and loading conditions are designed to mimic the iconic cantilever Michell truss solution.
There is only one load case for this design $\left(L^C=1\right)$.
The design space is discretized with the ground structure shown in Fig. \ref{fig:Result_Michell_Cantilever_Steel_1_Design_Space+Ground_Structure}.
It has $9 \times 7$ nodes and a total connectivity mesh composed of 1,234 candidate members.
The elongation bounds used for the McCormick constraints are set to $v^{\min}_{ij} = -\sigma^c_j l_i \:/\:E_j$ and $v^{\max}_{ij} = \sigma^t_j l_i \:/\:E_j$ for all candidate members $i$ and materials $j$.

\subsubsection{Steel Design} \label{sec:steel_cantilever_design}

\begin{figure}[h!]
    \centering
    \begin{subfigure}[b]{0.4\textwidth}
        \centering
        \includegraphics[width=\linewidth, trim={65cm 25cm 65cm 25cm}, clip]{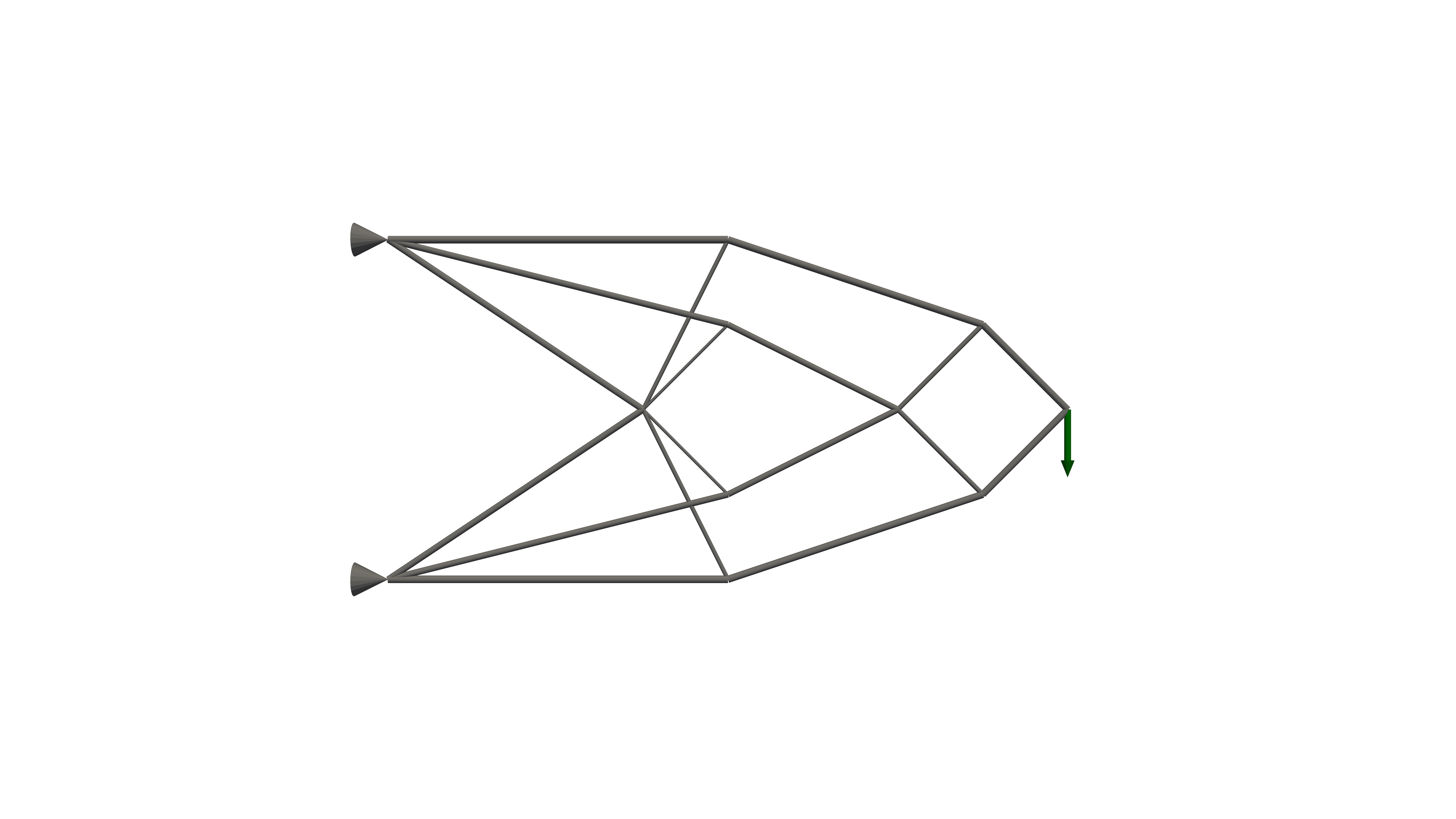}
        Embodied Carbon $= 2,375$ kgCO$_2$e \\
        Mass $= 1,638$ kg
        \caption{}
        \label{subfig:Result_Michell_Cantilever_Steel_1}
    \end{subfigure}
    \begin{subfigure}[b]{0.4\textwidth}
        \centering
        \includegraphics[width=\linewidth, trim={65cm 25cm 65cm 25cm}, clip]{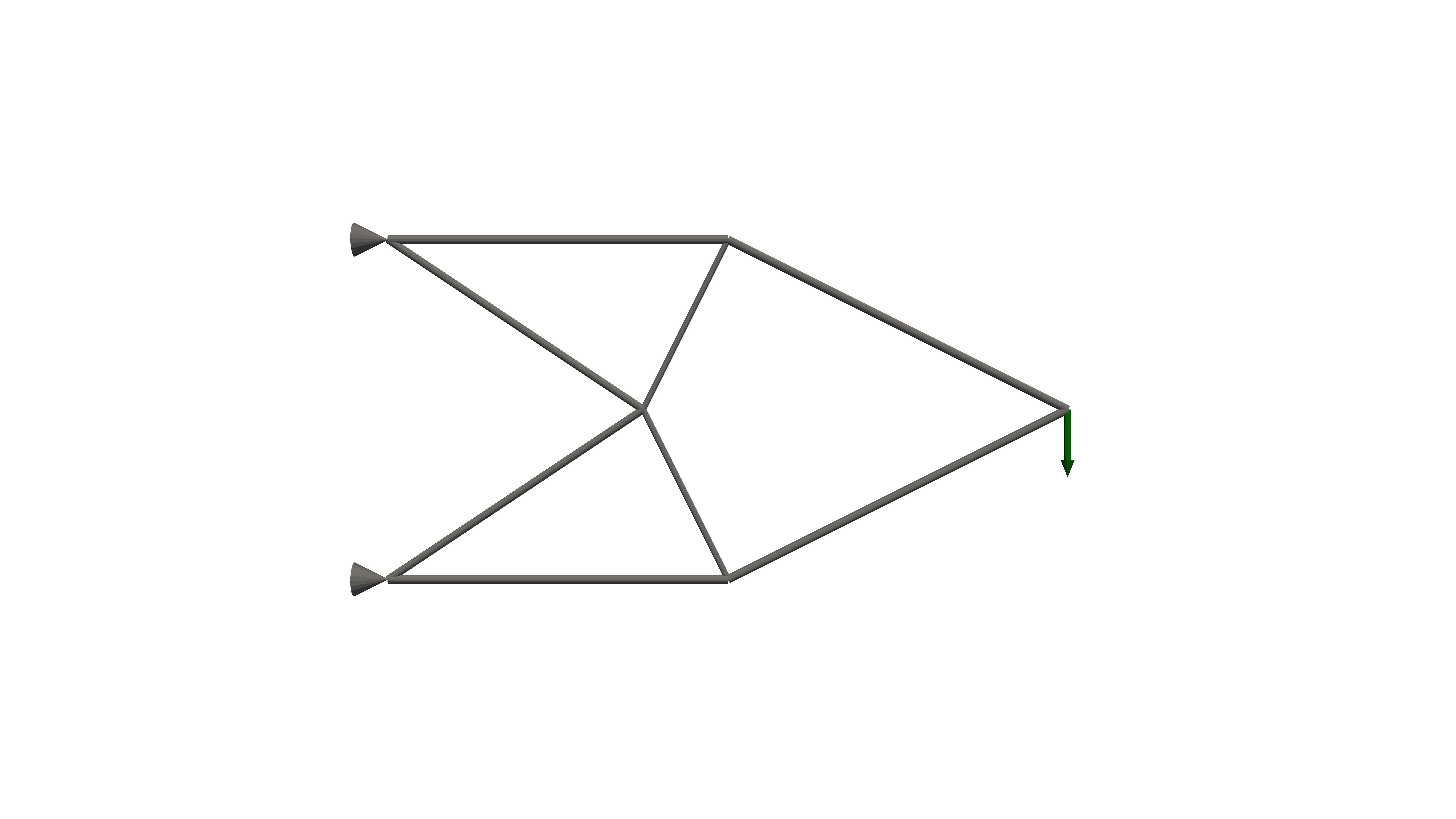}
        Embodied Carbon $= 2,392$ kgCO$_2$e \\
        Mass $=1,650$ kg
        \caption{}
        \label{subfig:Result_Michell_Cantilever_Steel_2}
    \end{subfigure}
    \caption{
    Minimum embodied carbon design results for Michell cantilever in steel;
    (a) without, and (b) with constructability constraints obtained by solving Eq. \eqref{eq:BasicMILP_Compatibility} and Eq. \eqref{eq:MultiMaterialMILP}, respectively.}
    \label{fig:Result_Michell_Cantilever_Steel}
\end{figure}

The design results for all steel specimens are shown in Fig. \ref{fig:Result_Michell_Cantilever_Steel}.
The result in Fig. \ref{subfig:Result_Michell_Cantilever_Steel_1} is achieved by applying only a maximum area constructability constraint of $A^{max}_{ij}=1,000$ cm$^2$.
Since there is no minimum area imposed and there is only one material ($M=1$), the problem can be reduced to the linear program formulated in Eq. \eqref{eq:LinearForm} with 2,468 continuous variables.
However, Eq. \eqref{eq:BasicMILP_Compatibility} is used to allow the solver the freedom to generate indeterminate solutions if there are no viable determinate solutions that satisfy the elastic strength and elongation requirements under the maximum area requirement.
This increases the variable count to 3,891 continuous variables and 1,234 binary variables.
The design is obtained in 3 minutes and 49 seconds.
As expected, the obtained design solution is symmetric.
Notice that the design has fewer members than the traditional minimum compliance Michell cantilever problem.
This is because the objective in this work is to minimize the embodied carbon, which, for this example with $M=1$, is equivalent to minimizing the material consumption.
Although the design does not appear overly complex from a total-member perspective, some of the connections will be challenging to fabricate.
This is particularly true for the center-left node, where three large cross-sections are converging to one node at sharp angles.
In the center left of the design, another complex node joins 6 members, some of which are separated by only 14.0$^\circ$.
Additionally, because the design is statically determinate and made of a single material, the design has 0\% error in the McCormick force and elongation approximations for all members.

To address some of the challenging complexities of the solution in Fig. \ref{subfig:Result_Michell_Cantilever_Steel_1}, the design is resolved using the MILP formulation in Eq. \eqref{eq:MultiMaterialMILP} considering constructability constraints.
The optimization problem now has 5,125 continuous variables and 1,234 binary variables.
As before, the maximum area requirement is kept at $A^{max}_{ij}=1,000$ cm$^2$. 
Additionally, a minimum area of $A^{min}_{ij}=10$ cm$^2$ is enforced, along with a maximum of $\eta_p^{max}=6$ members per joint and a minimum angle of separation of $\gamma^{\min} = 22.5^\circ$ between members.
The design result is the global optimum and is obtained in 5 minutes and 25 seconds and shown in Fig. \ref{subfig:Result_Michell_Cantilever_Steel_2}.
Similar to before, the solution is statically determinate and made of 1 material, making the McCormick envelope produce exact force and elongation results.
When comparing the two topologies with and without the added constructability constraints, one aesthetic similarity is that both designs are symmetric.
It is a known artifact of MILP truss topology optimization, however, that design results may not be symmetric \cite{stolpe2007reformulation}.
Therefore, this result cannot always be expected when enforcing the constructability constraints, even if designers are working with an isotropic material with identical tension and compression properties. 
Designers wishing to have symmetric designs with the MILP formulation must explicitly address this by applying a symmetry constraint or only modeling half of the design domain. 

A noticeable difference between the two topologies is that the constructability constrained design in Fig. \ref{subfig:Result_Michell_Cantilever_Steel_2} has fewer members and, more importantly, the joint complexity is reduced.
The most complicated joint connects 4 members and the smallest angle between members is 33.7$^\circ$.
The support nodes maintain multiple large cross-sections converging to one point, however they are now separated by significantly larger angles.
All this comes at the cost of a very slight 0.70\% increase in both the embodied carbon and weight of the design.

\subsubsection{Timber Design} \label{sec:timber_cantilever_design}

\begin{figure}[h!]
    \centering
    \begin{subfigure}[b]{0.4\textwidth}
        \centering
        \includegraphics[width=\linewidth, trim={65cm 25cm 65cm 25cm}, clip]{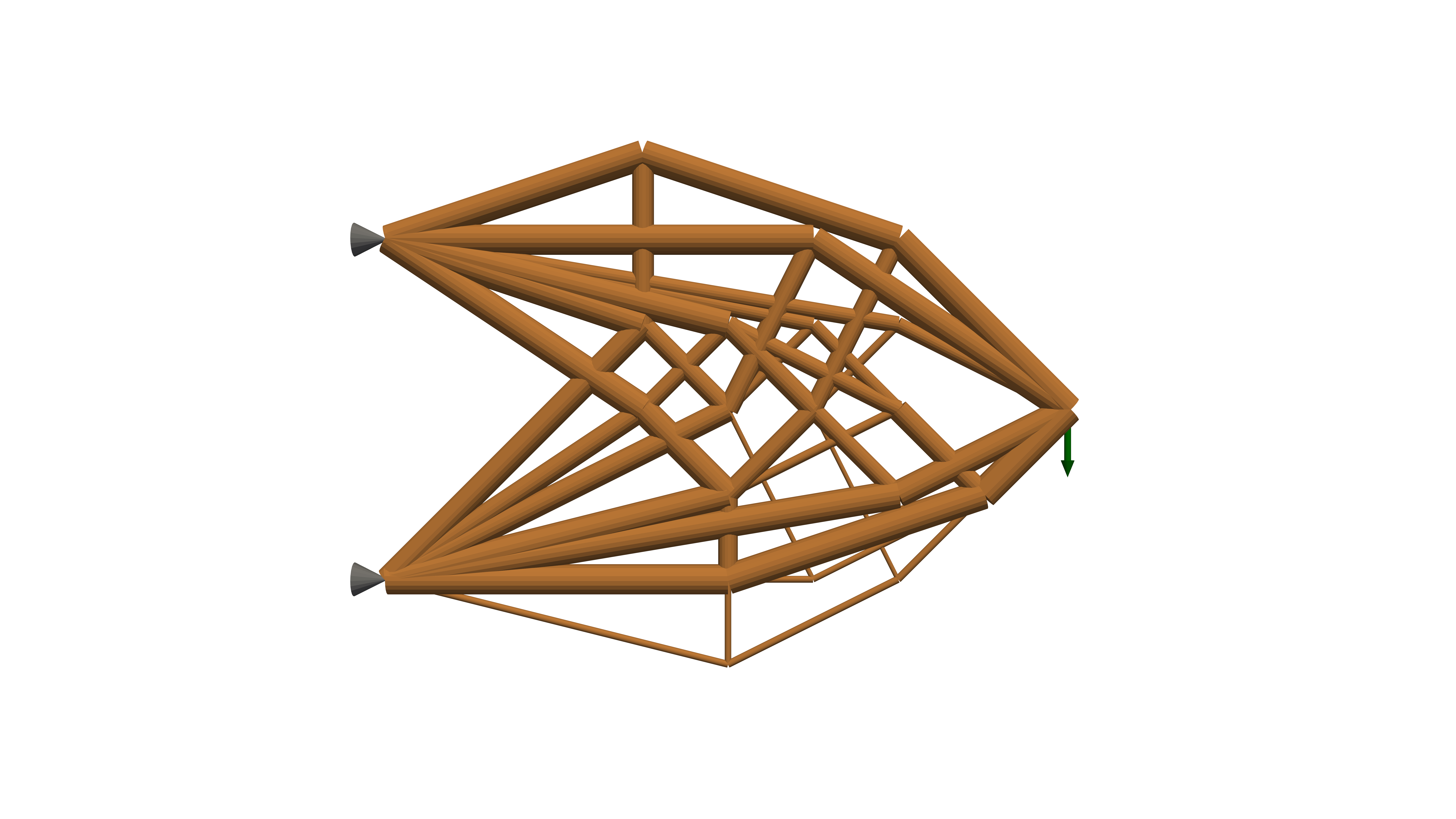}
        Embodied Carbon $= 1,891$ kgCO$_2$e \\
        Mass $= 3,578$ kg
        \caption{}
        \label{subfig:Result_Michell_Cantilever_Timber_1}
    \end{subfigure}
    \begin{subfigure}[b]{0.4\textwidth}
        \centering
        \includegraphics[width=\linewidth, trim={65cm 25cm 65cm 25cm}, clip]{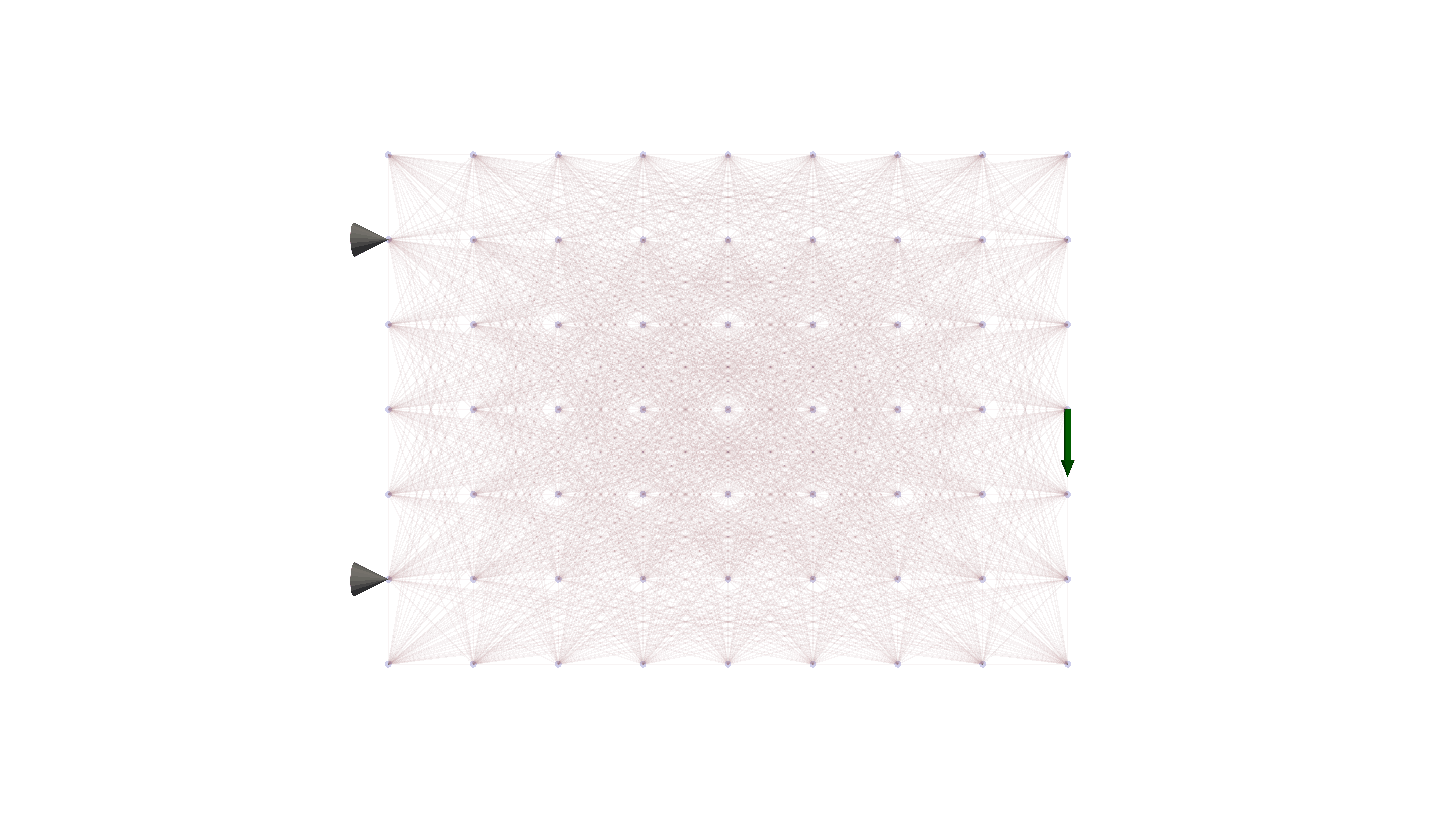}
        Embodied Carbon $=$ Infeasible \\
        Mass $=$ Infeasible
        \caption{}
        \label{subfig:Result_Michell_Cantilever_Timber_2}
    \end{subfigure}
    \caption{
    Minimum embodied carbon design results for Michell cantilever in timber;
    (a) without, and (b) with constructability constraints obtained by solving Eq. \eqref{eq:BasicMILP_Compatibility} and Eq. \eqref{eq:MultiMaterialMILP}, respectively.}
    \label{fig:Result_Michell_Cantilever_Timber}
\end{figure}

The same optimization scheme is run for exclusively timber designs and the results are shown in Fig. \ref{fig:Result_Michell_Cantilever_Timber}.
Fig. \ref{subfig:Result_Michell_Cantilever_Timber_1} shows the solution generated by solving the problem formulated in Eq. \eqref{eq:BasicMILP_Compatibility} with a maximum area of $A^{max}_{ij}=1,000$ cm$^2$.
Similar to the steel truss example above, using this equation instead of the linear formulation allows for indeterminate solutions if there are no viable determinate solutions satisfying elastic strength and elongation requirements under the maximum area constraint.
Identical to the steel truss, this problem has 3,891 continuous variables and 1,234 binary variables.
The solution is obtained in 3 minutes and 36 seconds.
Unlike the steel unconstrained steel solution shown in Fig. \ref{subfig:Result_Michell_Cantilever_Steel_1}, this design is not symmetric.
This is expected, however, as the compressive strength of timber is larger than its tensile strength.
The requires tension members to use more material than compression members under the same load and makes compression more attractive on a strength per material basis.
Additionally, this design is statically indeterminate with variable cross-sectional areas, producing some error in the force calculation under the McCormick constraints.
The extent of the inaccuracies is evaluated by recalculating the exact member forces for the final design.
The median absolute difference between the correct force and the estimated forces is 1.97\%.
Finally, this design looks very difficult to fabricate.
The connections at the supports have 7 members converging to one point, with some only being separated by 1.8$^\circ$.

The constructability constraints in Eq. \eqref{eq:MultiMaterialMILP} are applied to the problem with the aim to resolve difficult connections.
Identical to the steel example, a maximum of $\eta_p^{max}=6$ members per joint and a minimum angle of separation of $\gamma^{\min} = 22.5^\circ$ between members is enforced.
A minimum area of $A^{min}_{ij}=100$ cm$^2$ is enforced to provide a more suitable minimum area bound for timber.
The optimization problem now has 5,125 continuous variables and 1,234 binary variables.
However, applying this combination of realistic constraints makes the design problem infeasible, as shown in Fig. \ref{subfig:Result_Michell_Cantilever_Timber_2}.
Infeasibility is proved in 5 minutes and 14 seconds.
Because there is no solution that satisfies the constructability constraints and the unconstrained solution is not readily constructible, an all timber design may appear very unappealing for this design problem.

\subsection{Two-Material Design with Constructabiltiy Constraints} \label{subsec:Results_2_Material_Design}

The same design space, loading, boundary conditions, and ground structure are used to design new trusses where all components can either be timber or steel. 
As noted in Tab. \ref{tab:mat_probs}, timber has a much lower embodied carbon coefficient than steel.
It also has a substantially decreased elastic modulus and maximum stress capacity, with the maximum stress capacity varying for tension and compression.
The designs are obtained by setting $M=2$.

\begin{figure}[h!]
    \centering
    \begin{subfigure}[b]{0.3\textwidth}
        \centering
        \includegraphics[width=\linewidth, trim={65cm 25cm 65cm 2cm}, clip]{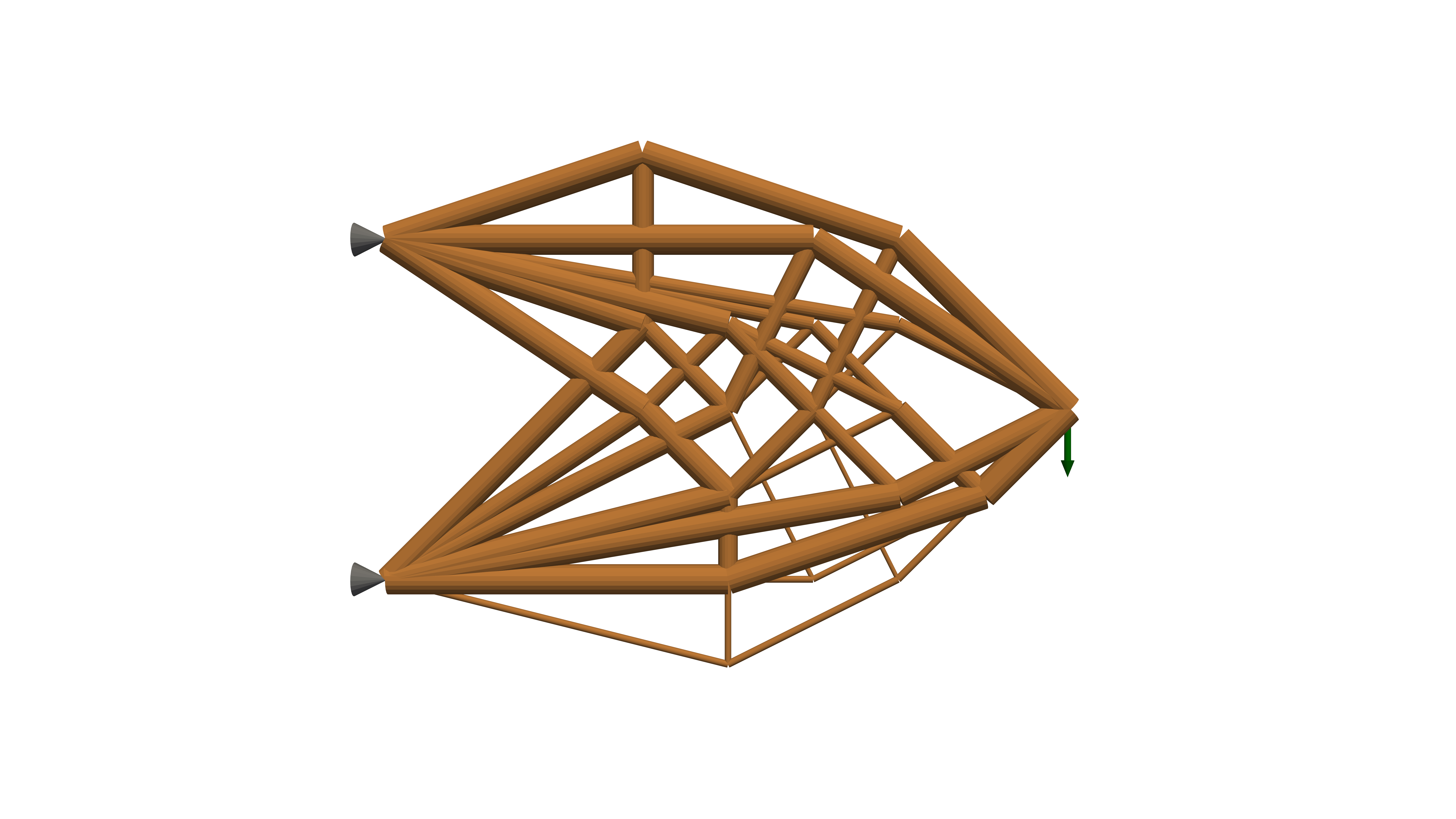}
        Embodied Carbon $= 1,891$ kgCO$_2$e \\
        Mass $= 3,578$ kg \\
        Timber Volume $= 100$\%
        \caption{}
        \label{subfig:Result_Michell_Cantilever_Hybrid_1}
    \end{subfigure}
    \begin{subfigure}[b]{0.3\textwidth}
        \centering
        \includegraphics[width=\linewidth, trim={65cm 25cm 65cm 25cm}, clip]{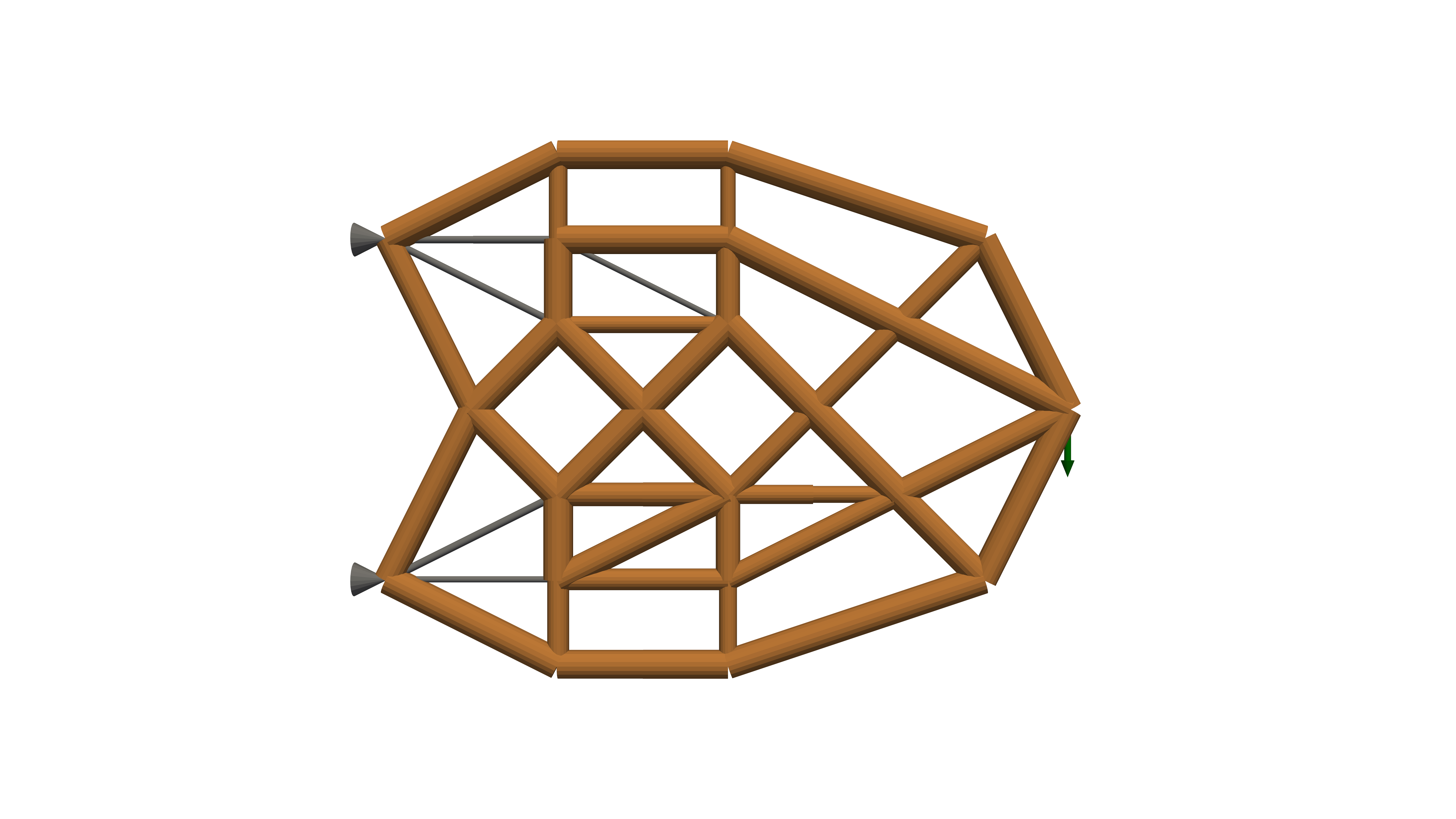}
        Embodied Carbon $= 2,163$ kgCO$_2$e \\
        Mass $= 3,306$ kg \\
        Timber Volume $= 99.0$\%
        \caption{}
        \label{subfig:Result_Michell_Cantilever_Hybrid_2}
    \end{subfigure}
    \begin{subfigure}[b]{0.3\textwidth}
        \centering
        \includegraphics[width=\linewidth, trim={65cm 25cm 65cm 25cm}, clip]{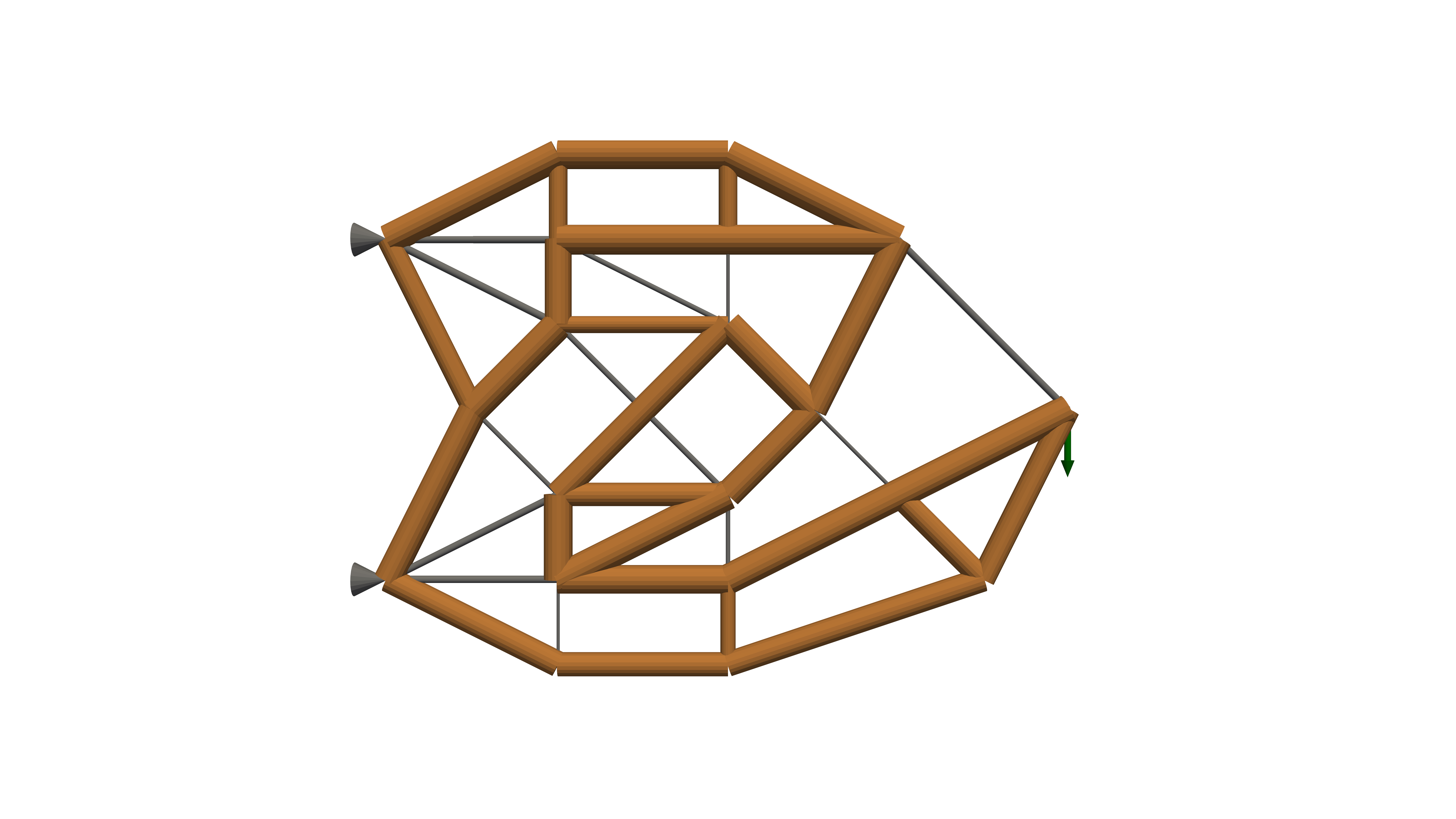}
        Embodied Carbon $= 2,196$ kgCO$_2$e \\
        Mass $= 2,869$ kg \\
        Timber Volume $= 97.8$\%
        \caption{}
        \label{subfig:Result_Michell_Cantilever_Hybrid_3}
    \end{subfigure}
    \caption{Minimum embodied carbon design results for timber-steel Michell cantilevers (a) without constructability constraints (Eq. \eqref{eq:BasicMILP}), (b) with constructability constraints (Eq. \eqref{eq:MultiMaterialMILP}), and (c) with an additional limit on the number of timber members per node.}
    \label{fig:Result_Michell_Cantilever_Hybrid}
\end{figure}

\begin{figure}[h!]
    \centering
    \begin{subfigure}[b]{0.3\textwidth}
        \centering
        \includegraphics[width=\linewidth, trim={65cm 25cm 65cm 25cm}, clip]{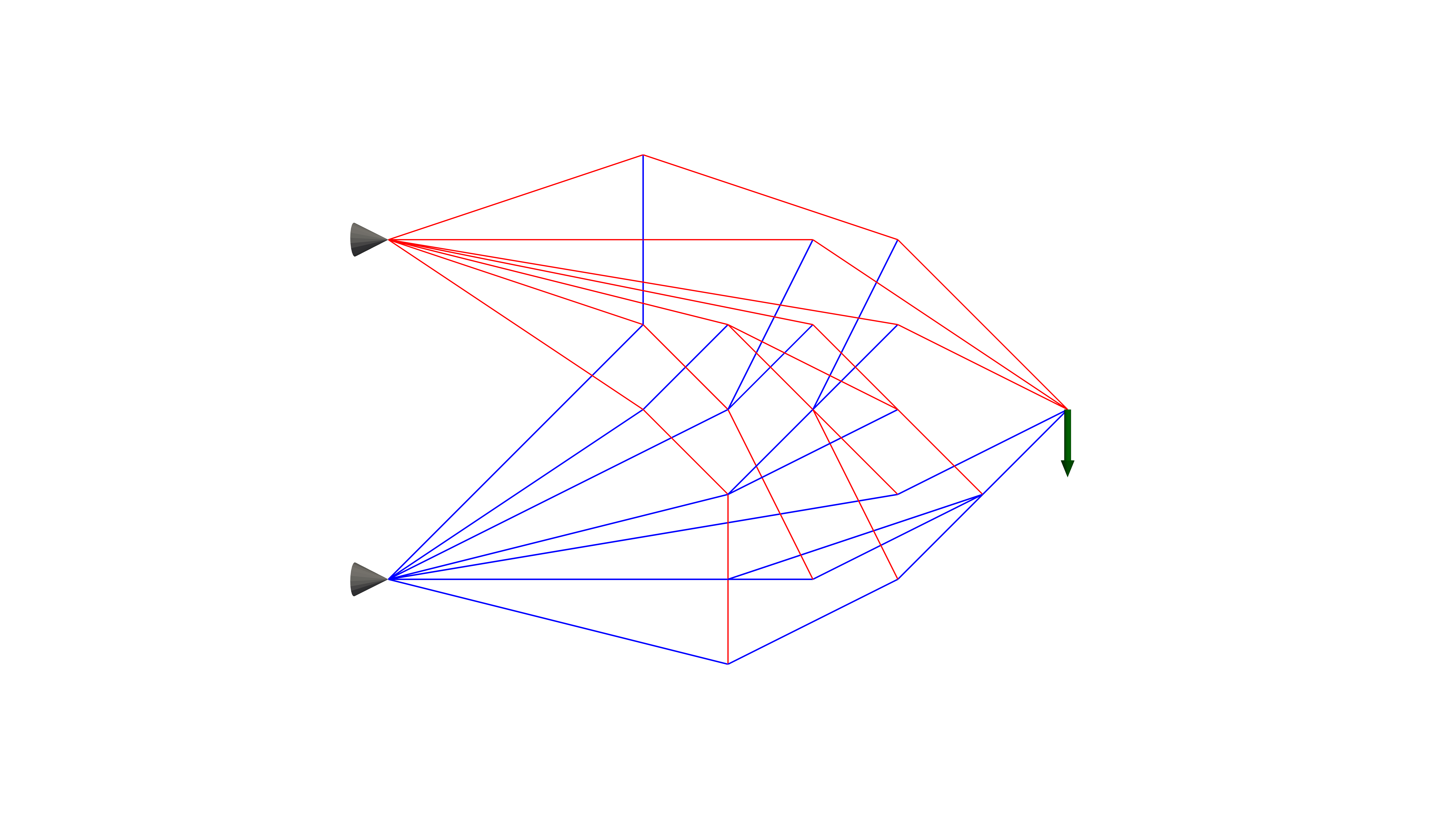}
        \caption{}
        \label{subfig:Result_Michell_Cantilever_Hybrid_1_Stresses}
    \end{subfigure}
    \begin{subfigure}[b]{0.3\textwidth}
        \centering
        \includegraphics[width=\linewidth, trim={65cm 25cm 65cm 25cm}, clip]{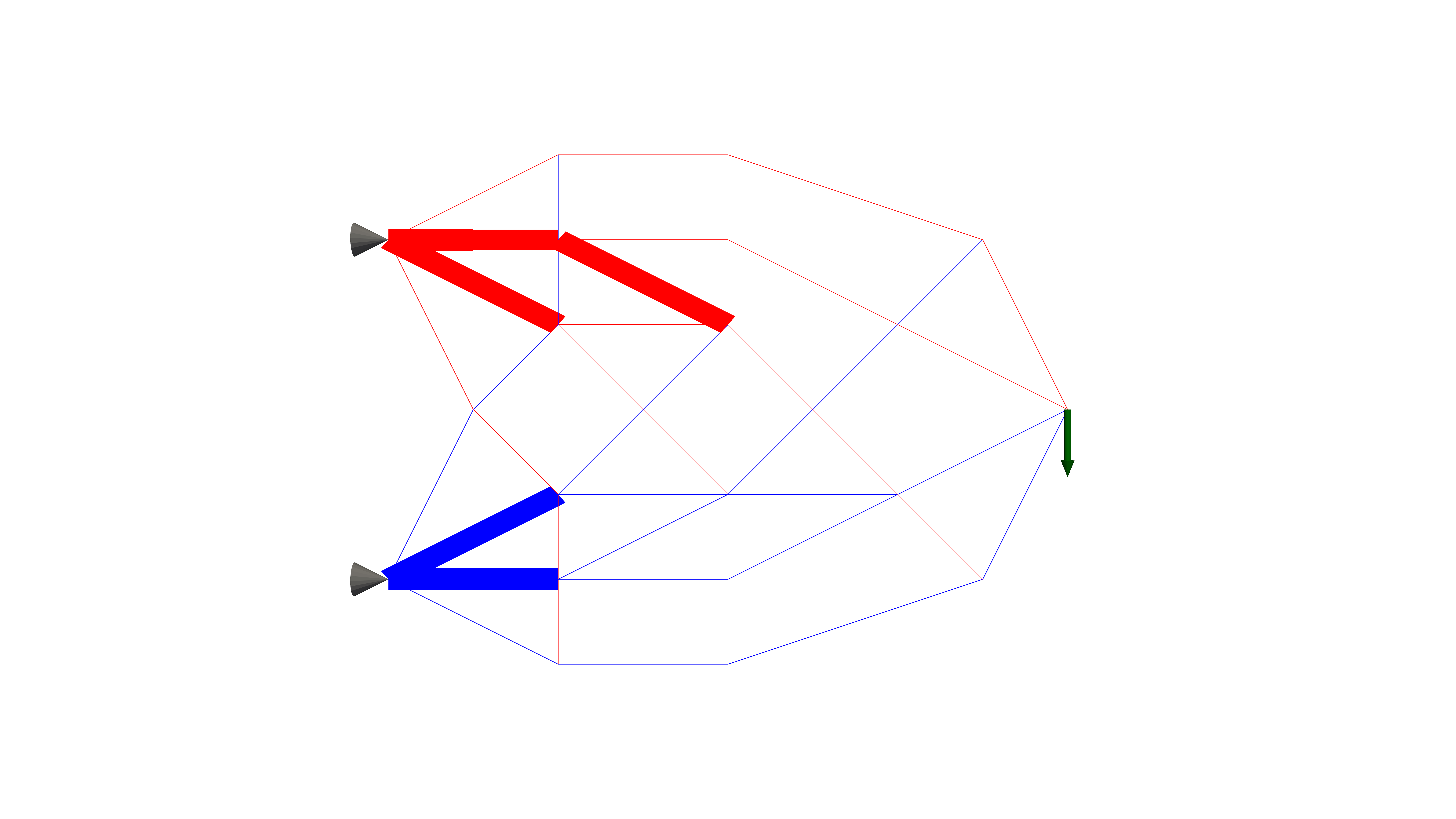}
        \caption{}
        \label{subfig:Result_Michell_Cantilever_Hybrid_2_Stresses}
    \end{subfigure}
    \begin{subfigure}[b]{0.3\textwidth}
        \centering
        \includegraphics[width=\linewidth, trim={65cm 25cm 65cm 25cm}, clip]{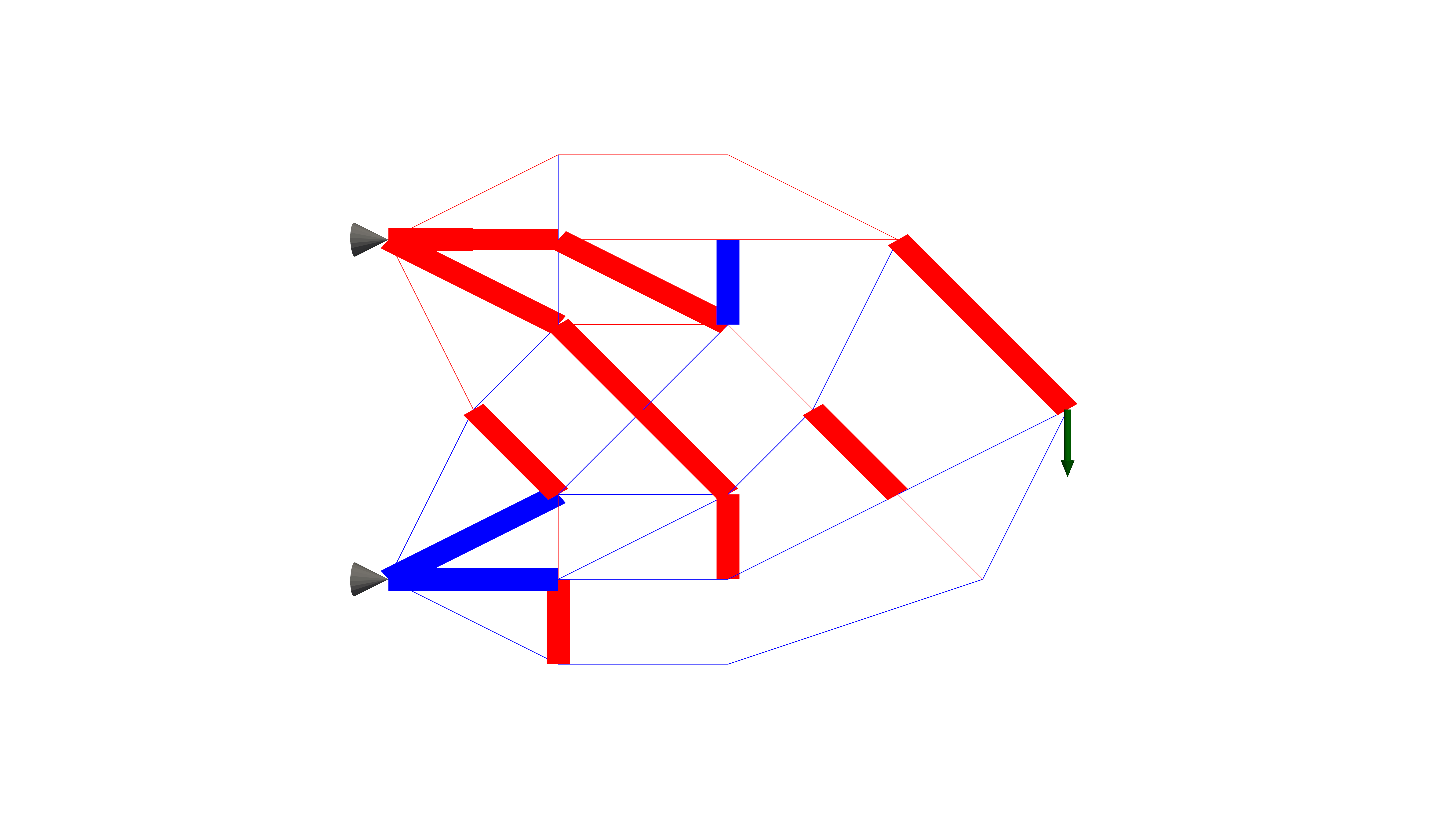}
        \caption{}
        \label{subfig:Result_Michell_Cantilever_Hybrid_3_Stresses}
    \end{subfigure}
    \caption{Member stresses for design results obtained in Fig. \ref{fig:Result_Michell_Cantilever_Hybrid} for designs (a), (b) and (c), respectively.
    In the stress plots, blue indicates compressive and red tensile stresses.
    The stress magnitude is scaled by the line thickness.
    (b) and (c) illustrate how the steel members carry much higher stresses than the timber members.}
    \label{fig:Result_Michell_Cantilever_Hybrid_Stresses}
\end{figure}

The design results for timber-steel hybrid structures under three different constraint scenarios are shown in Fig. \ref{fig:Result_Michell_Cantilever_Hybrid}.
The corresponding stress distributions are given in Fig. \ref{fig:Result_Michell_Cantilever_Hybrid_Stresses}.
The results in Fig. \ref{subfig:Result_Michell_Cantilever_Hybrid_1} give the solution to the simple MILP formulation in Eq. \eqref{eq:BasicMILP_Compatibility} without constructability constraints.
Eq. \eqref{eq:BasicMILP_Compatibility} requires 5,125 continuous and 2,468 binary variables for the given ground structure.
In addition to respecting the stress limits, the only constraint on the design is a maximum cross-sectional area bound of $A^{max}=1,000$ cm$^2$, regardless of the material.
This design is generated in 6 minutes and 12 seconds.
As indicated in the figure, the globally optimal solution has 0\% steel by volume.
It can also be observed that the solution produced here is identical to the solution from Fig. \ref{subfig:Result_Michell_Cantilever_Timber_1}.
This happens because its is preferable to use large quantities of timber for an embodied carbon objective.
The solution is thus using timber as much as possible while still satisfying the stress constraints.
As mentioned before, this structure can be considered very complex from a construction standpoint.
There are a total of 56 members, where multiple are at least partially overlapping or directly intersecting, making joints extremely complicated.

Fig. \ref{subfig:Result_Michell_Cantilever_Hybrid_2} shows the solution obtained when applying additional constructability constraints and solving Eq. \eqref{eq:MultiMaterialMILP}.
Introducing the minimum area bound increases the total number of continuous variables to 7,593.
The number of binary variables remains the same at 2,468.
A minimum cross-sectional area for timber and steel is enforced at $A_{ij}^{\min}=100$ cm$^2$ and $A_{ij}^{\min} = 10$ cm$^2$, respectively.
Joints are limited to $\eta_p^{\max}=6$ connecting members.
As there are no separate restrictions for timber and steel joint complexity, Eq. \eqref{eq:MultiMaterialMILPNodePerMaterial} is not applied for this example.
The minimum angle of separation between members is set to $\gamma^{\min} =22.5^\circ$ and no intersecting members are allowed to co-exist.
The design is obtained in 15 hours and 14 minutes.
The force inaccuracy is slightly under the McCormick relaxation higher for this problem, but the median absolute difference is still below 5\% (4.77\%).
Compared to the design without constructability constraints, 
the result in Fig. \ref{subfig:Result_Michell_Cantilever_Hybrid_2} is substantially easier to construct.
It has much fewer members and the most complicated joint only connects 6 members with the narrowest angle of separation being $\gamma = 26.6^\circ$.
Note that this is higher than the allowed $\gamma^{\min}$.
To reduce the joint complexity, the algorithm replaces a large number of weaker timber members with a small number of stronger steel members.
The stress plot in Fig. \ref{subfig:Result_Michell_Cantilever_Hybrid_2_Stresses} shows that the steel is substituted near the supports where multiple large forces are converging to a single spot.
It also shows that steel is slightly favored for replacing tension members.
The changes to the design resulting from the consideration of the constructability constraints causes a minor increase in the relative steel volume within the design from 0\% to 1.0\%, and consequently, the embodied carbon is increased by 14.4\%.
This happens while structural weight is reduced by 7.61\%.

Finally, a redesign is conducted with the additional constraint that only $\eta_{p,timber}^{\max}=3$ timber members are allowed to co-exist at a joint.
Typically, timber connections require more detailing than connections in steel, which can make it difficult to join a large number of timber members at one node.
This slight change results in the design in Fig. \ref{subfig:Result_Michell_Cantilever_Hybrid_3} when solving Eq. \eqref{eq:MultiMaterialMILP} in which Eq. \eqref{eq:MultiMaterialMILPNodePerMaterial} is now included.
The computation time for this example increases to 79 hours and 5 minutes.
The median absolute difference between the forces calculated under the McCormick relaxation and the corrected forces is 3.59\%.
While the design carries resemblance to the solution in Fig. \ref{subfig:Result_Michell_Cantilever_Hybrid_2}, there are some notable differences.
The total number of elements in the structure has gone down, but the new design increases the total number of steel members to 13.
This causes the steel volume to increase to 2.2\%.
The stress plot in Fig. \ref{subfig:Result_Michell_Cantilever_Hybrid_3_Stresses} shows that most steel elements are taking tensile forces.
Although the addition of the final constructability constraint causes noticeable changes to the topology, it only increases the embodied carbon by 1.54\% compared to the solution in Fig. \ref{subfig:Result_Michell_Cantilever_Hybrid_2}.
The structural weight is further reduced by 19.8\% compared to the initial design in Fig. \ref{subfig:Result_Michell_Cantilever_Hybrid_1}.
The most complex node connects 5 members, where 3 are made of timber and 2 are made of steel.
No nodes connects more than three timber members.
The minimum observed angle between members remains unchanged at 26.6$^\circ$.

The two timber-steel designs with constructability constraints, pictured in Fig. \ref{subfig:Result_Michell_Cantilever_Hybrid_2} and Fig. \ref{subfig:Result_Michell_Cantilever_Hybrid_3}, show a 9.57\% and 8.19\% decrease in embodied carbon compared to the all steel design with constructability constraints pictured in Fig. \ref{subfig:Result_Michell_Cantilever_Steel_2}.
The weight of these two designs, however, are 100.4\% and 73.9\% heavier.
The overall designs, while satisfying all constructability constraints, are also more complex than the 8 member steel cantilever in Fig. \ref{subfig:Result_Michell_Cantilever_Steel_2}.
For this Michell cantilever example, the embodied carbon reduction does not necessarily make the multi-material solutions clear winners.

\subsection{Effects of Individual Constructability Constraints} \label{subsec:Results_4_Constraint_Layers}

\begin{figure}[htp]
    \centering
    \includegraphics[width=0.7\linewidth, trim={0cm 0cm 0cm 0cm}, clip]{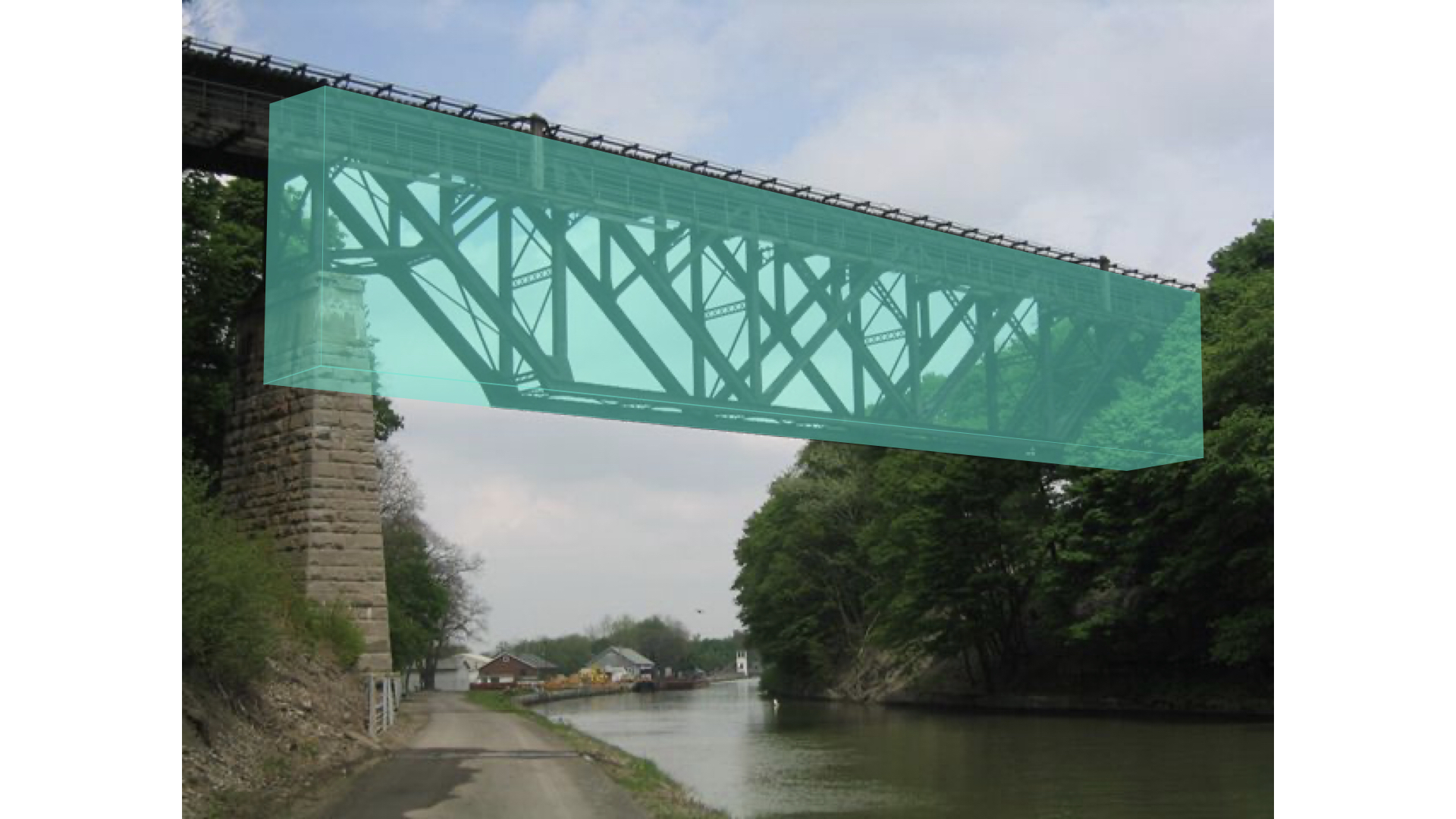}
    \caption{Lockport truss bridge passing over the Erie Canal near Buffalo, New York.
    The design space used in the following examples that aims to mimic this structure is highlighted in teal blue.}
    \label{fig:LockportTrussBridge}
\end{figure}

The results in Sections \ref{subsec:Results_1_Material_Design} and \ref{subsec:Results_2_Material_Design} show a dramatic difference between designs obtained with and without all the herein defined constructability constraints.
This section explores the effects of application of the individual considerations.
This is done using a 3D design example inspired by the Lockport rail truss bridge, shown in Fig. \ref{fig:LockportTrussBridge}, that crosses the Erie Canal near Buffalo, NY.
The dimensions of the design space are taken as $30 \times 5 \times 10$ m$^3$.

All of the following designs are obtained by solving versions of Eq. \eqref{eq:MultiMaterialMILP}.
To ensure that the bridge deck is properly supported, the top chords of the truss are forced to exist. 
Since the material of the top chords is not predetermined herein, their existence in the design solution is guaranteed by applying the following additional constraint: 
\begin{align} \label{eq:StringerConnectivity}
    & \sum_{j=1}^{M} z_{ij} = 1
    && \forall \ i \in S
\end{align}
where $S$ is the set that contains all candidate members in both top chords of the ground structure. 

\begin{figure}[ht]
    \centering
    \begin{subfigure}[b]{0.3\textwidth}
        \centering
        \includegraphics[width=\linewidth, trim={15cm 15cm 120cm 30cm}, clip]{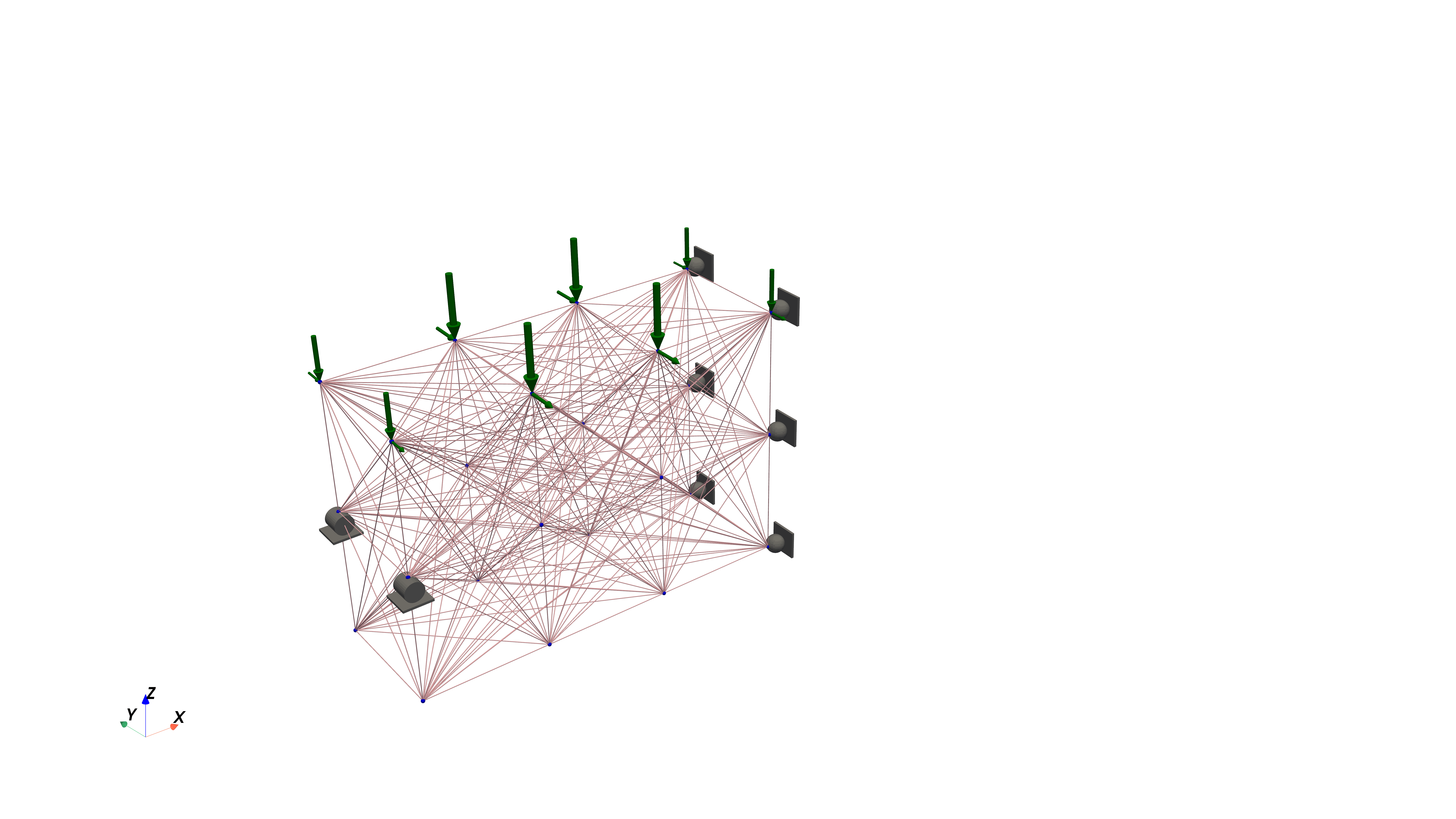}
        \caption{}
        \label{subfig:Result_Constraint_Layer_1_Design_Space_1}
    \end{subfigure} 
    \hspace{1cm}
    \begin{subfigure}[b]{0.3\textwidth}
        \centering
        \includegraphics[width=\linewidth, trim={15cm 15cm 120cm 30cm}, clip]{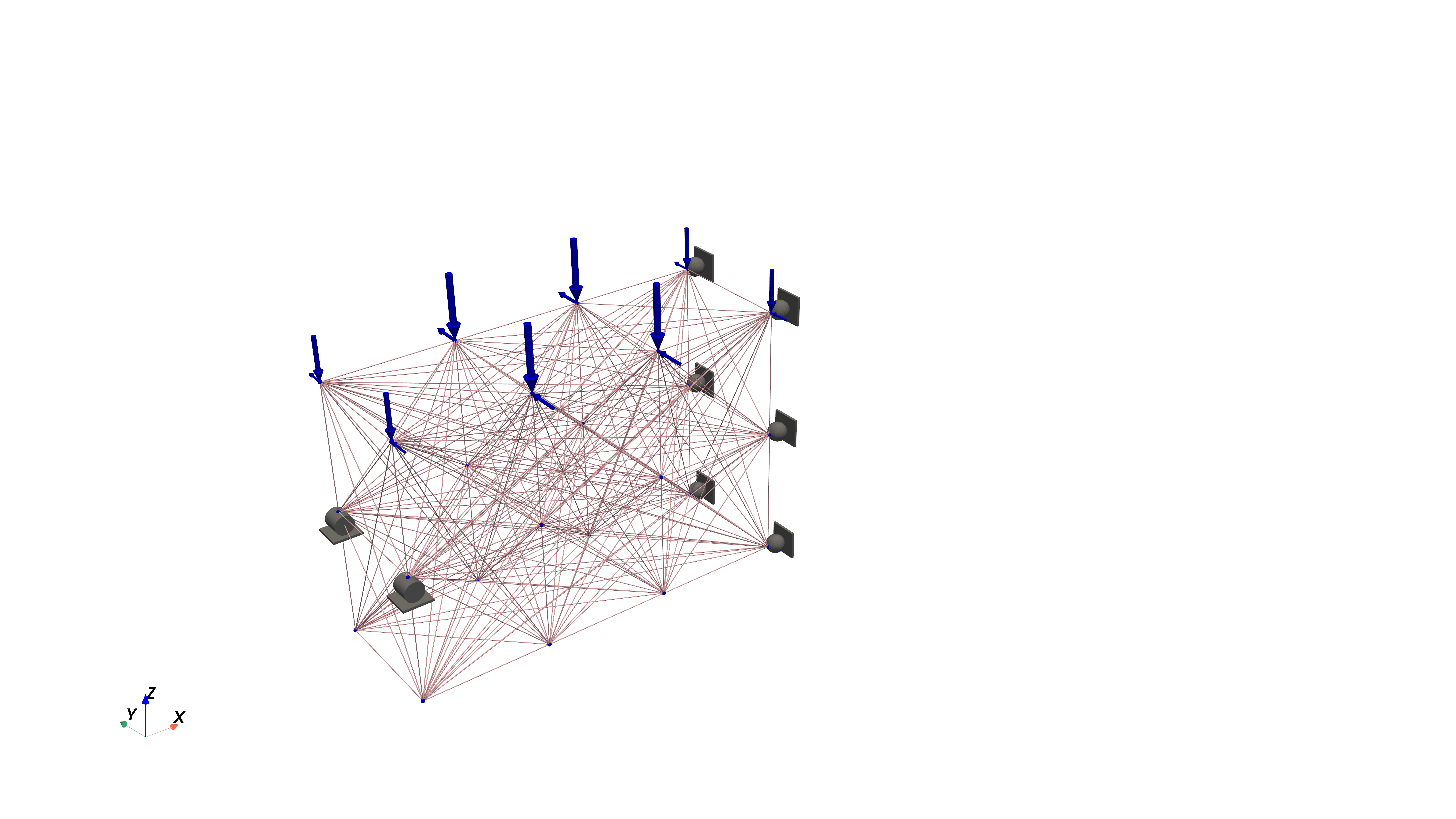}
        \caption{}
        \label{subfig:Result_Constraint_Layer_1_Ground_Structure_2}
    \end{subfigure}
        \caption{3D ground structure used for the design space mimicking the Lockport truss bridge.
        The design includes two load cases where the top chords are subjected to large vertical forces in both cases and a small line load that acts (a) horizontally out of the page, and (b) horizontally into the page.}
    \label{fig:Result_Constraint_Layer_GS}
\end{figure}

For this example, the design result is desired to be symmetric. 
As mentioned, symmetry is not guaranteed when solving MILPs unless it is explicitly enforced.
Therefore, this work exploits the symmetry of the design space by only modeling half of the bridge with the additional benefit of reducing the computational complexity. 
As before, the design space is discretized with a ground structure as shown in Fig. \ref{fig:Result_Constraint_Layer_GS}.
For this example, a $5\times 1 \times 2$ nodal discretization is used and meshed with 242 candidate members.
The elongation bounds used for the McCormick constraints are set to $v^{\min}_{ij} = -\sigma^c_j l_i \:/\:E_j$ and $v^{\max}_{ij} = \sigma^t_j l_i \:/\:E_j$ for all candidate members $i$ and materials $j$.
Two separate load cases $\left(L^C=2\right)$ are considered for the design of the bridge.
Both load cases are subjected to a distributed load $w=0.1$ MPa from rail traffic acting vertically downward along the top face of the design space.
This distributed load is applied as point loads on the nodes at the top of the ground structure with a magnitude of $P = 1250$ kN at the center nodes and $P = 625$ kN at the corner nodes.
Additionally, small horizontal line loads with magnitude $p=25$ kN/m are acting horizontally along the top chord  to mimic the static lateral loads caused by the vibration of traffic.
These are applied as point loads with $P=125$ kN at the center nodes and $P=62.5$ kN at the corner nodes.
As seen in Fig. \ref{subfig:Result_Constraint_Layer_1_Design_Space_1} and Fig.  \ref{subfig:Result_Constraint_Layer_1_Ground_Structure_2}, the direction of the line loads is different for the two load cases. 
The ground structures is supported by partial rollers that are restrained in the $y-$ and $z-$directions placed centrally along the depth of the bridge to imitate the location of the stone pier in Fig. \ref{fig:LockportTrussBridge}.
Symmetry is enforced by only applying rollers that are restrained in the $x-$direction at all nodes along the right face of the design space.
Note that all point load magnitudes are adjusted on nodes at the symmetry line and at corner nodes.

\subsubsection{Single-Material Design} \label{sec:singlematdesign}

\begin{figure}[h!]
    \centering
        \centering
        \includegraphics[width=0.45\textwidth, trim={48cm 24cm 64cm 40cm}, clip]{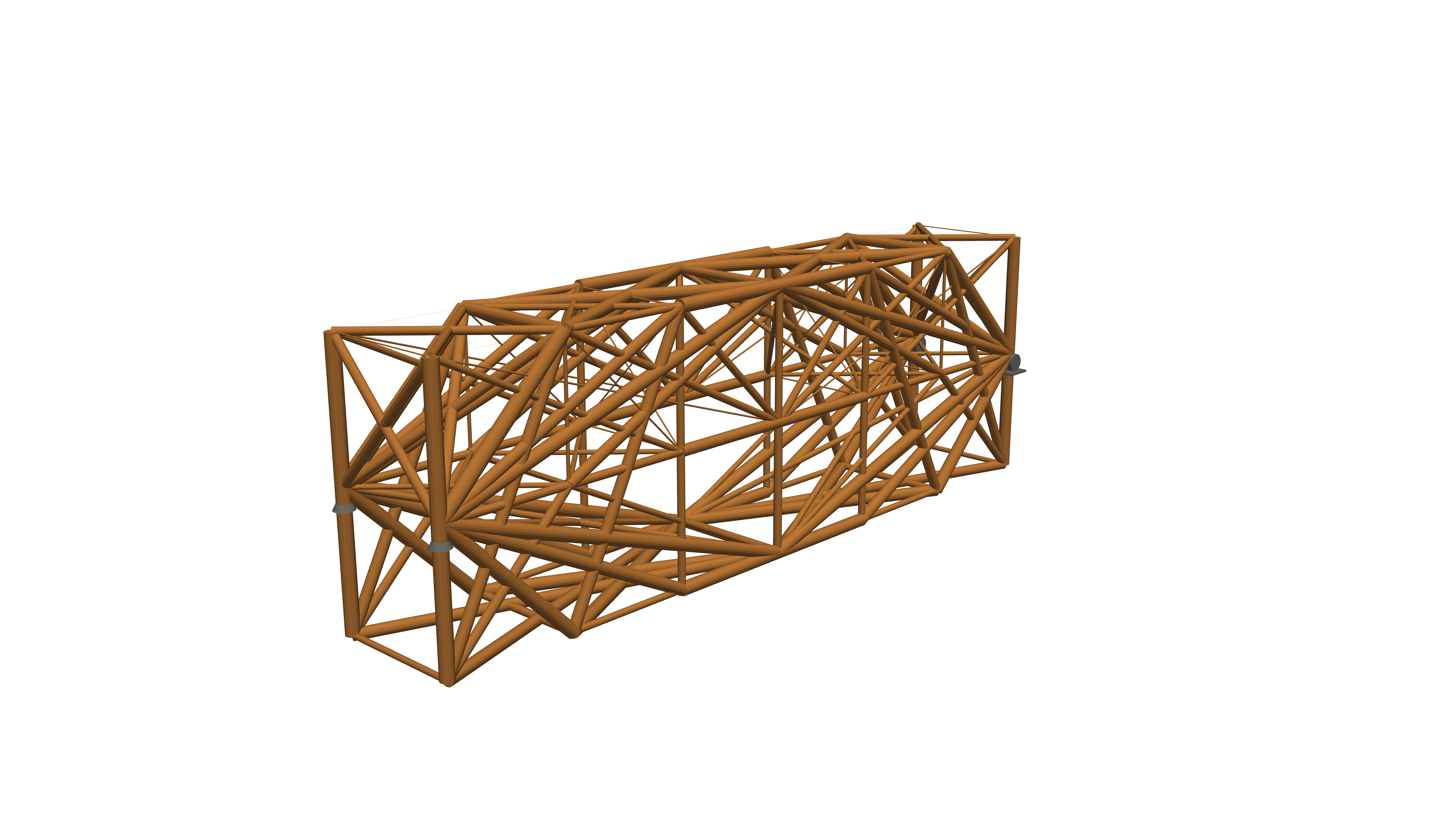}\\
        ${A^{\max}}$
    \caption{
    Result of timber-only designs of the rail bridge example considering  a constraint on the maximum cross-sectional area ($A^{\max}$). When other constraints are applied to the all-timber design, there are no feasible solutions.}
    \label{subfig:Result_Constraint_Layer_Timber_1_Isometric}
\end{figure}

\paragraph{Timber-Only Design}
The first design attempts to produce a bridge using only timber elements.
However, the only feasible solution occurs when applying the maximum allowable cross-sectional area $A_{timber}^{\max}=2,000$ cm$^2$.
Layering in a constraint to prevent intersecting members makes the problem infeasible.
This means there are also no feasible solutions when the problem is further restricted with the constructability constraints that are applied in Eq. \eqref{eq:MultiMaterialMILP}.
The unconstrained design is pictured in Fig. \ref{subfig:Result_Constraint_Layer_Timber_1_Isometric} and has an embodied carbon value of 4.01$\times10^4$ kgCO$_2$e.
This value is larger than the unconstrained hybrid timber-steel design discussed later due to the quantity of timber being used.
To generate a feasible solution, there is also a large relaxation with the structural member forces.
This solution is statically indeterminate and has a median error of 26.1\% when calculating the internal forces under the McCormick envelope.
Correcting some of these forces eats into the initial factor of safety applied on the timber strength limits or results in resizing some of the members.

\begin{figure}[h!]
    \centering
    \begin{subfigure}[b]{0.45\textwidth}
        \centering
        \includegraphics[width=\linewidth, trim={48cm 24cm 64cm 40cm}, clip]{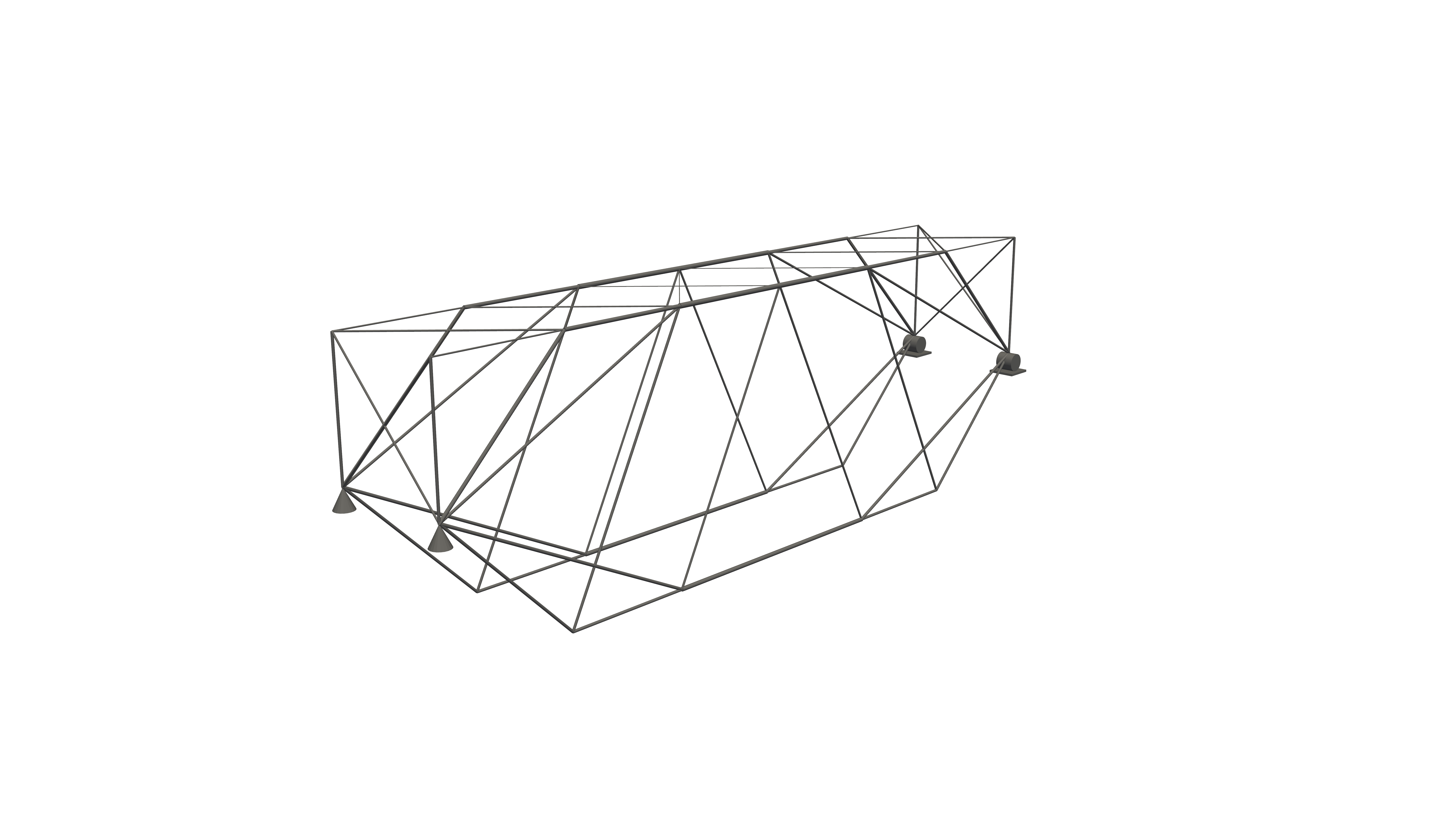}
        \caption{${A^{\max}}$}
        \label{subfig:Result_Constraint_Layer_Steel_1_Isometric}
    \end{subfigure}
    \begin{subfigure}[b]{0.45\textwidth}
        \centering
        \includegraphics[width=\linewidth, trim={48cm 24cm 64cm 40cm}, clip]{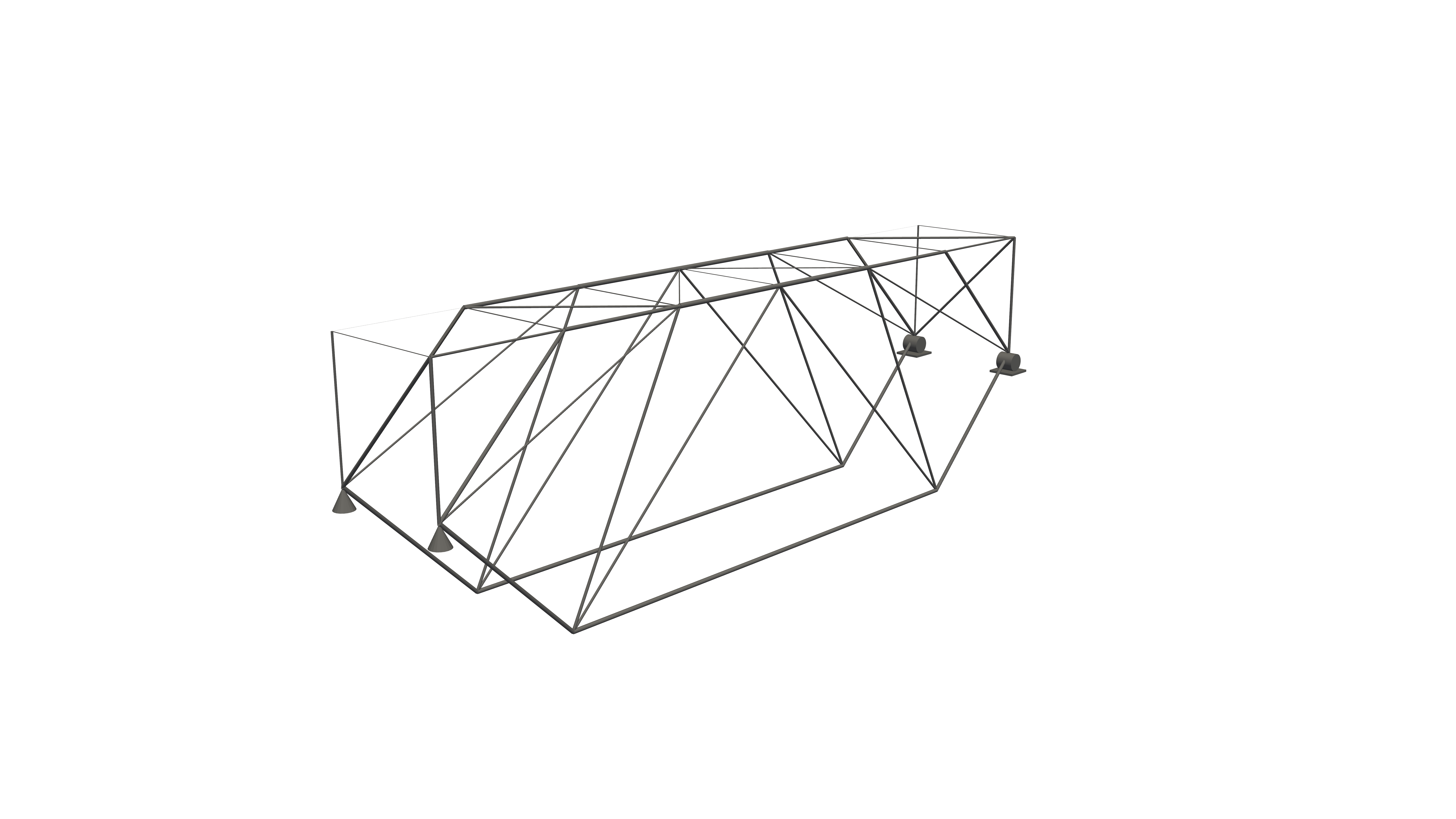}
        \caption{$+$No Intersect}
        \label{subfig:Result_Constraint_Layer_Steel_2_Isometric}
    \end{subfigure}

    \begin{subfigure}[b]{0.45\textwidth}
        \centering
        \includegraphics[width=\linewidth, trim={48cm 24cm 64cm 40cm}, clip]{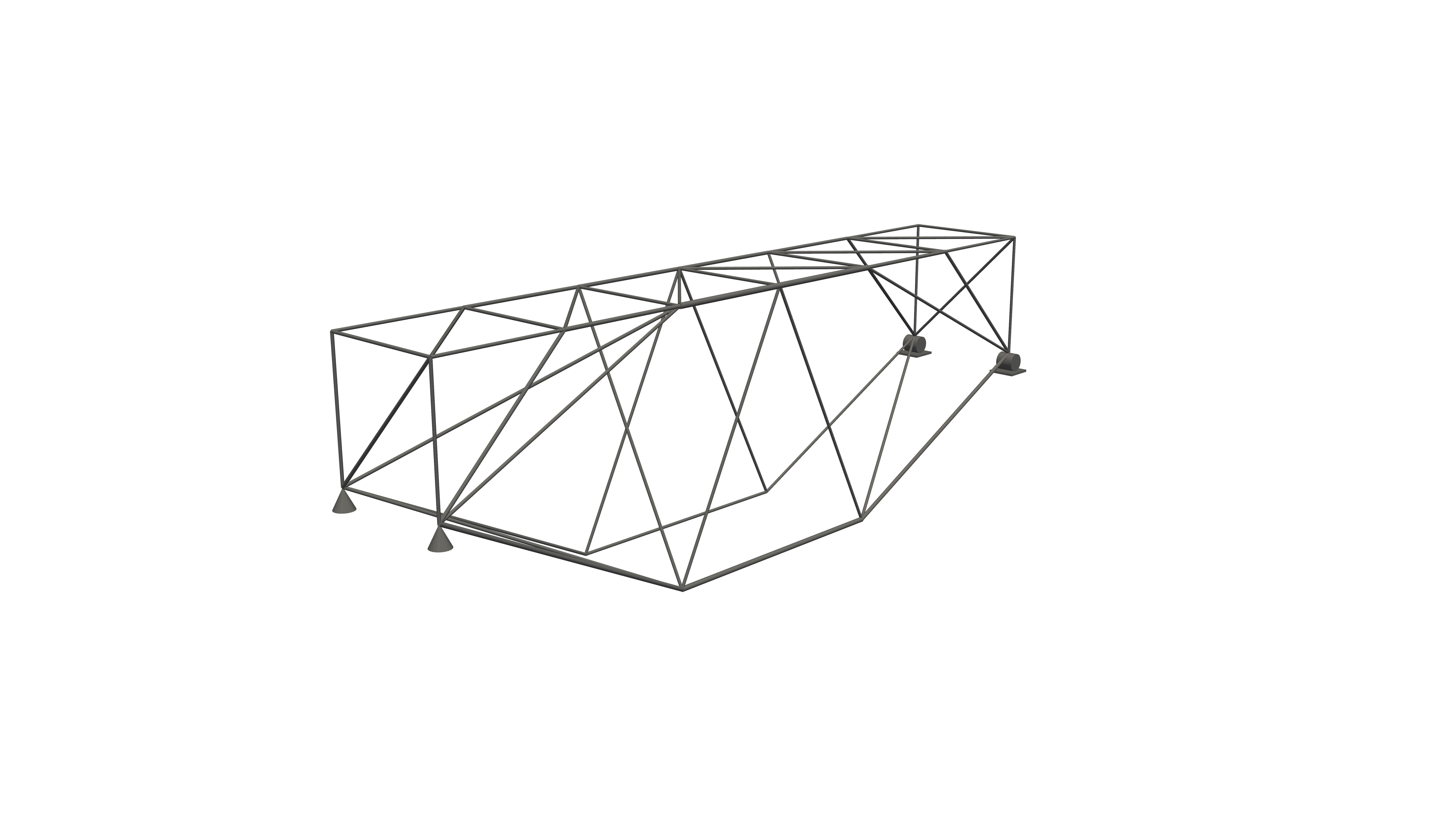}
        \caption{$+{A^{\min}}$}
        \label{subfig:Result_Constraint_Layer_Steel_3_Isometric}
    \end{subfigure}
    \begin{subfigure}[b]{0.45\textwidth}
        \centering
        \includegraphics[width=\linewidth, trim={48cm 24cm 64cm 40cm}, clip]{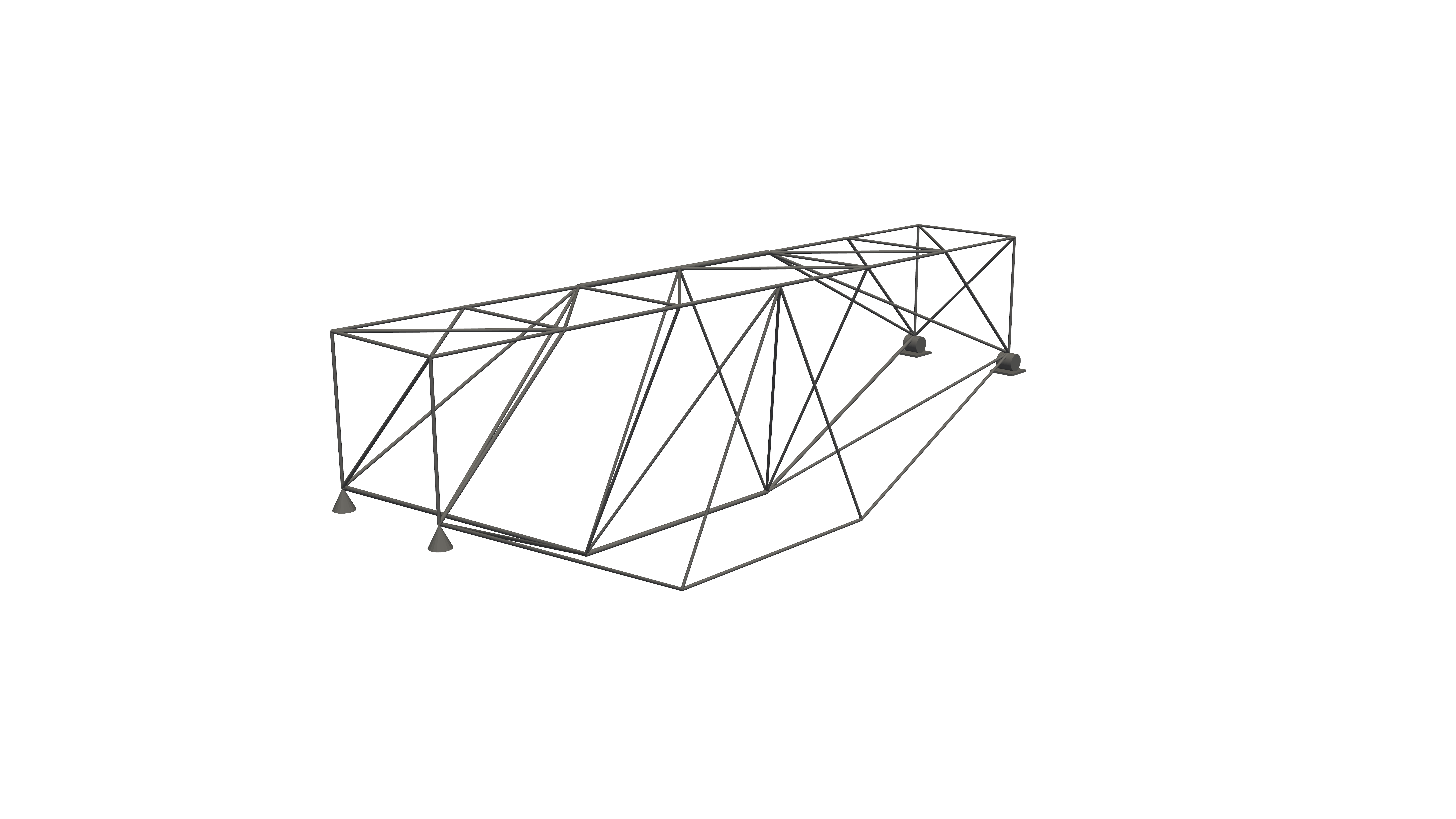}
        \caption{${+\eta_p^{\max}}$}
        \label{subfig:Result_Constraint_Layer_Steel_4_Isometric}
    \end{subfigure}

    \begin{subfigure}[b]{0.45\textwidth}
        \centering
        \includegraphics[width=\linewidth, trim={48cm 24cm 64cm 40cm}, clip]{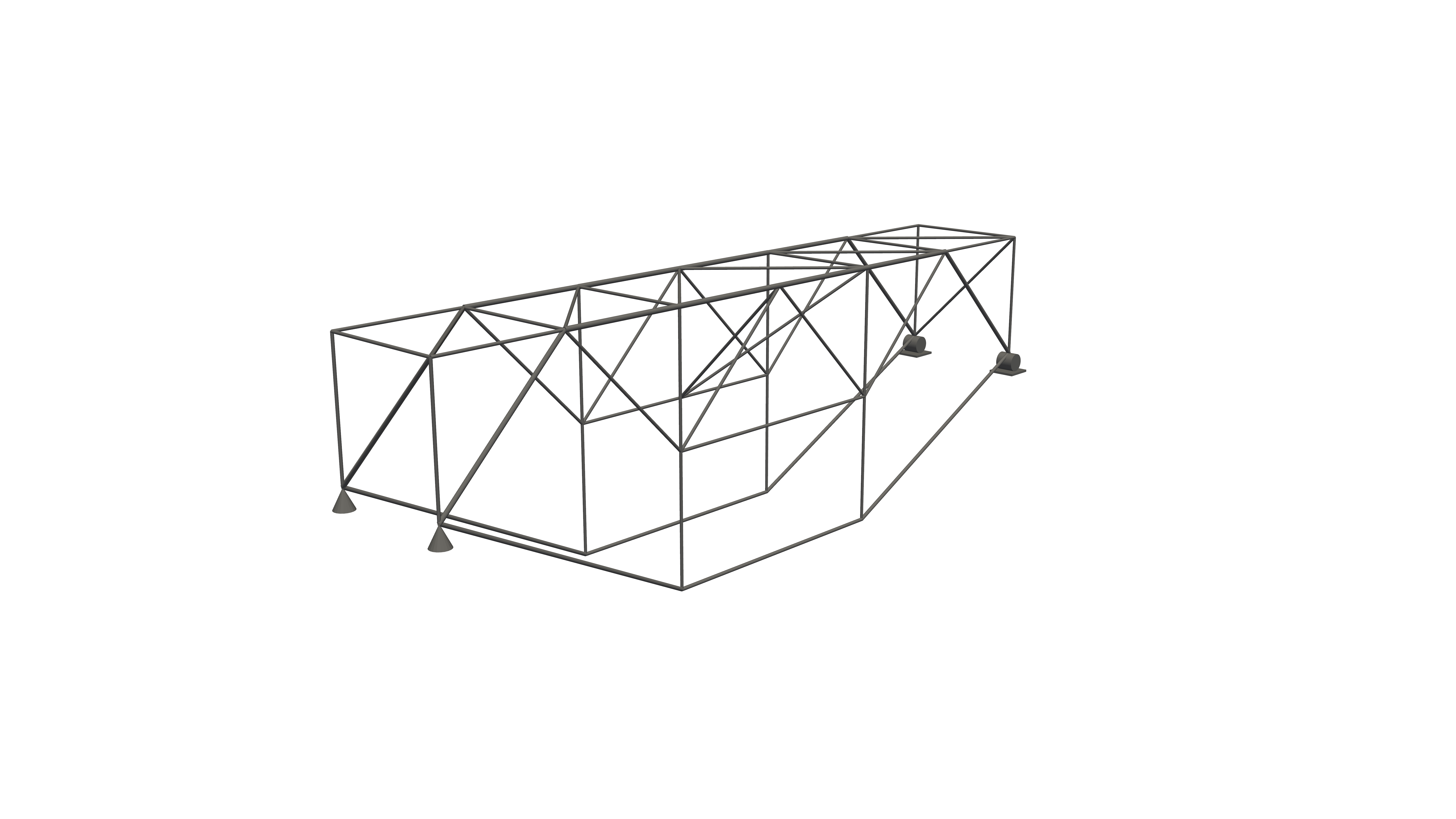}
        \caption{${+\gamma^{\min}}$}
        \label{subfig:Result_Constraint_Layer_Steel_5_Isometric}
    \end{subfigure}
    \caption{
    Results of steel-only designs of the rail bridge example considering (a) only a constraint on the maximum cross-sectional area ($A^{\max}$), (b) preventing the co-existence of intersecting elements ($A^{\max}$ and $+$No Intersect), (c) enforcing a minimum cross-sectional area ($A^{\max}$, $+$No Intersect, and $A^{\min}$), (d) constraining the number of elements per node ($A^{\max}$, $+$No Intersect, $A^{\min}$, and $\eta_p^{\max}$), and (e) also ensuring a minimum angle between elements at nodes ($A^{\max}$, $+$No Intersect, $A^{\min}$, $\eta_p^{\max}$, and $\gamma^{\max}$).
    }
    \label{fig:Result_Constraint_Layer_Steel_Full_Evolution}
\end{figure}

\paragraph{Steel-Only Design}
Next, the bridge is designed using only steel elements.
This is initially done by only constraining the maximum cross-sectional area of the elements ($A^{\max}$). 
Subsequently, intersecting members are prevented ($+$ No Intersect), a minimum cross-sectional area is enforced ($+A^{\min}$), the number of elements per node is limited ($+\eta_p^{\max}$), and, finally, the angle between elements at nodes is also constrained ($+\gamma^{\min}$).
All obtained designs are shown in Fig. \ref{fig:Result_Constraint_Layer_Steel_Full_Evolution} and compared in Tab. \ref{tab:steel_constraints}.
Because all of the steel solutions in this section are statically determinate and made from a single material, there is a 0\% error for all internal forces when calculated using the McCormick envelope.

When considering only the maximum bound on the cross-sectional area of $A^{\max}= 500$ cm$^2$ the design result in Fig. \ref{subfig:Result_Constraint_Layer_Steel_1_Isometric} is obtained in roughly 18 seconds.
It has an embodied carbon value of 4.27$\times10^4$ kgCO$_2$e.
The steel elements in the final design have cross-sectional areas varying from 6.40 cm$^2$ to 258 cm$^2$ and are thus not hitting the maximum bound.
At first glance, the result does not look overly complex apart from the appearance of elements with very small cross-sectional areas. 
The design has at total of 64 elements, where 44 have cross-sectional areas smaller than 100 cm$^2$. 
As listed in Tab. \ref{tab:steel_constraints}, the maximum number of members at any given node is $\eta_p = 6^*$ and the minimum angle of separation between members $\gamma =18.4^{\circ*}$.
However, these numbers are here given asterisks because the design does not prohibit the existence of intersecting members that create new nodes once the design is post-processed.
These new nodes are not accounted for in the presented numbers.

To prevent members from intersecting, as seen in the bottom corner of Fig. \ref{subfig:Result_Constraint_Layer_Steel_1_Isometric}, constraint \eqref{eq:MultiMaterialMILPAngle} is activated.
The design result has an embodied carbon value of 4.33$\times10^4$ kgCO$_2$e. 
It is pictured in Fig. \ref{subfig:Result_Constraint_Layer_Steel_2_Isometric} and is obtained in 23 seconds.
Compared to the initial design, the embodied carbon and weight increase by 1.53\%.
The total number of members joining at a node increases to 7 and the angle of separation remains 18.4$^{\circ}$.
However, these numbers no longer need asterisks because prohibiting intersecting members prevents the creation of new nodes during the design process and decreases post-processing.
While preventing intersections gives explicit control over all members in the design, there are still issues present.
Two of the members in one of the top chords, which are required to exist all the way across the span, take on a negligible cross-sectional area of 0.1 cm$^2$, which is only enforced for numerical stability.
The algorithm identifies a load path that does not use these members and essentially removes them in an effort to save material while respecting the continuity constraint. 

Next, the design problem is resolved with an explicit limit on the minimum cross-sectional area of $A^{\min}=100$ cm$^2$.
The addition of this constraint produces the design result in Fig. \ref{subfig:Result_Constraint_Layer_Steel_3_Isometric}, which is obtained in just over 13 minutes and has an embodied carbon value of 5.78$\times10^4$ kgCO$_2$e.
While the embodied carbon and structural weight are 33\% higher than the previous design, the new design now has steel elements with cross-sections ranging from 100 cm$^2$ to 353 cm$^2$.
Notice that both top chords are now continuous with non-negligible areas across the entire span.
As seen in Tab. \ref{tab:steel_constraints}, the nodes have not become more complex than the example that enforces the maximum area constraint and prevents intersecting.
However, this requirement is not explicitly defined and the nodes may still be more complex than desired.

To ensure that the nodes connect a limited number of elements, the design is resolved by activating the constraints in Eq. \eqref{eq:MultiMaterialMILPNode}.
The maximum number of elements at a given node is now strictly enforced to be $n_p^{\max} = 6$.
The design result in Fig. \ref{subfig:Result_Constraint_Layer_Steel_4_Isometric} is obtained in just under 92 minutes with an embodied carbon score of 6.08$\times10^4$ kgCO$_2$e.
This is a jump of 42\% compared to the initial design in Fig. \ref{subfig:Result_Constraint_Layer_Steel_1_Isometric}.
However, all members in the design have sufficiently large cross-sectional areas, there are no intersecting members that need to be dealt with in post-processing, and the design is guaranteed to have a limited number of elements connected at each node.
In the back bottom right of the mid-span of the bridge, however, it can be noted that 6 members are joining at sharp angles, some of which measure $\gamma=18.4^\circ$.

Finally, the entries in the violation matrix, $\Gamma_{xi}$, in Eq. \eqref{eq:MultiMaterialMILPAngle} are updated to prevent sharp angle connections.
Designs containing members separated by less than $\gamma^{\min}=30^\circ$ are prohibited.
The design result in Fig. \ref{subfig:Result_Constraint_Layer_Steel_5_Isometric} is obtained in about 106 minutes with an objective value of 6.31$\times10^4$ kgCO$_2$e.
This design is the most computationally expensive and increases the objective value by 47.6\% compared to the initial unconstrained design.
However, this design satisfies all major constructability concerns listed herein.
Member cross-sections range from 100 cm$^2$ to 293 cm$^2$, intersecting members are prevented, joints connect a maximum of 6 members, and the sharpest angle between any two members is $\gamma=35.3^\circ$.

\begin{table}[h!]
    \centering
\begin{tabular}{l c c c c c} \hline
\pmb{Constructability constraints} & $\pmb{A^{\max}}$ & \pmb{$+$No Intersect} & $\pmb{+A^{\min}}$ & $\pmb{+\eta_p^{\max}}$ & $\pmb{+\gamma^{\min}}$\\ \hline
    Runtime                           & 18s & 23s & 13m 18s & 91m 52s & 105m 40s \\
    Min steel size (cm$^2$)           & 6.4 & 0.1 & 100 & 100 & 100 \\
    Max steel size (cm$^2$)           & 258 & 240 & 353 & 326 & 293 \\
    Max members at joint              & 6$^*$ & 7 & 7 & 6 & 6 \\
    Min angle between members         & 18.4$^{\circ*}$ & 18.4$^{\circ}$ & 18.4$^\circ$ & 18.4$^\circ$ & 35.3$^\circ$ \\
    Mass (kg)                       & 29,461 & 29,912 & 39,829 & 41,927 & 43,486 \\
    Embodied carbon (kgCO$_2$e) & 42,719 & 43,373 & 57,752 & 60,794 & 63,055 \\ \hline 
\end{tabular}
    \caption{Performance properties for the steel-only rail bridge truss designs with increasing levels of constructabilty constraints in Fig. \ref{fig:Result_Constraint_Layer_Steel_Full_Evolution}.
    *Asterisks refers to designs where existence of intersecting elements is not prohibited and new nodes with other complexities therefore might appear once the design is post-processed.}
    \label{tab:steel_constraints}
\end{table}

When examining the relative embodied carbon values for the current example in Tab. \ref{tab:steel_constraints}, the most significant difference between designs is found when imposing the minimum cross-sectional area constraint.
As mentioned, it increases by 33\%.
The additional consideration of limiting the number of elements at joints has less of a relative effect (5.3\% increase) and requiring a minimum angular spacing causes only small changes (3.7\% further increase).
The consideration of the constructability constraints is seen to have a direct effect on the computational time that is approx. 1.3, 44, 306 and 352 times higher than the unconstrained design.
The computational cost for this example is seen to have two significant jumps; the first when adding the minimum cross-sectional area constraint and the second when constraining the maximum number of elements per node. 
There is a further increase in computational time associated with also enforcing a minimum angular spacing, however it is less pronounced. 
In summary, the total expense of going from a hard-to-construct unconstrained steel truss design solution to a result that fulfills all herein defined constructability constraints is a 47.6\% increase in embodied carbon and an additional 105 minutes of compute time.

\subsubsection{Timber-Steel Design} \label{sec:two_mat_bridge_design}
\begin{figure}[h!]
    \centering
    \begin{subfigure}[b]{0.45\textwidth}
        \centering
        \includegraphics[width=\linewidth, trim={48cm 24cm 64cm 40cm}, clip]{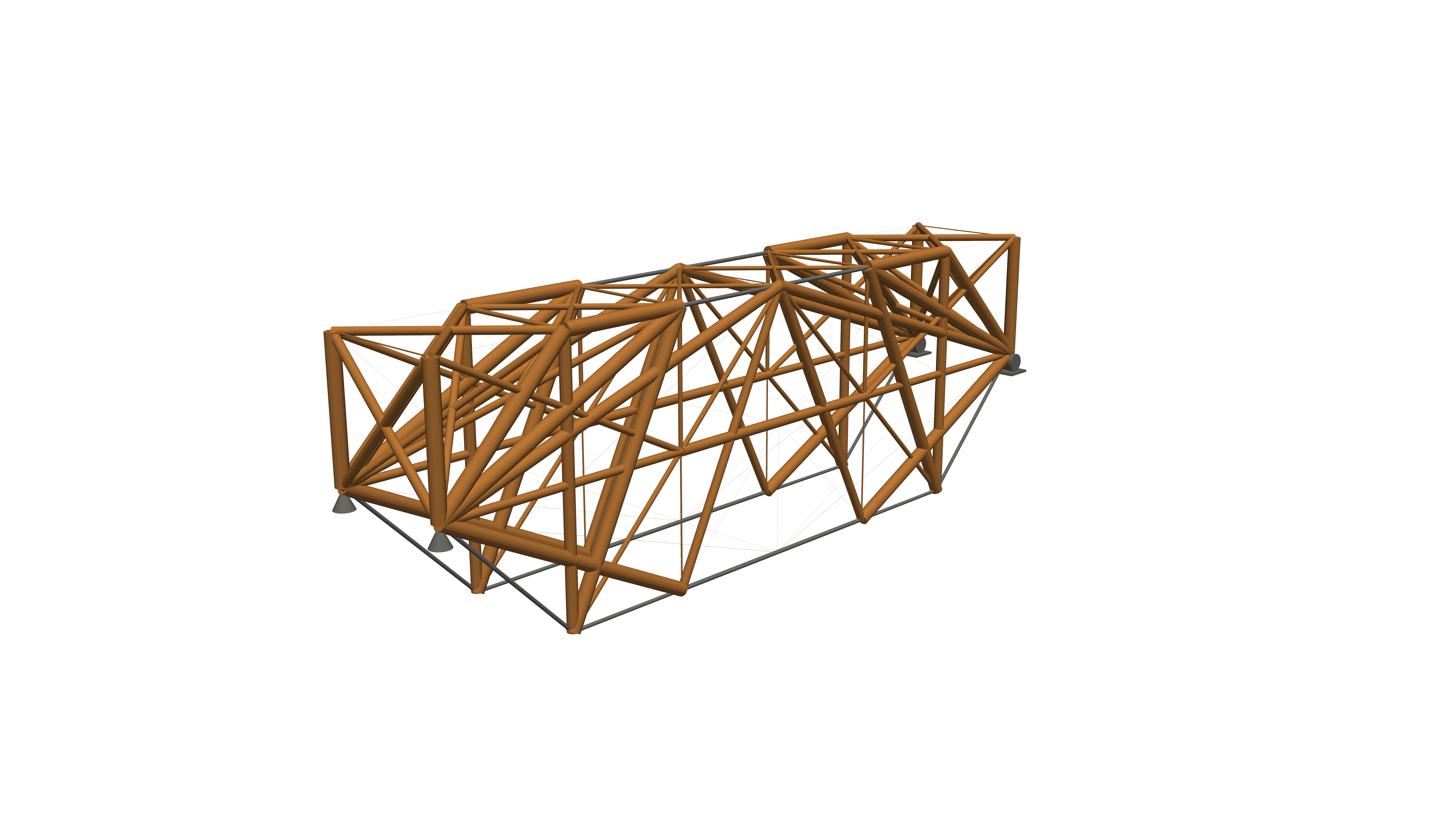}
        \caption{${A^{\max}}$}
        \label{subfig:Result_Constraint_Layer_1_Isometric}
    \end{subfigure}
    \begin{subfigure}[b]{0.45\textwidth}
        \centering
        \includegraphics[width=\linewidth, trim={48cm 24cm 64cm 40cm}, clip]{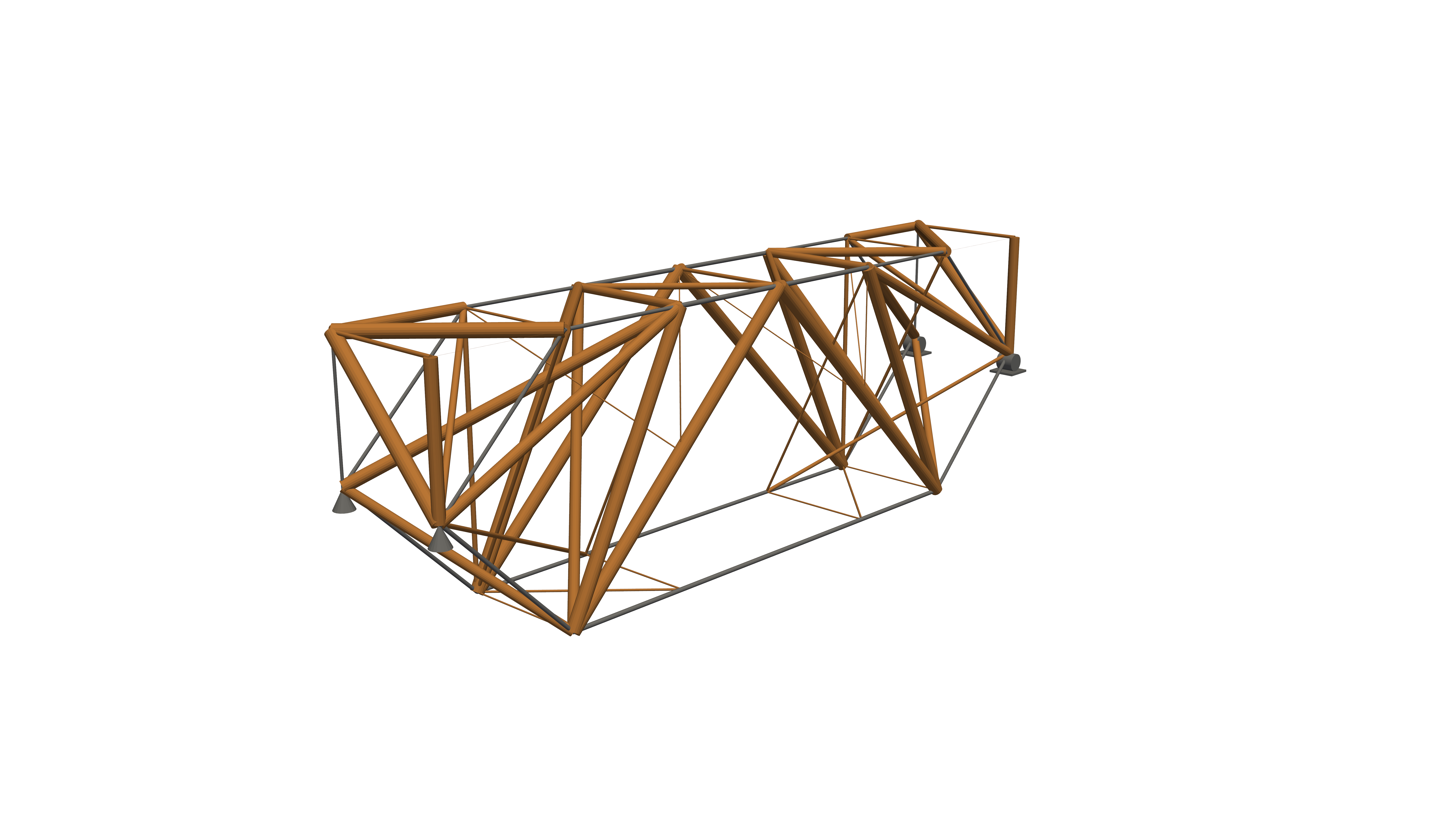}
        \caption{$+$No Intersect}
        \label{subfig:Result_Constraint_Layer_2_Isometric}
    \end{subfigure}

    \begin{subfigure}[b]{0.45\textwidth}
        \centering
        \includegraphics[width=\linewidth, trim={48cm 24cm 64cm 40cm}, clip]{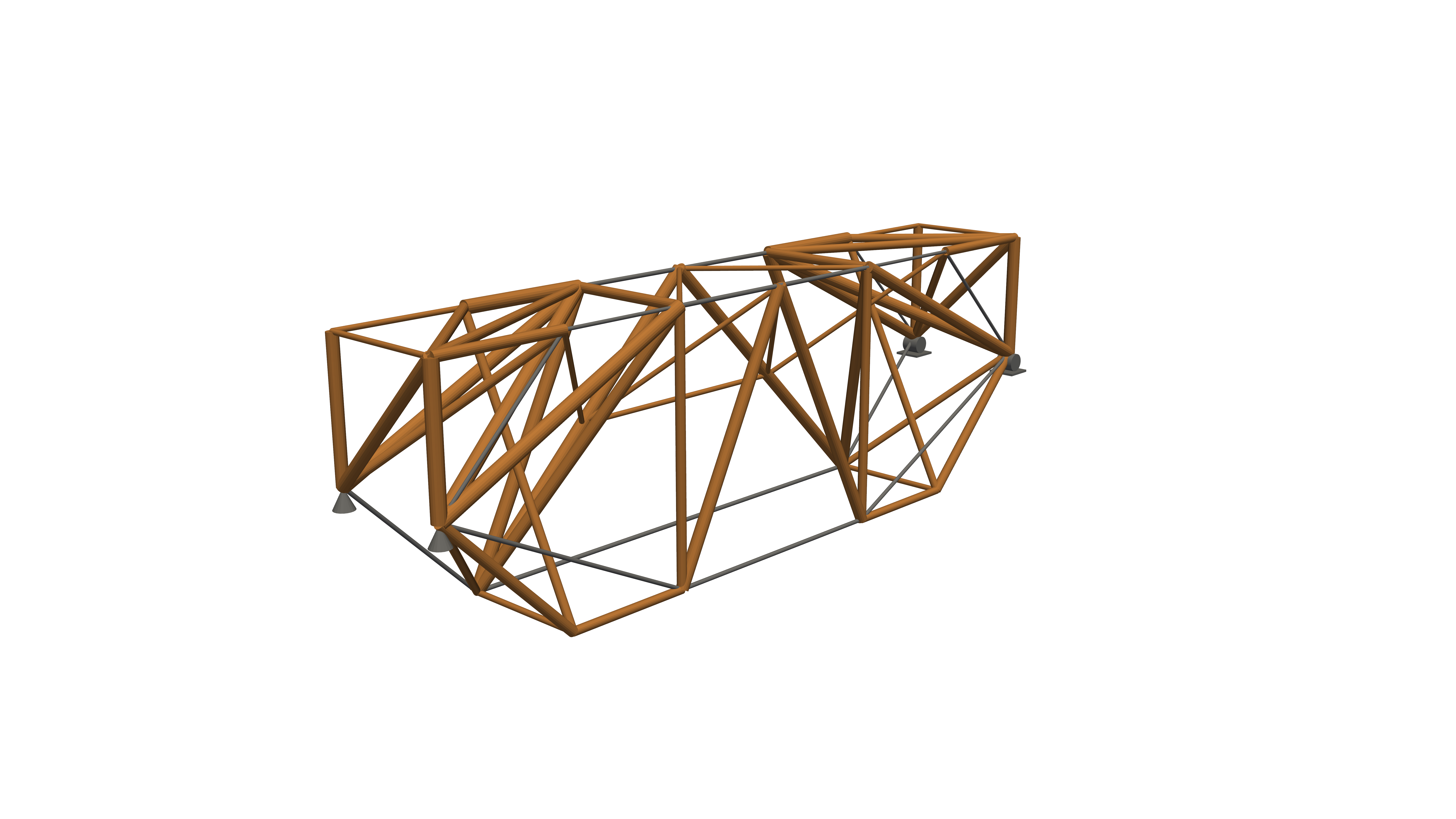}
        \caption{$+{A^{\min}}$}
        \label{subfig:Result_Constraint_Layer_3_Isometric}
    \end{subfigure}
    \begin{subfigure}[b]{0.45\textwidth}
        \centering
        \includegraphics[width=\linewidth, trim={48cm 24cm 64cm 40cm}, clip]{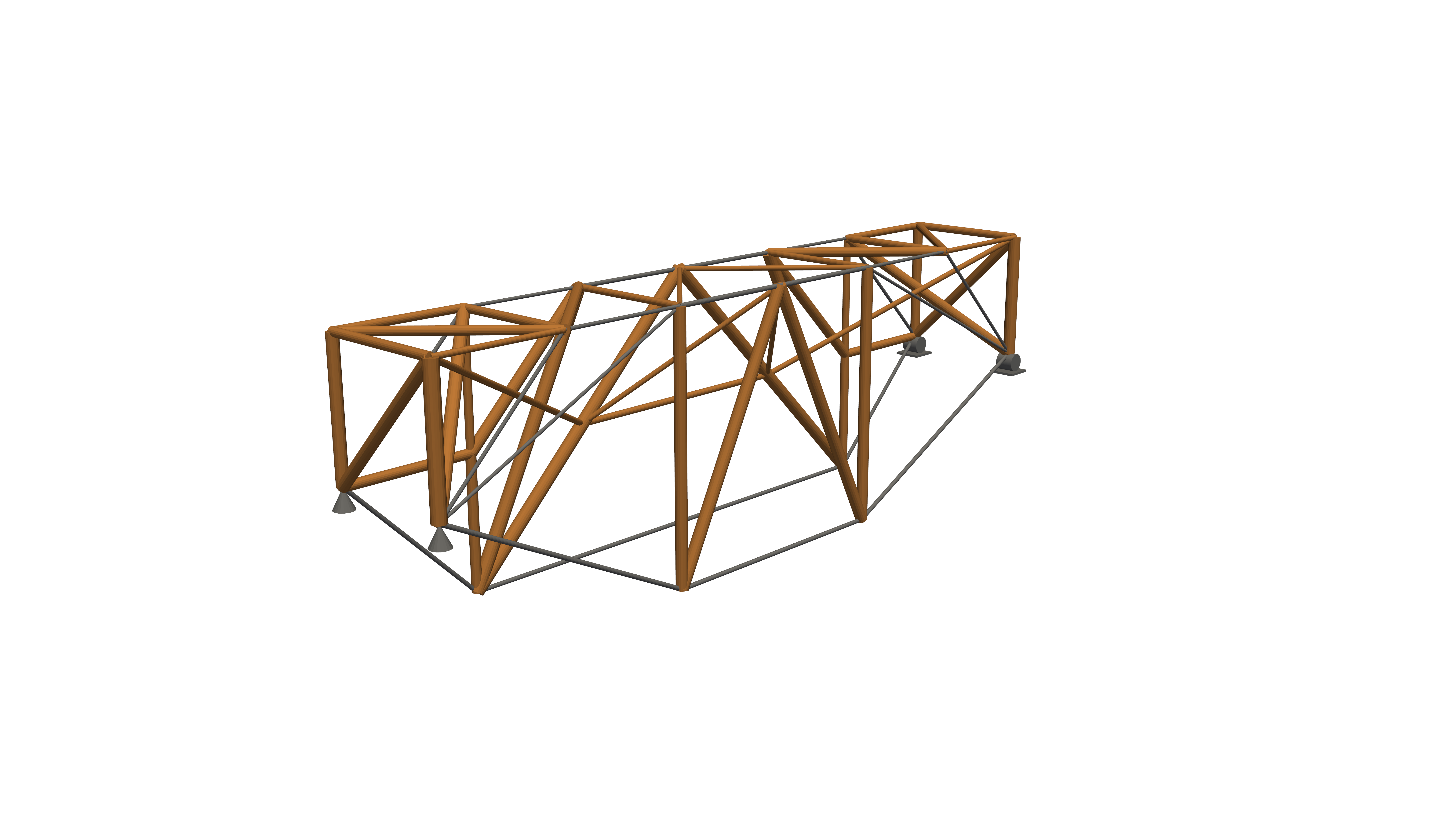}
        \caption{$+{\eta_p^{\max}}$}
        \label{subfig:Result_Constraint_Layer_4_Isometric}
    \end{subfigure}

    \begin{subfigure}[b]{0.45\textwidth}
        \centering
        \includegraphics[width=\linewidth, trim={48cm 24cm 64cm 40cm}, clip]{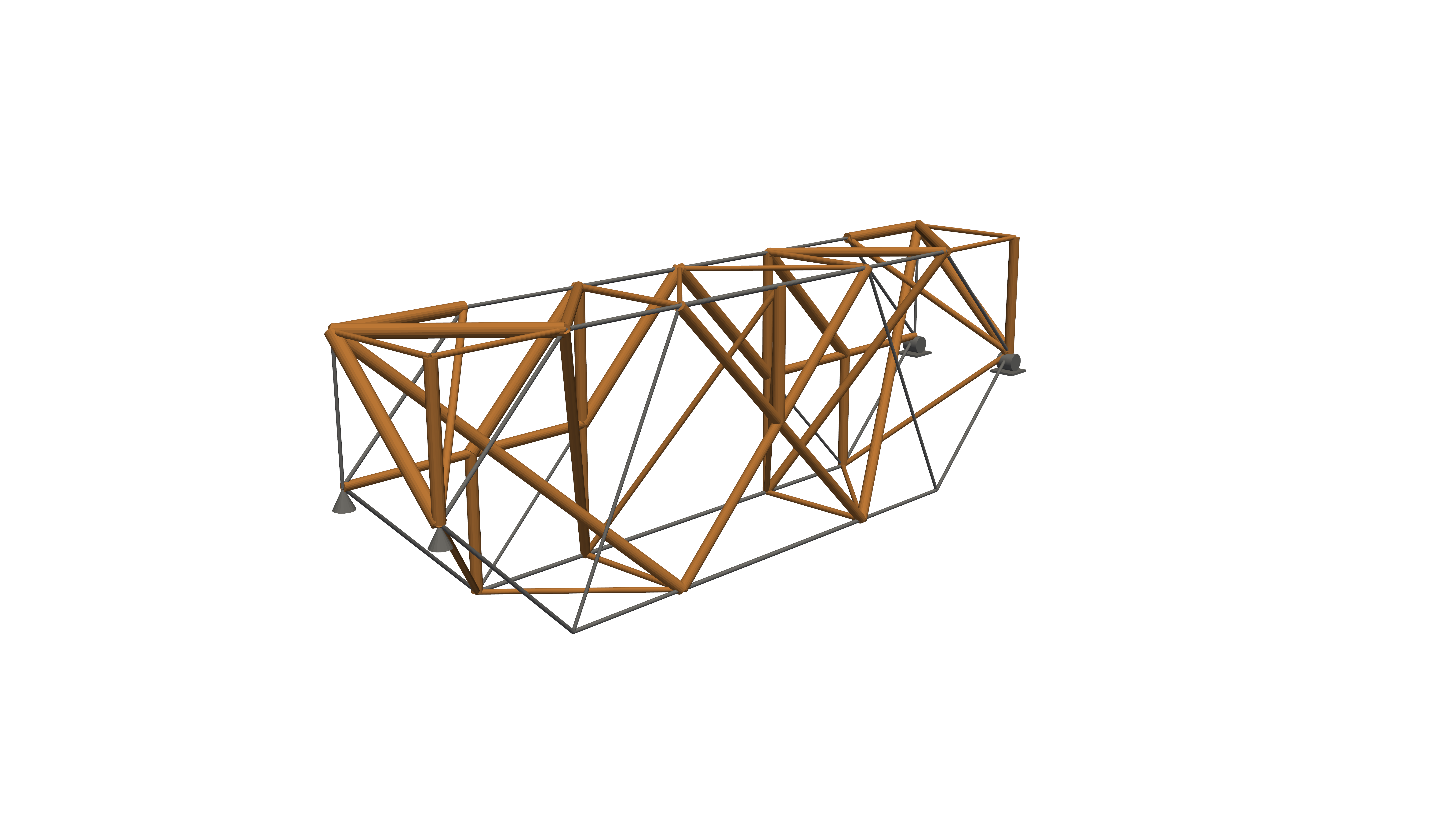}
        \caption{$+{\gamma^{\min}}$}
        \label{subfig:Result_Constraint_Layer_5_Isometric}
    \end{subfigure}
    \caption{
    Results of timber-steel designs of the rail bridge example considering (a) only a constraint on the maximum cross-sectional area ($A^{\max}$), (b) preventing the co-existence of intersecting elements ($A^{\max}$ and $+$No Intersect), (c) enforcing a minimum cross-sectional area ($A^{\max}$, $+$No Intersect, and $A^{\min}$), (d) constraining the number of elements per node ($A^{\max}$, $+$No Intersect, $A^{\min}$, and $\eta_p^{\max}$), and (e) also ensuring a minimum angle between elements at nodes ($A^{\max}$, $+$No Intersect, $A^{\min}$, $\eta_p^{\max}$, and $\gamma^{\max}$).
    }
    \label{fig:Result_Constraint_Layer_Full_Evolution}
\end{figure}

The rail truss bridge design problem is now resolved with $M=2$, with timber and steel as material options.
The results for solutions with different constraints applied are shown in Fig. \ref{fig:Result_Constraint_Layer_Full_Evolution}.
All designs are obtained by solving the MILP in Eq. \eqref{eq:MultiMaterialMILP} and compared in Table \ref{tab:timber_steel_constraints}.

Similar to the steel-only design, a maximum cross-sectional area bound is initially applied as the only constructability constraint.
The maximum allowable area for steel and timber members is set to $A_{steel}^{\max}=500$ cm$^2$ and $A_{timber}^{\max}=2,000$ cm$^2$.
The design result is obtained in 47 seconds and is given in Fig. \ref{subfig:Result_Constraint_Layer_1_Isometric}.
The McCormick median error for this result is 40.6\%, which is very high.
The large area bounds and minimal constructability constraints largely contribute to this.
As seen in the Michell truss example in Section \ref{subsec:Results_2_Material_Design}, the design converges to a solution that uses a large amount of timber.
Steel only makes up 1.88\% of the structural volume.
Timber is heavily favored for compressive elements and steel elements mostly occur in tension.
The primary use of steel is found in the bottom tensile chords of the truss.
This design has numerous constructability issues.
The minimum timber sections, at the ends of the top chords and converging in the middle of the steel bottom chord, have cross-sectional areas of only 0.1 cm$^2$.
Additionally, joints near the supports are complex, as they connect many elements at sharp angles just above 8$^{\circ}$.

Next, the constraint preventing members from intersecting is applied.
The result is shown in Fig. \ref{subfig:Result_Constraint_Layer_2_Isometric} and increases the embodied carbon value by 6.1\% compared to the initial design.
The median error in the McCormick calculation of the internal forces drops to 16.7\%.
While the area bounds have not tightened, preventing intersecting members reduces the convolution in the force flow and provides less freedom for the McCormick envelope.
Adding this constraint also increases the steel volume to 5.26\% for this design.
Although these are marked improvements over the initial design and the total number of members is greatly reduced, there are still plenty of small cross-section members, such as the members at the end of the top chord, and difficult connections to deal with.

Applying a minimum area bound for steel and timber of $A_{steel}^{\min}=100$ cm$^2$ and $A_{timber}^{\min}=300$ cm$^2$ gives the design in Fig. \ref{subfig:Result_Constraint_Layer_3_Isometric}.
It is obtained in just under 92 minutes with an 8.5\% increase in the objective value compared to the baseline solution in Fig. \ref{subfig:Result_Constraint_Layer_1_Isometric}.
The steel volume is increased to 3.94\% of the design and Median McCormick error in the internal force calculation is again below 5\% (4.52\%).
In this design, the small timber sections in the top chords take on a non-negligible cross-sectional size.
Some thin timber members near the center of the span have also been removed.
With these small adjustments, the design still appears highly complex with timber heavily favored for the structural components. 
There are a number of complex joints, such as near the center of the top chord that connects 7 large timber members, or the support nodes that have members connecting at angles just above 18$^{\circ}$.

The problem is now redesigned where intersecting elements are prohibited from co-existing and a maximum joint constraint of $\eta_p^{\max}=6$ elements is enforced.
This produces the design in Fig. \ref{subfig:Result_Constraint_Layer_4_Isometric} after just over 76 minutes.
The design has an increased embodied carbon objective of 10.4\% compared to the unconstrained timber-steel design.
The steel volume within the design is increased further to 6.87\% and the median McCormick error is reduced to 2.17\%.
Importantly, the design now appears significantly less complex.
The most complex node joins 6 elements and the minimum angle of separation between elements is 18.4$^{\circ}$.
It is also interesting to note that this design is lighter compared to the previous solutions. 
Even though the structure has only increased its steel volume to 6.87\%, the structural weight is reduced by 20.0\% compared to the unconstrained solution.
This indicates that the unconstrained design uses many tensile timber elements to establish the same tensile capacity as can be achieved with relatively little steel. 
It also reinforces the notion raised by e.g. Feickert and Mueller \cite{feickert2024thin} that it is important to consider the substructure when pursuing low-carbon designs as the heavier unconstrained solution will need a larger foundation, which is not accounted for herein. 

Finally, applying a minimum angle of separation between elements of $\gamma_{\min}=30^\circ$ gives the design in Fig. \ref{subfig:Result_Constraint_Layer_5_Isometric}.
This solution is generated in 230 minutes and 8 seconds and increases the objective by 19.2\% compared to the baseline design.
The median McCormick error in the internal force calculation remains low for the design at 2.40\%.
At 7.46\%, it has the greatest steel volume of any hybrid timber-steel design.
While there are not drastic changes from the previous solution in Fig. \ref{subfig:Result_Constraint_Layer_4_Isometric}, the minimum angle between members in this design almost doubled to 35.3$^{\circ}$.
Perhaps unexpectedly, the structural weight is increased slightly.
However, it is still significantly lower than for the unconstrained design.

\begin{table}[http]
    \centering
\begin{tabular}{l c c c c c} \hline
\pmb{Constructability constraints} & $\pmb{A^{\max}}$ & \pmb{$+$No Intersect} & $\pmb{+A^{\min}}$ & $\pmb{+\eta_p^{\max}}$ & $\pmb{+\gamma^{\min}}$\\ \hline
    Runtime                           & 47s & 1m 51s & 91m 46s & 76m 1s & 230m 8s \\
    Min timber size (cm$^2$)          & 0.1 & 0.1 & 300 & 300 & 300 \\
    Max timber size (cm$^2$)          & 2,000 & 2,000 & 2,000 & 2,000 & 1,961 \\
    Min steel size (cm$^2$)           & 0.1 & 84.9 & 100 & 109 & 107 \\
    Max steel size (cm$^2$)           & 205 & 241 & 242 & 243 & 255 \\
    Max members at joint              & 12$^*$ & 8 & 7 & 6 & 6 \\
    Min angle between members         & 8.13$^{\circ*}$ & 18.4$^{\circ}$ & 18.4$^\circ$ & 18.4$^\circ$ & 35.3$^\circ$ \\
    Mass (kg)                       & 51,492 & 42,293 & 46,557 & 41,199 & 43,603 \\
    Embodied carbon (kgCO$_2$e) & 38,003 & 40,306 & 41,219 & 41,952 & 45,288 \\ \hline 
\end{tabular}
    \caption{Performance properties for the steel-only rail bridge truss designs with increasing levels of constructability constraints in Fig. \ref{fig:Result_Constraint_Layer_Full_Evolution}.
    *Asterisks refers to designs where existence of intersecting elements is not prohibited and new nodes with other complexities therefore might appear once the design is post-processed.}
    \label{tab:timber_steel_constraints}
\end{table}

As seen in Sec. \ref{sec:singlematdesign}, a constructible single material solution is only possible when designing with steel for this rail bridge example. 
When comparing the unconstrained single and multi-material truss designs in Fig. \ref{subfig:Result_Constraint_Layer_Timber_1_Isometric}, Fig. \ref{subfig:Result_Constraint_Layer_Steel_1_Isometric} and Fig. \ref{subfig:Result_Constraint_Layer_1_Isometric}, the steel-only solution appears less complex and more easily constructible. 
For the unconstrained designs, this apparent reduction of complexity is associated with a carbon cost, as the steel-only design has an embodied carbon objective that is 12.4\% higher than the timber-steel alternative. 
However, opposite to the Michell truss design example, the benefit of pursuing a hybrid design significantly increases as the constructability constraints are applied. 
When all the herein considered constructability requirements are applied, the timber-steel solution outperforms the steel-only truss by reducing the embodied carbon by 28.2\%. 
The weight of the two structures are comparable, with the timber-steel design being just 117 kg heavier. 
It is interesting to note that the computation time associated with the two-material designs vary much more than the single material solutions.
This illustrates how the solution time of MILPs is highly problem dependent and difficult to predict.
Sometimes adding constraints helps the computation time and sometimes it hurts the computation time.

\subsection{Three-Component Tensegrity Design} \label{subsec:Results_3_Material_Design}
The herein presented design framework can be used to design with any $M$ number of candidate materials or components. 
To demonstrate its capability with $M=3$ candidate components, a tensegrity example is considered. 
The design has the options of using steel cables and steel or timber bars.
The timber and steel bars are designed with zero tensile strength to ensure that all tensile forces are taken by cables. 

\begin{figure}[h!]
    \centering
    \includegraphics[width=0.4\linewidth, trim={92cm 15cm 97cm 40cm}, clip]{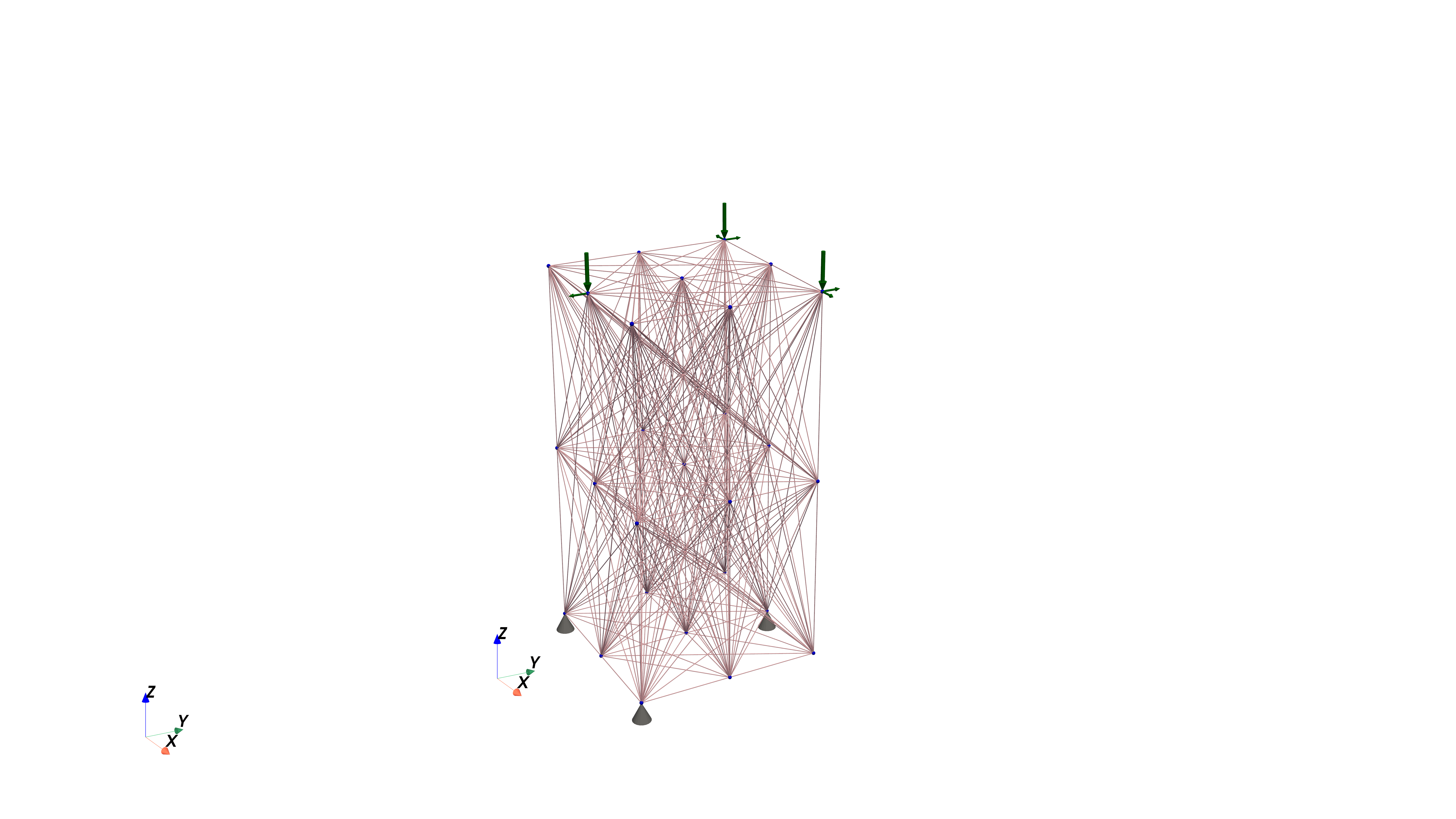}
    \caption{3D ground structure with 302 candidate members used for minimum embodied carbon tensegrity tower design.}
    \label{fig:Result_3_Material_Ground_Structure}
\end{figure}

The ground structure used for the tensegrity design example is shown in Fig. \ref{fig:Result_3_Material_Ground_Structure} and has $2 \times 2 \times 2$ nodes spanning 1.5 m $\times$ 1.5 m $\times$ 3 m.
The mesh is composed of 302 candidate members.
The design space is supported by three pins at the base, arranged in an isosceles triangular pattern.
There are three vertically applied loads at three of the nodes at the top of the design space, each with a magnitude of $P_z = -75$ kN.
Two of these vertical loads are located in the top corners that are opposite to the corner bases with pins.
At each of these nodes, there are two additional $P_x=P_y = \pm10$ kN forces acting outwards from the structure in the horizontal plane.
The third vertical load is placed at the center of the top edge opposite of the single pin at the base.
Here, a lateral force of $P_y = -10$ kN is also applied.

The elongation bounds used for the McCormick constraints are set to $v^{\min}_{ij} = -0.001l_i$ and $v^{\max}_{ij} = 0.001l_i$ for all candidate members $i$ and materials/components $j$.
This is different from before because the struts are given zero tensile capacity and the cables are given zero compression capacity.
Using the same values from previous examples that are based on maximum allowable compressive stress would prevent a cable from elastically shrinking under applied loads.
However, a prestressed cable can shrink under applied loads and remain in tension if the initial prestressing force is large enough.
Prestressed structures designed using the McCormick relaxation may need to be post-processed before being constructed in the real world.
The introduction of prestressing force variables in addition to member force variables from applied external loads does not necessarily provide a clean way to quantify error in the relaxation.
If the total forces are relaxed too much, it is up to the designer to determine if they wish to alter the initial prestressing forces, check if the corrected total force still falls below an acceptable value, or resize the members to balance out any inaccuracies in the final force calculations.

In addition to the tensegrity constraints, constructability constraints are applied for this example.
The minimum allowable area for steel cables, steel bars, and timber bars is set to $A_{steel\:cable}^{\min}=1$ cm$^2$, $A_{steel\:bar}^{\min}=30$ cm$^2$, and $A_{timber\:bar}^{\min}=30$ cm$^2$.
The maximum allowable area for steel cables, steel bars, and timber bars is set to $A_{steel\:cable}^{\max}= 30$ cm$^2$, $A_{steel\:bar}^{\max}=200$ cm$^2$, and $A_{timber\:bar}^{\max}=200$ cm$^2$.
The maximum number of members allowed at a joint is restricted to be $\eta_p^{\max}=8$.
The minimum angle of separation between members at joint is limited by $\gamma^{\min}=10^{\circ}$.
The nodal complexity constraints are notably more relaxed than in previous examples.
Recall here that the tensegrity constraints ensure that there will only be one compression element at each node, and that nodes will thus primarily connect cables.
The minimum initial prestressing force for all tension and compression members is set to $q^{0_t} = q^{0_c} 5$ kN

\begin{figure}[h!]
    \centering
    \begin{subfigure}[b]{0.4\textwidth}
        \centering
        \includegraphics[width=\linewidth, trim={92cm 15cm 97cm 40cm}, clip]{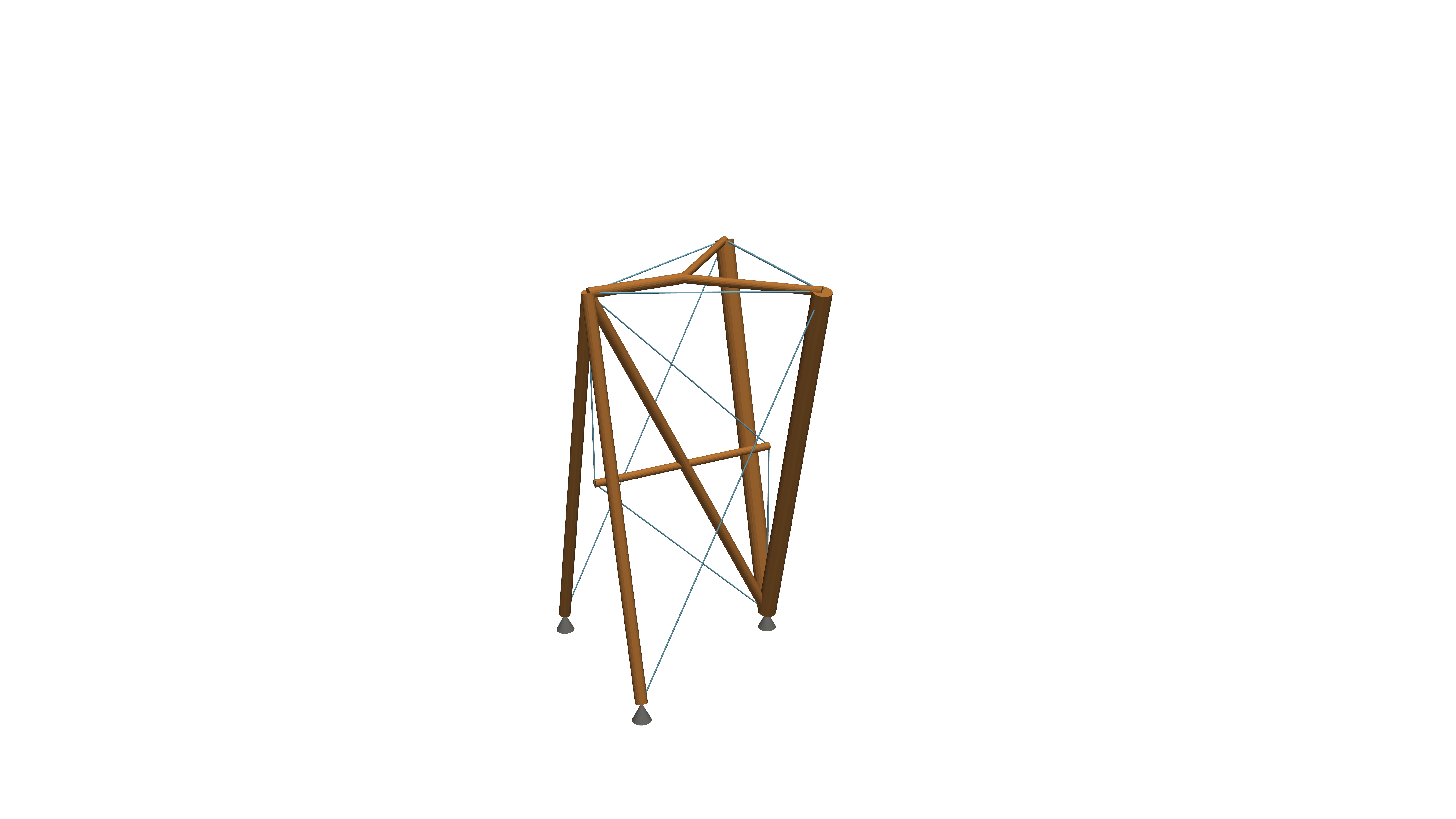}
        \caption{}
        \label{subfig:Result_3_Material_No_Tensegrity}
    \end{subfigure}
    \begin{subfigure}[b]{0.4\textwidth}
        \centering
        \includegraphics[width=\linewidth, trim={92cm 15cm 97cm 40cm}, clip]{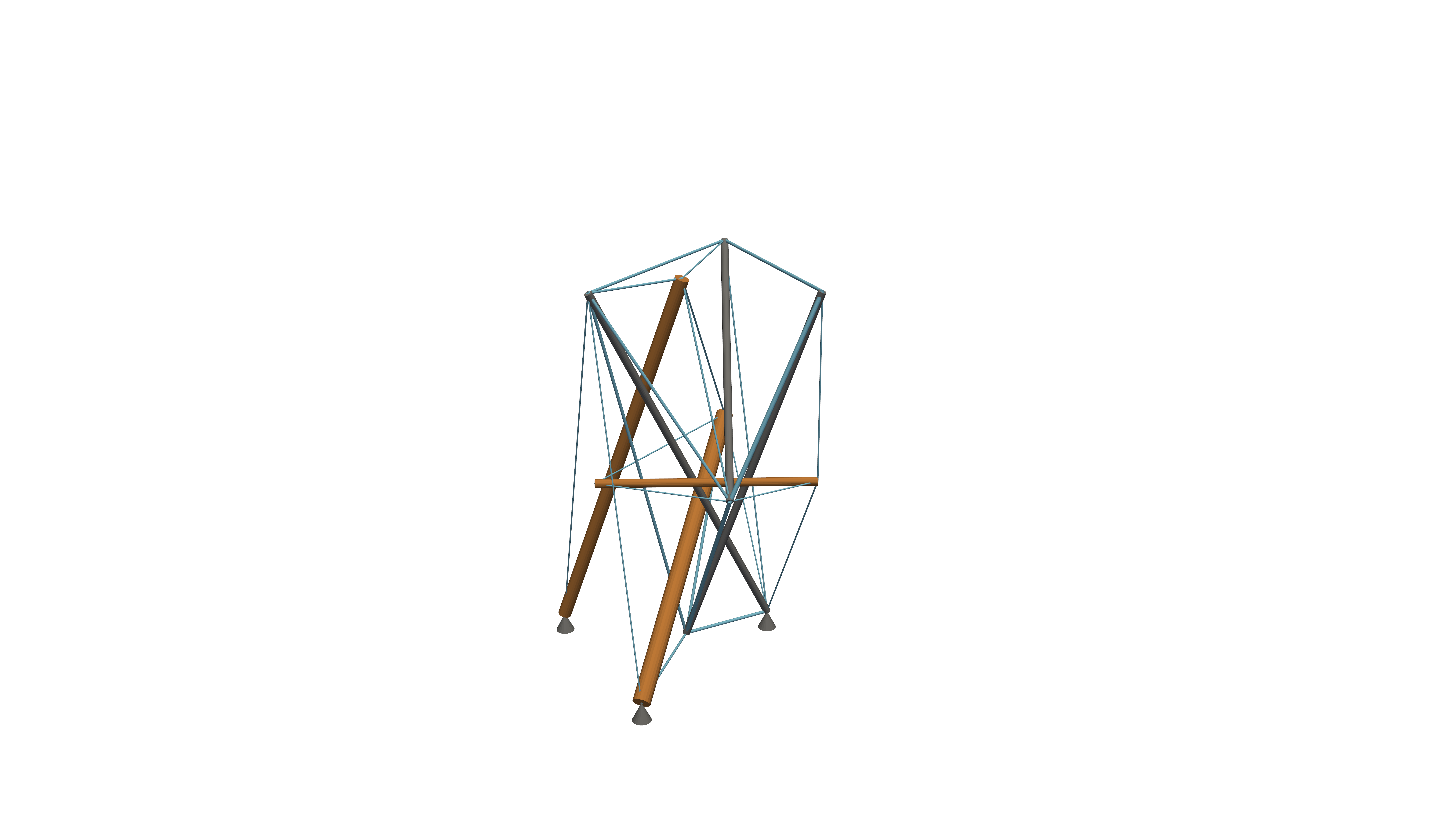}
        \caption{}
        \label{subfig:Result_3_Material_Tensegrity}
    \end{subfigure}
    \caption{Results for three-component designs of (a) a prestressed truss, and (b) a tensegrity structure.
    The color of each element denote the component type where cyan refers to steel cables, gray elements are compressive steel bars, and brown elements are timber.}
    \label{fig:Result_3_Material_Compare}
\end{figure}

To provide a baseline result, a prestressed truss design without the tensegrity constraints is obtained by solving Eq. \eqref{eq:Tensegrity_Full_Formulation} without Eq. \eqref{eq:Tensegrity_Full_Tens_1_Mem} or Eq. \eqref{eq:Tensegrity_Full_Tens_2_Mem}.
The design problem has 3,101 continuous variables and 1,208 binary variables.
The resulting design is shown in Fig. \ref{subfig:Result_3_Material_No_Tensegrity}
and some of the properties of the design are listed in the first column of Tab. \ref{tab:tensegrity_results}.
As seen in the table, the design is obtained in 70 minutes, has an embodied carbon value of 76.0 kg CO$_2$e, and has a mass of 115 kg.
Notice how the structure does not need any steel bars even though they are provided as an option.
Timber bar sections range from 30 cm$^2$ $\leq A_{timber\:bar} \leq$ 181 cm$^2$ and steel cable sections range from 1 cm$^2$ $\leq A_{steel\:cable} \leq$ 2.1 cm$^2$.
The design has nine steel cables and eleven compressive timber elements.
The most complex joint is located directly underneath the vertical load placed at the center of the top edge opposite of the single pin at the base, where four large timber members connect.
The vertical members are only separated by 14.0$^\circ$.
Because the structure is prestressed, all cables are in tension before the application of any loads.

\begin{table}[h!]
    \centering
\begin{tabular}{l c c } \hline
\pmb{Design Type} & \pmb{Prestressed Truss} & \pmb{Tensegrity} \\ \hline
    Runtime                           & 45m 23s & 255m 57s  \\
    Steel Cable Count                 & 9 & 22  \\
    Steel Bar Count                   & 0 & 3  \\
    Timber Bar Count                  & 11 & 3  \\
    Mass (kg)                         & 115 & 333  \\
    Embodied Carbon (kgCO$_2$e) & 76.0 & 447   \\ \hline 
\end{tabular}
    \caption{Performance properties for the prestressed truss and tensegrity designs in Fig. \ref{fig:Result_3_Material_Compare}. }
    \label{tab:tensegrity_results}
\end{table}

\begin{figure}[h!]
    \centering
    \begin{subfigure}[b]{0.31\textwidth}
        \centering
        \includegraphics[width=\linewidth, trim={92cm 15cm 97cm 40cm}, clip]{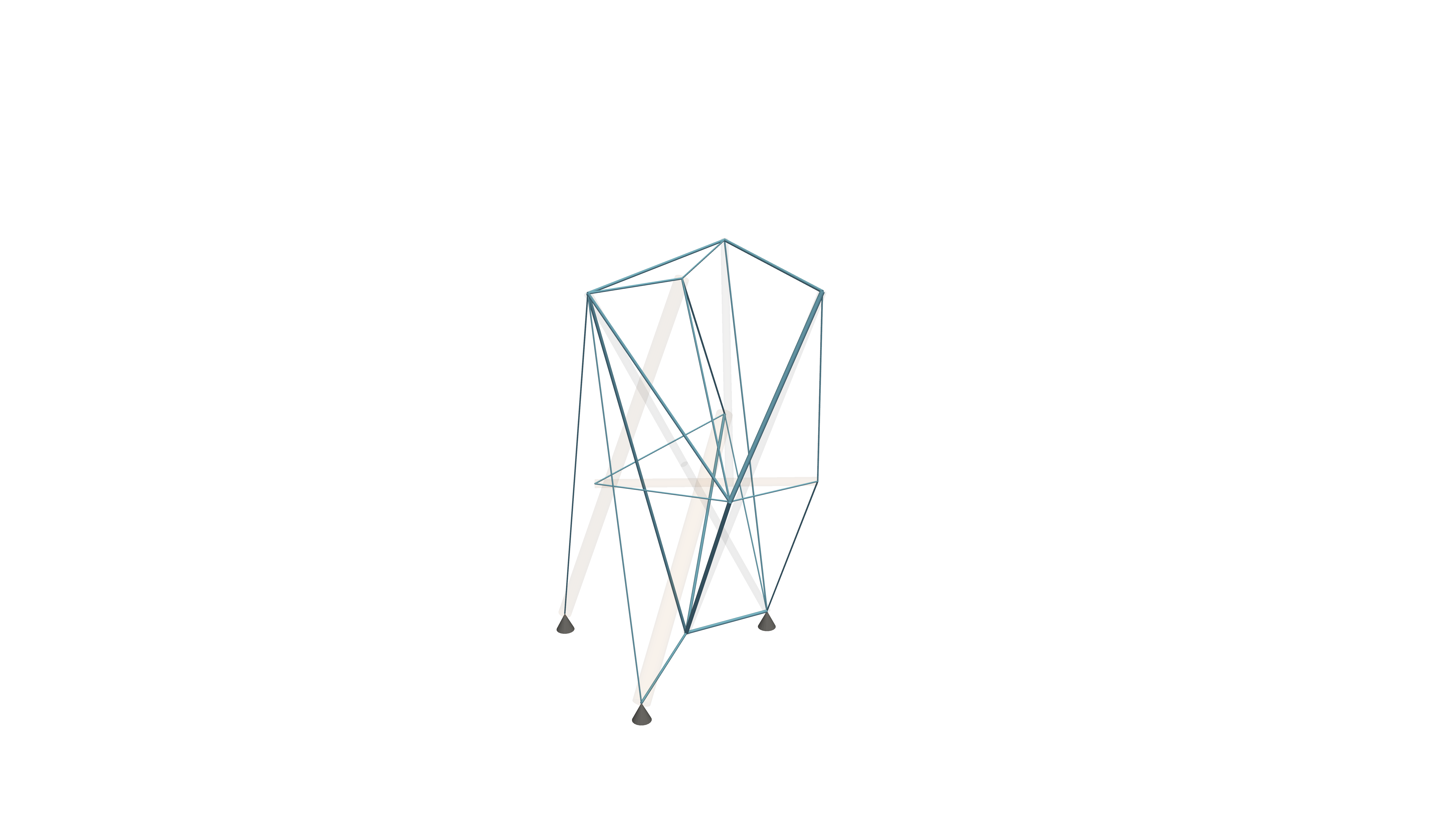}
        \caption{}
        \label{subfig:Result_3_Material_Isometric_Fade_1}
    \end{subfigure}
    \hfill
    \begin{subfigure}[b]{0.31\textwidth}
        \centering
        \includegraphics[width=\linewidth, trim={92cm 15cm 97cm 40cm}, clip]{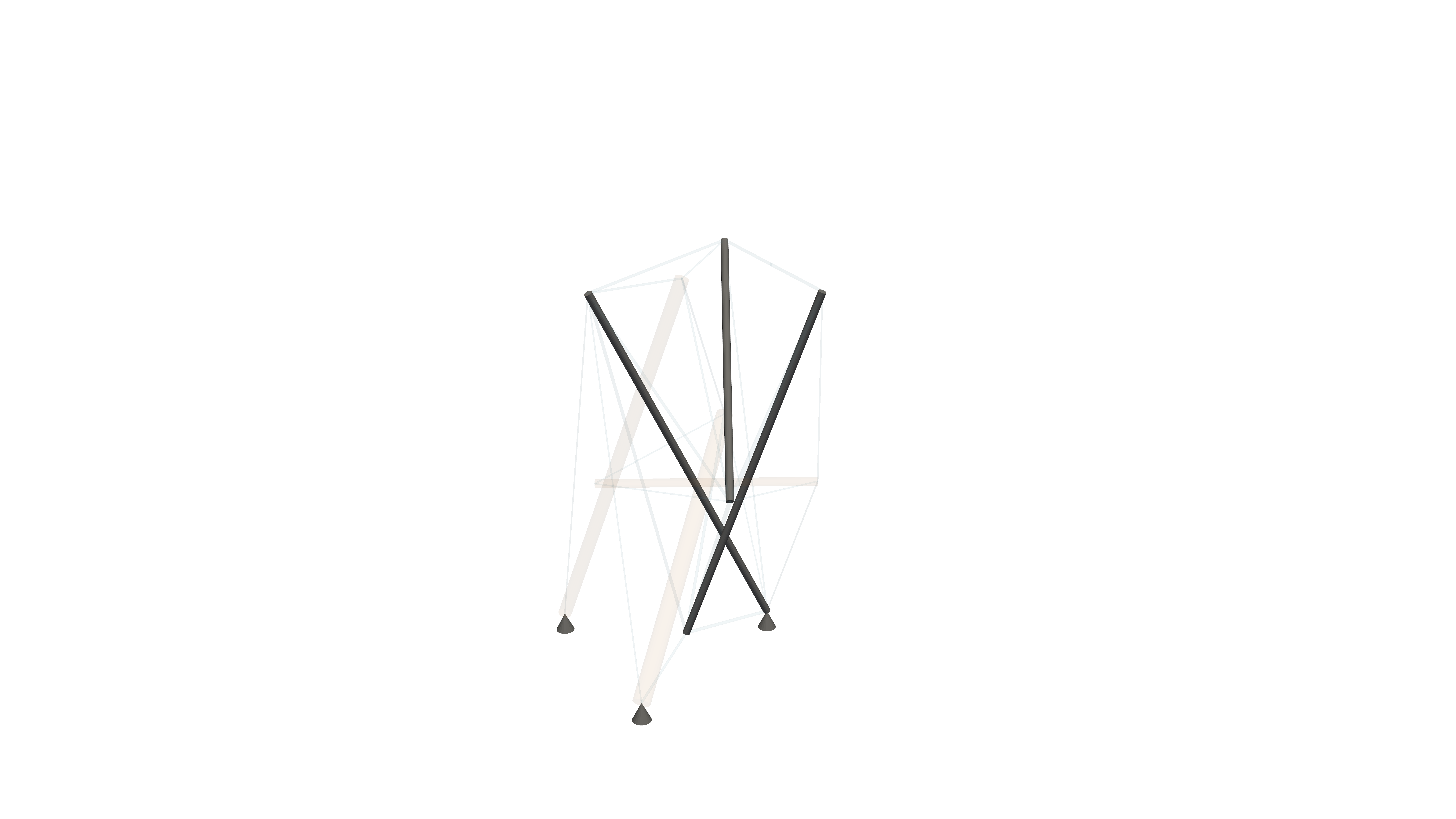}
        \caption{}
        \label{subfig:Result_3_Material_Isometric_Fade_2}
    \end{subfigure}
    \hfill
    \begin{subfigure}[b]{0.31\textwidth}
        \centering
        \includegraphics[width=\linewidth, trim={92cm 15cm 97cm 40cm}, clip]{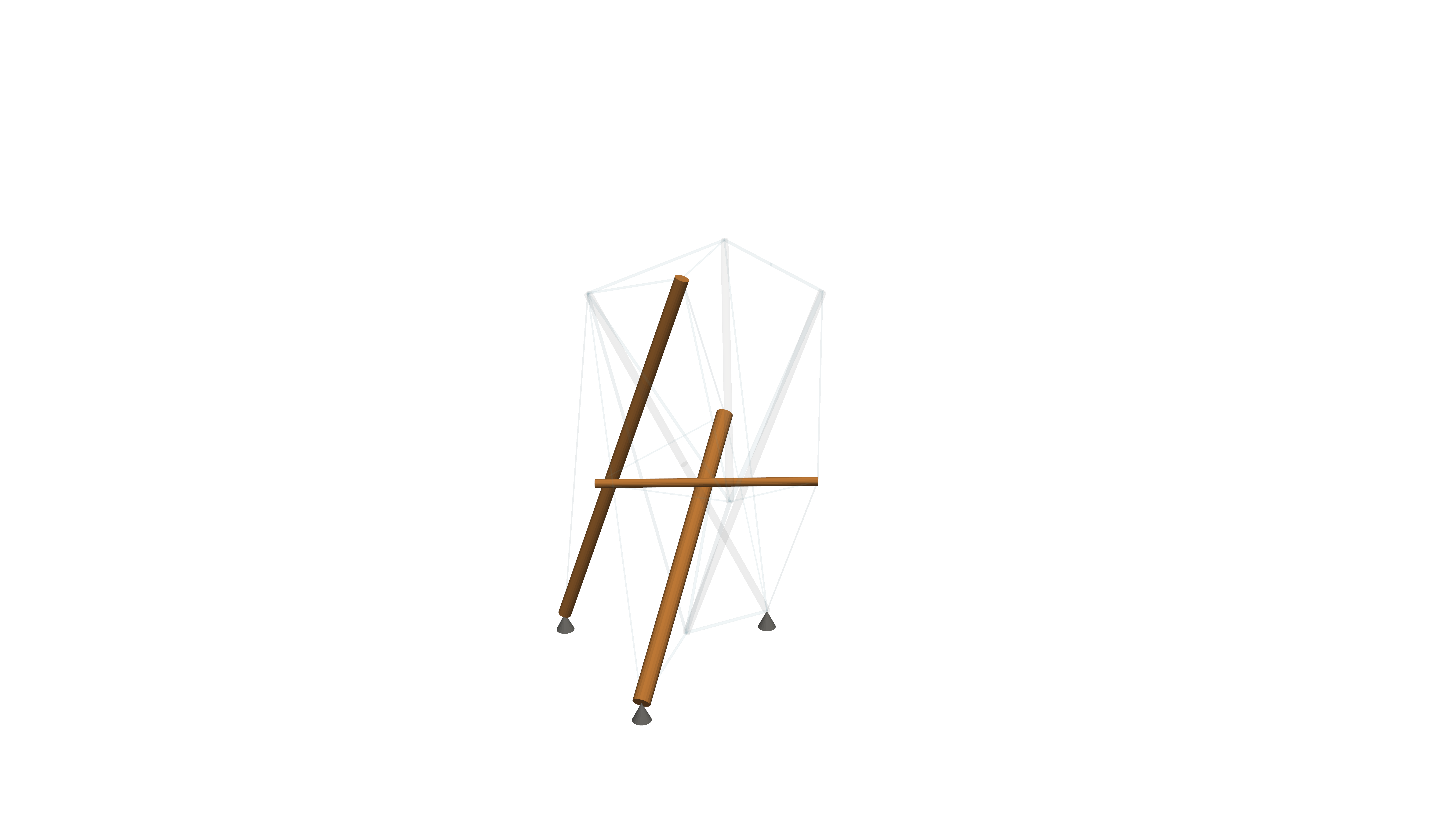}
        \caption{}
        \label{subfig:Result_3_Material_Isometric_Fade_3}
    \end{subfigure}
    \caption{Tensegrity design from Fig. \ref{subfig:Result_3_Material_Tensegrity} that highlights the different component types in the final design, namely; (a) the steel cables, (b) the compressive steel element, and (c) the compressive timber elements.}
    \label{fig:Result_3_Material_Design_Fades}
\end{figure}

When solving the design problem with all constraints in Eq. \eqref{eq:Tensegrity_Full_Formulation}, 
the design in Fig. \ref{subfig:Result_3_Material_Tensegrity} is obtained. 
This problem uses the same number of continuous and binary variables.
The computation time to achieve the presented result is nearly 256 minutes, which is a noticeable increase from the result in Fig. \ref{subfig:Result_3_Material_No_Tensegrity}.
The embodied carbon and mass of the design increase by 488\% and 190\%, respectively.
This aligns with the findings by Nanayakkara et al. \cite{nanayakkara2020simple} that tensegrity structures have less efficient load paths, requiring more material and increasing the embodied carbon.
Notice how the design solution now makes use of all three component options.
Fig. \ref{fig:Result_3_Material_Design_Fades} shows the design where the three different component types are individually highlighted.
In Fig. \ref{subfig:Result_3_Material_Isometric_Fade_1}, the abundant number of small-section steel cables can be seen creating a continuous network of tensile forces tying multiple nodes together.
There are over three times as many tensile cable elements as there are compressive bar elements.
The cables vary in cross-sectional area between 1 cm$^2$ $\leq A_{steel\:cable} \leq$ 10.5 cm$^2$.
Fig. \ref{subfig:Result_3_Material_Isometric_Fade_2} shows that the design only has three long compressive steel bars. 
All of them have a cross-sectional area of 30 cm$^2$.
Finally, Fig. \ref{subfig:Result_3_Material_Isometric_Fade_3} shows the placement of remaining compressive elements, namely the three timber bars.
The timber bars are larger, with cross sectional areas in ranging from 35.5 cm$^2$ $\leq A_{timber\:bar} \leq$ 154 cm$^2$.
When overlaying Fig. \ref{subfig:Result_3_Material_Isometric_Fade_2} and Fig. \ref{subfig:Result_3_Material_Isometric_Fade_3}, it is clear that the compressive elements do not share nodes or intersect.
Thus, the tensegrity condition is fulfilled.
The most complex node joins seven elements with the smallest angle of separation between members in the final design being $12.5^{\circ}$.
This joint is at the center left of the structure, halfway up between the two pins at the base and the load applied on the left side.

\section{Conclusion}
\label{sec:Conclusion}
This contribution presents novel mixed-integer linear programs (MILPs) to design minimum embodied carbon truss and tensegrity structures with multiple candidate materials or component types.
The herein-developed formulation integrates constructability constraints allowing designers to restrict the minimum and maximum cross-sectional area bounds for each material or component type, explicitly limit the number of members connecting at a joint, and ensure all members are separated by a minimum angle.
When designing with multiple materials or components, the designer can also limit the number of a given component type at each joint. 
The design approach is shown capable of respecting all herein-defined constructability constraints for both single, multi-material, and multi-component designs.
The problem formulation is shown to be extendable to other design constraints, such as prestressing and tensegrity characteristics.

As expected, this contribution demonstrates that the interplay between environmental and constructability considerations is complex. 
The constructability constraints are shown to have a project-specific and highly varied effect on the total embodied carbon of truss designs.
For single-material structures, the embodied carbon is increased within the large range of 1--48\% for the case studies considered herein. 
For multi-material trusses (herein shown for timber-steel design), both case studies see an increase in carbon cost of approximately 16--19\% when all constructability constraints are applied.
These numbers are not generalizable as the effect of including constructability considerations depends heavily on the project-specific requirements for compatible section sizes and desired connection design complexity.
When comparing constructible single and multi-material truss design solutions, the case studies herein indicate that general design recommendations cannot be made.
The benefit of introducing a second material has a smaller impact for the 2D Michell truss design example (8.2\%), whereas combining timber and steel elements is shown to improve the environmental performance by over 28\% for the 3D rail bridge design case.

The ability to include constructability constraints using a MILP formulation tends to be associated with an increase in computational time. 
This tendency is also seen herein, where the computation time is shown to be problem-dependent without a rigorous way to predict the required computational effort for a given problem beforehand.
For the truss problems in the current work, the computational time varies from seconds (for unconstrained problems) to hours in the most extremely constrained cases.
The most computationally intensive design case considered in this work is the two-material Michell cantilever subjected to constructability constraints, which requires about 79 hours to identify the global minimum. 
However, the continuous improvement of MILP solvers is an active and ongoing research area that might result in future decreases in the computational expense of using MILP for structural design.

Additionally, this work implements a McCormick relaxation of the constitutive material law.
Doing so enables design capabilities for indeterminate and multi-material structures and keeps the problem formulation linear, even when cross-sectional areas are kept as a continuous design variable.
This does come with two primary associated costs.
First, including compatibility conditions and the constitutive law in the formulation generally increases the computation time when compared to neglecting these conditions and assuming an elastic-perfectly plastic material response.
This is because the problem sees an increase in both variables and constraints.
Second, by applying a linear relaxation of the nonlinear constitutive law, the calculation of structural forces may be inaccurate.
While most inaccuracies are small (below 5\%) and can be accounted for with a factor of safety on the maximum allowable material stress, there are designs where this error can grow unacceptably large.
As seen in the unconstrained timber and timber-steel bridge designs in Section \ref{subsec:Results_4_Constraint_Layers}, large errors tend to occur in highly indeterminate solutions where load paths have a large amount of flexibility.
The McCormick envelope exploits this flexibility to improve the objective function by both overestimating and underestimating the member forces.
Because constructability constraints generally reduce the possibility of highly indeterminate truss designs, they tend to aid the McCormick constraints in producing more accurate member force approximations.
Future work could look into tightening the bounds for these McCormick constraints to improve accuracy while maintaining a balance of preventing overly conservative designs.

The current contribution has some further limitations in that buckling is not included in the design formulation. 
However, buckling- and stability-constrained MILP truss design for single material trusses has previously been suggested \cite{2014_mela,cai2024topology}. 
Although not tested, the existing literature approaches should be compatible with the problem formulation within this contribution. 
Additionally, this work only minimizes the embodied carbon of the structural elements associated with the product stage of the design's life cycle \cite{de2017measuring}.
Other relevant considerations for the environmental impact of a structural design are not included, such as the embodied carbon associated with fabricating connections or with the foundation. 
Possible extensions of the current work can capture these effects using multi-objective optimization that simultaneously minimizes the carbon contributions from the structural elements, the carbon fabrication cost of the total number of connections, and the structural weight that directly influences the material quantities needed for the foundation design.

\newpage
\printbibliography
\end{document}